\documentclass[letter]{emulateapj}

\usepackage{natbib}
\usepackage{amssymb}
\usepackage{amsmath}
\usepackage{color}
\usepackage{enumitem}
\usepackage{graphicx}

\definecolor{myred}{rgb}{0.8,0.0,0.0}

\begin{document}

\title{A Measurement of the Kinetic Sunyaev-Zel'dovich Signal Toward MACS J0717.5+3745}

\author{
    J.~Sayers\altaffilmark{1,2,12}, 
    T.~Mroczkowski\altaffilmark{1,3,4},
    M.~Zemcov\altaffilmark{1,3},
    P.~M.~Korngut\altaffilmark{3},
    J.~Bock\altaffilmark{1,3},
    E.~Bulbul\altaffilmark{5},
    N.~G.~Czakon\altaffilmark{1},
    E.~Egami\altaffilmark{6},
    S.~R.~Golwala\altaffilmark{1},
    P.~M.~Koch\altaffilmark{7},
    K.-Y.~Lin\altaffilmark{7},
    A.~Mantz\altaffilmark{8},
    S.~M.~Molnar\altaffilmark{9},
    L.~Moustakas\altaffilmark{3},
    E.~Pierpaoli\altaffilmark{10},
    T.~D.~Rawle\altaffilmark{6},
    E.~D.~Reese\altaffilmark{11},
    M.~Rex\altaffilmark{6},
    J.~A.~Shitanishi\altaffilmark{10},
    S.~Siegel\altaffilmark{1},
    \& K.~Umetsu\altaffilmark{7}
 }
 \altaffiltext{1}
   {Division of Physics, Math, and Astronomy, California Institute of Technology, 
     1200 East California Blvd, Pasadena, CA 91125}
 \altaffiltext{2}
   {Norris Foundation CCAT Postdoctoral Fellow}
 \altaffiltext{3}
   {Jet Propulsion Laboratory, 4800 Oak Grove Drive, Pasadena, CA 91109}
 \altaffiltext{4}
   {NASA Einstein Postdoctoral Fellow}
 \altaffiltext{5}
   {Harvard-Smithsonian Center for Astrophysics, 60 Garden Street, Cambridge, MA 02138}
 \altaffiltext{6}
   {Steward Observatory, University of Arizona, 933 North Cherry Avenue,
     Tucson, AZ 85721}
 \altaffiltext{7}
   {Institute of Astronomy and Astrophysics, Academia Sinica, 
     P.O. Box 23-141, Taipei 10617, Taiwan}
 \altaffiltext{8}
   {Kavli Institute for Cosmological Physics, University of Chicago, 5640 South Ellis Avenue, Chicago, IL 60637}
 \altaffiltext{9}
   {LeCosPA Center, National Taiwan University, 
     Taipei 10617, Taiwan}
 \altaffiltext{10}
   {Department of Physics and Astronomy, University of Southern California, 
     3620 McClintock Avenue, Los Angeles, CA 90089}
 \altaffiltext{11}
   {Department of Physics and Astronomy, University of Pennsylvania, 
     209 South 33rd Street, Philadelphia, PA 19104}
 \altaffiltext{12}
   {jack@caltech.edu}

\begin{abstract}

  We report our analysis of MACS J0717.5+3745
  using 140 and 268~GHz Bolocam data collected at
  the Caltech Submillimeter Observatory.
  We detect extended Sunyaev-Zel'dovich (SZ) effect signal at high significance 
  in both Bolocam bands, and we employ {\it Herschel}-SPIRE
  observations to subtract the signal from dusty
  background galaxies in the 268~GHz data.
  We constrain the two-band SZ surface brightness
  toward two of the sub-clusters of MACS J0717.5+3745:
  the main sub-cluster (named C), and a sub-cluster
  identified in spectroscopic optical data to have
  a line-of-sight velocity of $+3200$~km s$^{-1}$ (named B).
  We determine the surface brightness in two separate ways:
  via fits of parametric models and via direct
  integration of the images.
  For both sub-clusters, we find consistent 
  surface brightnesses from both analysis methods.
  We constrain spectral templates 
  consisting of relativistically corrected thermal and kinetic SZ signals,
  using a jointly-derived electron temperature from {\it Chandra} and {\it XMM-Newton}
  under the assumption that each sub-cluster is isothermal.
  The data show no evidence for a kinetic SZ signal toward sub-cluster C,
  but they do indicate a significant kinetic SZ signal toward sub-cluster B.
  The model-derived surface brightnesses for sub-cluster B
  yield a best-fit line-of-sight velocity of {$v_z = +3450 \pm 900$}~km s$^{-1}$, with 
  $(1 - \textrm{Prob}[v_z \ge 0]) = 1.3 \times 10^{-5}$ 
  ($4.2\sigma$ away from 0 for a Gaussian distribution).
  The directly integrated sub-cluster B SZ surface
  brightnesses provide a best-fit {$v_z = +2550 \pm 1050$}~km s$^{-1}$, with
  $(1 - \textrm{Prob}[v_z \ge 0]) = 2.2 \times 10^{-3}$ ($2.9\sigma$).

\end{abstract}
\keywords{
galaxies: clusters: intracluster medium ---
galaxies: clusters: individual: (MACS J0717.5+3745)
}

\section{Introduction}

  Measurements of large-scale peculiar velocities provide
  a direct probe of cosmological models and can be used to place
  constraints on parameters that are highly degenerate
  and/or unconstrained via other cosmological probes,
  such as measurements of primary CMB fluctuations 
  \citep{bennet12, hinshaw12, planck13_cmb1, planck13_cmb2}
  and supernovae distance measurements 
  \citep{conley11, suzuki12}.
  Specifically, these peculiar velocities depend on the properties
  and distributions of large-scale structure, along with
  the characteristics of dark energy and the behavior
  of gravity on the corresponding length scales.
  Consequently, peculiar velocity measurements
  for large numbers of objects can
  probe the redshift evolution of the properties
  of dark energy \citep{bhattacharya08} and also
  distinguish between dark energy and modified gravity
  models \citep{kosowsky09}.
  In addition, measurements of an extremely large peculiar velocity
  for a single object (e.g., 1E0657-56, also
  known as the bullet cluster) can be used 
  to directly test the validity of standard cosmological models
  \citep{hayashi06, lee10, thompson12}.

  In the local universe, line-of-sight peculiar velocities can be 
  measured using a combination of spectroscopy
  and distance measurements via the extragalactic
  distance ladder, generally using the relation
  described by \citet{tully77}.
  Such measurements have been used to constrain
  cosmological parameters like the total matter
  density $\Omega_m$ and the normalization of 
  density fluctuations $\sigma_8$, generally finding
  good agreement with other cosmological probes
  (e.g., \citealt{feldman10, nusser11, ma12}).
  Unfortunately, uncertainties in the extragalactic
  distance ladder are proportional to distance,
  therefore preventing the application of these
  methods outside the local universe.
  In contrast, the kinetic Sunyaev-Zel'dovich (SZ)
  effect provides a direct measurement of the
  line-of-sight peculiar velocity of the distribution of
  hot electrons within galaxy clusters 
  \citep[][See Section~\ref{sec:sz}]{sunyaev72}.
  In addition, the surface brightness of the kinetic
  SZ signal is independent of redshift, depending only
  on the electron optical depth and line-of-sight
  peculiar velocity.
  Consequently, many groups have performed detailed studies
  of the cosmological constraints that would be possible
  with large-scale peculiar velocity surveys using the 
  kinetic SZ signal
  (e.g., \citealt{bhattacharya08, kosowsky09, mak11}).

  Despite the great promise of kinetic SZ surveys, 
  measurements of the kinetic SZ 
  signal have proven to be a significant observational
  challenge.
  Over the past two decades, several attempts have been
  made to detect the kinetic SZ signal toward
  a variety of individual massive clusters.
  These observational efforts have used a range of instrumentation,
  including: the dedicated multi-band
  photometer SuZIE and its successors \citep{holzapfel97, benson03}, 
  multi-band data collected from a range of facilities
  \citep{kitayama04},
  the moderate resolution spectroscopic receiver Z-Spec
  \citep{zemcov12},
  and the two-band photometric imaging camera Bolocam
  \citep{mauskopf12, mroczkowski12}.
  None of these observations have made a high-significance detection
  of the kinetic SZ signal, and the derived uncertainties
  on the line-of-sight peculiar velocities have
  not improved significantly from the first
  measurements with SuZIE, at least in part because none
  of the subsequent measurements have used instrumentation
  specifically designed to detect the kinetic SZ signal.

  Recently, data from the {\it WMAP} and {\it Planck}
  satellites have been used to place upper limits
  on bulk flows and rms variations in peculiar 
  velocities via the kinetic SZ signal
  \citep{osborne11, planck13_kSZ}.
  In addition, \citet{hand12} used a combination of 
  Atacama Cosmology Telescope (ACT) and Sloan
  Digital Sky Survey III data to constrain the 
  mean pairwise momentum of clusters using a kinetic
  SZ signature that is inconsistent with noise
  at a confidence level of 99.8\%.
  Furthermore, upper limits on the kinetic SZ 
  power spectrum measured by the South Pole Telescope
  (SPT) have been used to inform cosmological
  simulations and to place constraints on the 
  reionization history of the universe
  \citep{reichardt12, zahn12}.

  One of the strongest hints of a kinetic SZ detection
  was presented in \citet[][hereafter M12]{mroczkowski12}
  toward the massive cluster MACS J0717.5+3745
  using Bolocam measurements
  at 140 and 268~GHz.
  Motivated by this result, we have collected a significant
  amount of additional 268~GHz Bolocam data
  toward this cluster.
  The results
  we obtain using this additional, deeper data are presented in
  this manuscript, which is organized as follows.
  In {Section~\ref{sec:sz} we present the SZ effect and
  in Section~\ref{sec:m0717} we describe previous analyses
  of MACS J0717.5+3745.
  In Section~\ref{sec:reduction} we provide the details of our data reduction.}
  We describe our model of the SZ signal toward MACS J0717.5+3745
  in Section~\ref{sec:model} and provide the corresponding
  constraints on the two-band SZ surface brightnesses of 
  the cluster in Sections~\ref{sec:b} and \ref{sec:c}.
  In Section~\ref{sec:vpec} we give the line-of-sight
  peculiar velocity constraints derived from these
  surface brightnesses, in Section~\ref{sec:discussion}
  we put these results in a broader context,
  and in Section~\ref{sec:summary} we briefly summarize
  our analysis. 
  We also include an Appendix which fully details
  our treatment of the cosmic infrared background (CIB)
  in the 268~GHz data.

\section{The SZ Effect}
  \label{sec:sz}

  {When a massive galaxy cluster is moving with respect
  to the rest frame of the CMB, the Doppler-induced 
  spectral distortion of the CMB due to the bulk motion
  of the electrons in the intra-cluster medium (ICM)
  is described by the kinetic SZ effect
  (e.g., \citealt{sunyaev72, birkinshaw99, carlstrom02}).
  The change in CMB temperature due to the kinetic SZ
  effect is given by
  \begin{equation}
    \frac{\Delta T_{CMB}}{T_{CMB}} = - \frac{v_z}{c} \tau_e,
  \end{equation}
  where $v_z$ is the ICM peculiar velocity along the line-of-sight,
  $c$ is the speed of light, and
  $\tau_e$ is the total electron optical depth
  \begin{equation}
    \tau_e = \int n_e \sigma_T dl,
    \label{eq:tau}
  \end{equation}
  for an electron density $n_e$ integrated along the line
  of sight $dl$ ($\sigma_T$ is the Thompson cross section).
  We note that a positive peculiar velocity results in 
  a negative temperature change,
  under the convention that a Doppler shift toward higher
  redshift corresponds to a positive value of $v_z$.
  In addition, we note that there are small 
  relativistic corrections to the kinetic SZ
  signal (e.g., \citealt{nozawa98b, sazonov98, nozawa06, chluba12}).}

  {There is also a thermal SZ effect,
  which describes the Compton scattering of CMB photons off of high
  energy electrons in the ICM of massive clusters
  (e.g., \citealt{sunyaev72, rephaeli95, birkinshaw99, carlstrom02}).
  Specifically, the change in CMB temperature due to the
  thermal SZ effect is given by
  \begin{equation}
    \frac{\Delta T_{CMB}}{T_{CMB}} = f(\nu, T_e) y,
  \end{equation}
  where $f(\nu,T_e)$ encodes the frequency $\nu$ dependence,
  including relativistic corrections that depend
  on the electron temperature $T_e$ (e.g., 
  \citealt{rephaeli95b, itoh98, nozawa98, itoh04, chluba12})
  and
  \begin{equation}
    y = \int n_e \sigma_T \frac{k_B T_e}{m_e c^2} dl
    \label{eq:szy}
  \end{equation}
  ($k_B$ is Boltzmann's constant and $m_e$ is the electron mass).
  In the limit of an isothermal distribution, $y$ is 
  directly and linearly proportional
  to the total electron optical depth $\tau_e$.}

\section{Previous Analyses of MACS J0717.5+3745}
  \label{sec:m0717}

  MACS J0717.5+3745, located at $z = 0.55$,
  was discovered as part of the Massive Cluster Survey 
  \citep[MACS,][]{ebeling01, ebeling07},
  and is extremely massive and dynamically disturbed.
  As such, it has been the focus of many studies at a range of 
  wavelengths, and it has been chosen as part of the six-cluster
  Hubble Space Telescope Frontier Fields program.\footnote{
    http://www.stsci.edu/hst/campaigns/frontier-fields/}
  Radio observations have shown that MACS J0717.5+3745 hosts
  the most powerful radio halo known 
  \citep{edge03, vanweeren09, bonafede09}, and strong lensing
  data have shown that MACS J0717.5+3745 has the largest known
  Einstein radius
  \citep{zitrin09, meneghetti11, waizmann12}.
  From both the galaxy distribution \citep{ebeling04},
  and weak lensing studies \citep{jauzac12, medezinski13},
  MACS J0717.5+3745 also appears to be part of a large, extended
  filamentary structure.
  In addition, it has the highest X-ray temperature
  among all of the clusters in the MACS catalog
  \citep{ebeling07}.

  \begin{figure}
    \includegraphics[width=\columnwidth]{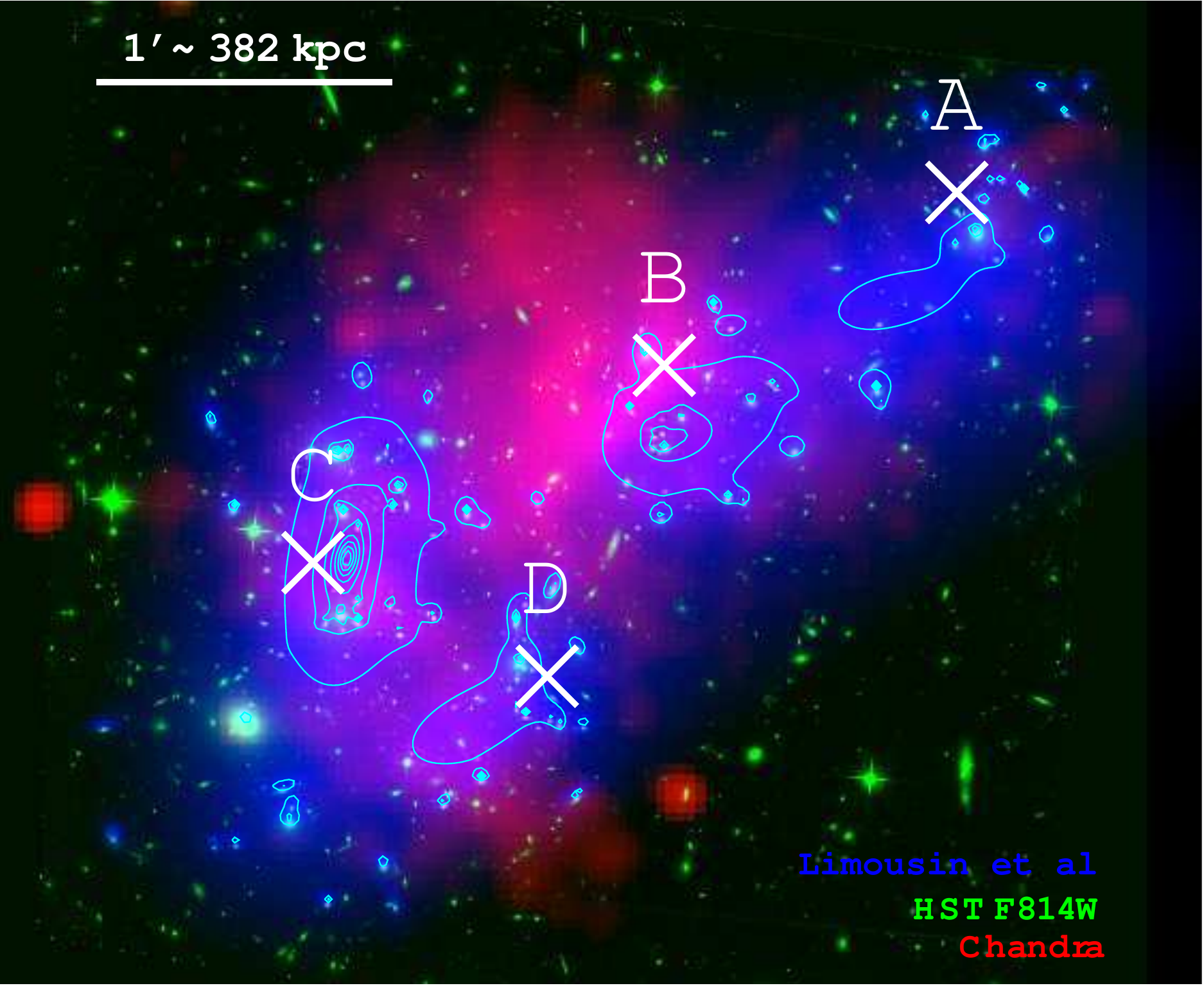}
    \caption{False-color composite image of MACS J0717.5+3745
     with the lensing results of \citet{limousin12} in blue,
     the Hubble Space Telescope image using the F814W filter
     in green, and the {\it Chandra} X-ray image in red.
     The blue contours show the \citet{limousin12} result
     on a linear scale, and clearly indicate the four
     sub-clusters labeled A through D,
     with white Xs marking the sub-cluster positions determined
     by \citet{ma09} from the galaxy distribution.}
    \label{fig:composite}
  \end{figure}

  \citet{ma09} performed
  a joint analysis using X-ray data, along with
  the measured galaxy positions and redshifts,
  and identified four distinct sub-clusters in
  MACS J0717.5+3745, from N to S labeled as A, B, C, and D
  (see Figure~\ref{fig:composite}).
  An independent strong lensing analysis described
  in \citet{limousin12} also identified four
  sub-clusters, with similar positions to
  the ones given in \citet{ma09}.
  Both analyses found sub-cluster C to be the most
  massive system,
  and \citet{ma09} determined that sub-cluster C
  is probably the highly disturbed core of the main system.
  Sub-clusters B and D are assumed to be relatively
  intact cores of systems that are merging along
  a direction close to the line-of-sight.
  In particular, sub-cluster B is coincident with
  an X-ray temperature that is colder than the surrounding
  regions, indicating that its core has not been
  highly disrupted by the merger.
  From the spectroscopic data, \citet{ma09}
  found that sub-cluster B has a line-of-sight velocity
  that differs from the other components by approximately
  3000~km s$^{-1}$.
  Further indications of this large line-of-sight velocity
  for sub-cluster B were presented in M12, who found
  a similar best-fit velocity by
  using X-ray and SZ measurements to constrain the
  kinetic SZ signal toward that sub-cluster,
  although the statistical significance of their kinetic SZ constraint
  on the velocity is modest ($\simeq 2\sigma$).
  This wide range of observational data toward MACS J0717.5+3745
  is therefore converging to what appears to be a 
  coherent picture of this complex system.

\section{Data Reduction}
 \label{sec:reduction}

 \subsection{Bolocam}
  \label{sec:bolocam}

  We observed MACS J0717.5+3745 with Bolocam from the 
  Caltech Submillimeter Observatory (CSO) for a total of 12.5 hours
  at 140~GHz and for a total of 27.3 hours at 268~GHz,
  where the effective band centers are quoted for a CMB spectrum.
  Compared to the previous Bolocam
  analysis presented in M12,
  this represents an additional 19.3 hours of data collected 
  at 268~GHz in 2012 December.
  In contrast to the original 8.0 hours of 268~GHz integration
  used in M12, much
  of which was collected in poor observing conditions
  with a 225~GHz optical depth $\tau_{225} > 0.10$, most of the
  additional 19.3~hours of 268~GHz integration was obtained
  with $\tau_{225} \simeq 0.05$.
  This additional data was therefore collected during
  the lowest opacity conditions generally
  available from the CSO.

  The Bolocam instrument has an $8\arcmin$ diameter circular 
  field of view (FOV), and point-spread
  functions (PSFs) {that are approximately Gaussian}
  with full-widths at half-maximums (FWHMs) 
  equal to $58\arcsec$ and $31\arcsec$ at 140 and 268~GHz, respectively
  \citep{glenn02, haig04}.
  All of our Bolocam observations of MACS J0717.5+3745
  involved scanning the CSO in a Lissajous pattern
  with an RMS velocity of approximately $4\arcmin$/sec.
  The details of our data reduction are given elsewhere
  \citep[][M12]{sayers11}, and we briefly summarize
  our procedure below.

  First, we obtain pointing corrections accurate to $5\arcsec$
  using frequent observations of nearby quasars, and we
  obtain an absolute flux calibration accurate to 5\% and 10\%
  at 140 and 268~GHz, respectively,
  using observations of Uranus and Neptune
  \citep{griffin93, sayers12_planet}.
  {We note that \citet{hasselfield13_planet} recently determined the
  brightness temperature of Uranus to be $106.7 \pm 2.2$~K
  at 149~GHz using ACT
  observations calibrated against
  the primary CMB anisotropies measured by the
  {\it WMAP} satellite.
  Also, \citet{planck13_calibration} recently determined the
  brightness temperature of Uranus to be $108.4 \pm 2.9$~K
  at 143~GHz based on {\it Planck} data.
  Our calibration model assumes a brightness temperature
  of $106.6 \pm 3.5$~K for the 140~GHz Bolocam bandpass,
  which was measured in \citet{sayers12_planet}
  by extrapolating the {\it WMAP}
  94~GHz brightness measurements presented in 
  \citet{weiland11} using the model of \citet{griffin93}.
  This model predicts the brightness
  temperature of Uranus to increase with decreasing
  frequency. As a result, the ACT and {\it Planck} measurements
  imply a best-fit 140~GHz brightness temperature
  that is approximately 2.5~K higher than our
  assumed value of 106.6~K.
  However, this difference is comparable to the
  ACT and {\it Planck} measurement uncertainties,
  and it is well below our estimated 5\%
  flux calibration uncertainty at 140~GHz.
  We therefore have not updated our calibration model.
  Furthermore, we note that the accuracy of the ACT and {\it Planck}
  Uranus brightness temperatures is $2-3$\%, which is only slightly smaller 
  than the 3.3\% accuracy of our assumed
  140~GHz brightness temperature.
  Furthermore, our 140~GHz calibration uncertainty receives an approximately
  equal contribution from our 3.1\% beam solid angle uncertainty.
  Revising our flux calibration using ACT and {\it Planck} would
  thus not have a significant effect on our overall calibration
  uncertainty, which itself is already sub-dominant to measurement
  uncertainties (see Table~\ref{tab:sz_brightness}).
  Finally, we note that our 10\% flux calibration at 268~GHz is limited
  largely by atmospheric fluctuations, and therefore a more accurate
  Uranus brightness temperature at that frequency would have no effect
  on our overall calibration uncertainty.}

  To remove atmospheric fluctuations from the data, we
  first subtract a template of the common mode signal
  over the FOV, and we then high-pass filter (HPF)
  the time-stream data at 250 and 500~mHz at 140 and 268~GHz, respectively.
  The large amplitude of the atmospheric fluctuations
  in the 268~GHz data necessitates this more aggressive
  HPF, and this filtering represents a 
  slight change from the M12 analysis, which
  used a 250 mHz HPF for both datasets.
  {We used a scan speed of $\simeq 4\arcmin$/sec
  for our observations, and the
  HPFs at 250 or 500~mHz therefore
  correspond to angular scales of $16\arcmin$ and
  $8\arcmin$, respectively.
  Consequently, the maximum angular scale preserved
  by our filtering
  is largely set by the common mode subtraction
  over Bolocam's $8\arcmin$ FOV.
  Because our processing removes astronomical signals 
  with angular sizes larger than the $8\arcmin$ FOV,
  we determine the map-space transfer function 
  at each wavelength by reverse-mapping and processing
  an image template through the entire reduction pipeline.}
  We estimate the instrumental and atmospheric noise
  in our images by forming 1000 separate jackknife
  realizations of the data, where a randomly selected
  subset of half the {single observations is multiplied by $-1$
  to remove all astronomical signals.
  There are 75 single 10-minute observations of MACS J0717.5+3745 at
  140 GHz and 164 single 10-minute observations of MACS J0717.5+3745
  at 268 GHz.}

  To account for noise from unwanted astronomical signals,
  we first add a different random realization of the
  primary cosmic microwave background (CMB) 
  fluctuations, using the power spectrum measurements 
  from the South Pole Telescope (SPT), to each jackknife 
  \citep{reichardt12, story12}.
  At 140 GHz, we add an additional random realization of the CIB, 
  again based on the measured SPT power spectra \citep{hall10},
  {under the assumption that the fluctuations are
  Gaussian. This assumption is not strictly true, but 
  the CIB fluctuations are more than an order of magnitude dimmer
  than the other noise fluctuations in the data, and therefore
  a breakdown of this assumption is not likely
  to have a noticeable effect on our results.}
  The CIB is significantly brighter at 268~GHz, and we therefore use
  a much more detailed model to account for it in those data,
  as described in the Appendix.
  Throughout this manuscript we refer to these 1000 jackknife plus
  astronomical noise realizations as ``noise realizations''.

  For the analyses described in this manuscript, we make use of
  the Bolocam data in two different ways.
  We use images of the processed data, which cover a maximum
  size of $14\arcmin \times 14\arcmin$, to constrain parametric
  models of the astronomical signals (see Section~\ref{sec:model}).
  This analysis involves convolving the model with the 
  signal transfer function of the data processing
  and the Bolocam PSF.
  To determine best-fit parameters for a given model,
  we use the generalized least squares fitting algorithm
  MPFITFUN \citep{markwardt09} under the simplifying assumption
  that the map noise covariance matrix is diagonal.
  We have demonstrated that this fitting method produces
  unbiased estimates of the best-fit parameter values \citep{sayers11}, 
  although in some cases it does produce
  a slightly biased estimate of the uncertainties on these best-fit parameters.
  Therefore, to fully account for all of the subtleties of our noise,
  we derive all of the parameter uncertainties
  via the spread of best-fit values we obtain from 
  applying the same fitting algorithm to a 
  sample of 1000 noise plus signal realizations.
  Each noise plus signal realization is generated
  by adding a noise realization to
  the best-fit model found for the real data.

  We also deconvolve the transfer function of the data processing
  to obtain unbiased images after first reducing the
  image to a maximum size of $10\arcmin \times 10\arcmin$ to prevent
  significant amplification of the noise on the 
  largest angular scales.
  One subtlety in this process is the fact that the 
  signal transfer function is equal to 0 at an 
  angular wavenumber of 0 (i.e., the DC signal level
  of the image is unconstrained).
  We therefore use the parametric model fits to constrain
  the DC signal level, as described in Section~\ref{sec:photometry}.
  {As a result, the deconvolved images
  have some model dependence.
  Consequently, to ensure that the uncertainties on the model
  accurately represent the underlying uncertainties on the
  data, both for the model fits alone and for the results
  derived from the deconvolved images, the model must 
  provide an acceptable fit quality.
  By requiring a model with an acceptable fit quality,
  we also ensure that the results derived from 
  model fits will be consistent with those derived
  from the deconvolved images.}
  To estimate the noise in the deconvolved images, we
  also deconvolve the transfer function from each
  of the 1000 noise realizations.

 \subsection{{\it Chandra}}
  \label{sec:Chandra}

  \begin{figure*}
    \centering
    \includegraphics[width=.7\textwidth]{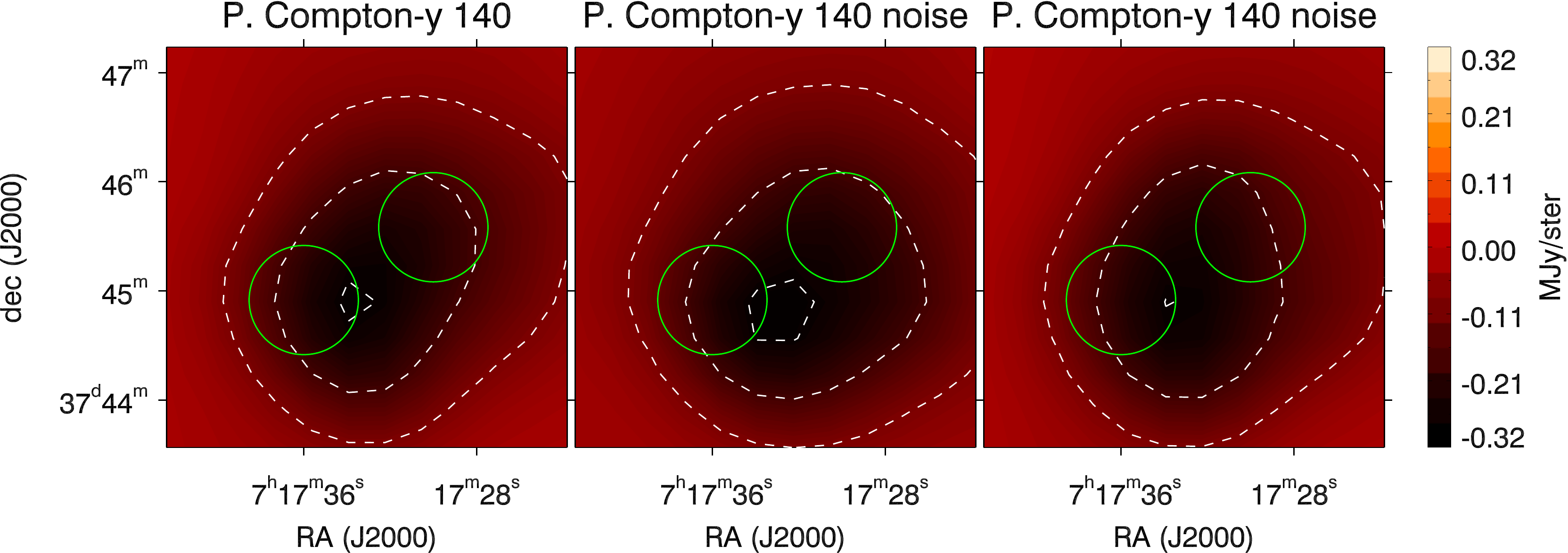}

    \vspace{15pt}

    \centering
    \includegraphics[width=.7\textwidth]{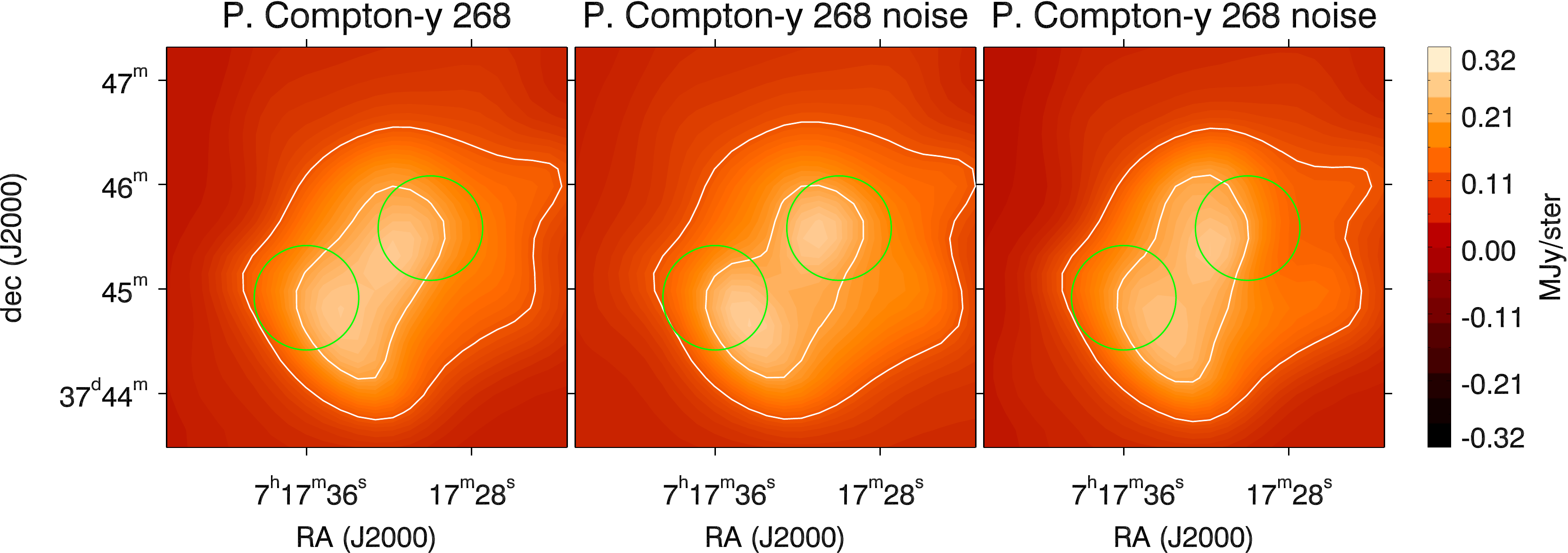}
    \caption{Thumbnails of the pseudo Compton-$y$ maps
    we derive from the {\it Chandra} X-ray data.
    From left to right, the thumbnails show the
    best-fit map and two realizations based on
    the X-ray measurement uncertainties.
    The top row shows the maps at 140~GHz, and the
    bottom row shows the maps at 268~GHz.
    In both cases, the maps are convolved with the
    Bolocam PSF at the respective frequency.
    The white contours are spaced by 0.10~MJy sr$^{-1}$,
    with solid representing positive values and
    dashed representing negative values.
    The $1\arcmin$ diameter apertures centered on
    sub-cluster C (lower left) and sub-cluster B (upper right)
    are shown in green.
    Note that the approximate conversion factor from MJy sr$^{-1}$
    to $y$ is $- 1 \times 10^{-3}$ at 140~GHz and
    $+ 1 \times 10^{-3}$ at 268~GHz.}
    \label{fig:pseudoy}
  \end{figure*}

  Our analysis of the {\it Chandra} X-ray exposures of MACS J0717.5+3745
  is nearly identical to the analysis described in M12, and we briefly
  summarize the main aspects below.
  As in M12, we utilize both
  {\it Chandra} ACIS-I X-ray observations of MACS~J0717.5+3745 
  (Obs IDs 1655 and 4200), for a total exposure time of 81~ksec
  (see \citet{reese10} for the reduction details).
  From these X-ray data we compute pseudo-pressure 
  \begin{equation}
    P_e = n_e k_B T_e \simeq \sqrt{\frac{4 \pi (1+z)^3 S_X}{l \Lambda_{ee}(T_e,Z)}} k_B T_e,
    \label{eq:pseudoPressure}
  \end{equation}
  where $S_X$ is the X-ray surface brightness
  \begin{equation}
  \label{eq:xray_sb}
    S_X = \frac{1}{4\pi (1+z)^3} \! \int \!\! n_e^2 \Lambda_{ee}(T_e,Z) \,dl,
  \end{equation}
  $l$ is the effective line-of-sight extent of the ICM,
  and $\Lambda_{ee}(T_e,Z)$ is the X-ray emissivity
  as a function of $T_e$ and metallicity $Z$.
  To generate pseudo-pressure maps from the {\it Chandra} images,
  we first bin the data using {\it contbin}
  \citep{Sanders2006}.
  We construct the pseudo-pressure maps from $T_e$ maps generated
  by computing $T_e$ within each bin and $n_e$ maps
  computed from the X-ray surface brightness
  (see Equation~\ref{eq:pseudoPressure}).
  {To rescale the pseudo-pressure maps to units
  of Compton-$y$, we need to determine the value of $l$
  (which in M12 was done using 31~GHz SZA data, but in practice
  is left as a free parameter in all of our fits).}
  This X-ray template for the thermal SZ signal, which is simply a 
  rescaling of the X-ray pseudo-pressure map, 
  is called a ``pseudo Compton-$y$ map'' throughout this manuscript.
  For consistency with M12, we employ the same pseudo Compton-$y$ map
  that was generated for that analysis.
  We note that this map was generated using CIAO version 4.3 and 
  calibration database (CALDB) version 4.4.5 \citep{Fruscione2006}. 

  New for this analysis compared to M12, we
  also generate 20 realizations of the 
  pseudo Compton-$y$ map that are fluctuated
  by the X-ray measurement uncertainties on $T_e$,
  which dominate the uncertainty of the pseudo-pressure maps
  (see Figure~\ref{fig:pseudoy}).
  We note that, in addition to measurement uncertainties, the pseudo Compton-$y$
  maps are also subject to possible systematic errors due
  to gas clumping within the ICM.
  Clumping is defined as
  \begin{equation}
	C = \frac{\langle n_e^2 \rangle}{\langle n_e \rangle^2},
	\label{eq:clumping}	
  \end{equation}
  and from the equations listed above, we see that the 
  pseudo-Compton-$y$ maps are sensitive to $\sqrt{C}$.
  Typical clumping factors of $C \simeq 1.1-1.2$ within $R_{500}$ are expected from simulations 
  (e.g., \citealt{Zhuravleva2013}).
  Assuming clumping is uncorrelated with temperature variations, clumping is
  sub-dominant to the variations included in the input temperature maps,
  and we therefore do not include any additional uncertainty from
  clumping in our 20 realizations of the pseudo Compton-$y$ maps.
  In addition, we note that
  the average systematic trend for X-ray surface brightness to be boosted by clumping is mitigated
  by the fact that the amplitude of the Compton-$y$ maps is constrained
  by the SZ data via the factor of $l$.

  We also use
  the {\it Chandra} data to constrain the electron temperature
  $T_e$ within two $1\arcmin$ diameter regions
  centered on sub-clusters B and C
  (these temperatures are applied to our analysis in Section~\ref{sec:vpec}).
  In contrast to the pseudo Compton-$y$ maps, which we
  obtain via the same reduction that was used in M12,
  we constrain these values of $T_e$
  using maps generated with CIAO version 4.5 and 
  calibration database (CALDB) version 4.5.6 \citep{Fruscione2006}. 
  We fit the temperatures and metallicities of the regions in XSPEC 
  \citep{Dorman2001} using the Astrophysical Plasma Emission Code 
  (APEC) model \citep{Smith2001}.  
  We find $T_e = 13.8^{+1.6}_{-1.3}$~keV for sub-cluster B and
  $T_e = 24.4^{+7.8}_{-3.8}$~keV for sub-cluster C, using
  the extended C-statistic to determine the 
  temperature likelihoods.\footnote{
    Our uncertainties on the {\it Chandra}-derived
    temperature of sub-cluster B are
    significantly lower than the values reported in Table 3 of M12.
    This is due to the fact that our current analysis
    uses a $1\arcmin$ diameter region, while the M12
    analysis used a $40\arcsec$ diameter region
    (both analyses use a $1\arcmin$ diameter region
    for sub-cluster C). There are additional $<5$\%
    differences due to the updated calibration
    we use for our current analysis.}

 \subsection{{\it XMM-Newton}}
   \label{sec:xmm}

  To better constrain the electron temperatures
  of sub-clusters B and C, we make use of $\simeq 200$~ksec
  of {\it XMM-Newton} X-ray data toward
  MACS J0717.5+3745.  These data became public in 2012 
  October (Obs Ids; 0672420101, 0672420301, 0672420301),
  and therefore were not included in M12.
  We perform the {\it XMM} MOS data processing and background modeling with
  the {\it XMM} Extended Source Analysis Software ({\it ESAS}) using 
  the methods reported in \citet{kuntz2008} and \citet{snowden2008}. 
  The details of our {\it XMM} analysis are described fully in \citet{bulbul2012},
  and we provide a summary and discuss important differences here.

  Our {\it XMM-Newton} data analysis includes production of the calibrated event files, 
  filtering for the high intensity soft proton flares, 
  and determination of the background intensity in each observation.
  The net exposure time after filtering the event files for good time intervals is 155~ksec.
  Given the superior spatial resolution of {\it Chandra}, we use both
  {\it Chandra} and {\it XMM} to identify regions contaminated by 
  extragalactic X-ray sources not associated with the cluster gas.
  Excluding these regions, we extract spectra using $1\arcmin$ diameter regions
  centered on sub-clusters B and C, identical to the regions
  we use for the {\it Chandra} analysis.
  The temperature gradient is not large, and so contamination
  by adjacent regions (e.g., other sub-clusters) due to
  the wider PSF of {\it XMM} should not affect the extracted temperature
  for each region.

  For each extracted spectrum, we model a superposition 
  of four main background components:
  quiescent particle background, soft X-ray background emission 
  (including solar wind charge exchange, 
  Galactic halo, local hot bubble, and extragalactic unresolved sources), and residual 
  contamination from soft protons \citep{kuntz2008}. 
  As in {\citet{snowden2008}}, we model the contamination due to unresolved point sources 
  using an absorbed power law
  component with spectral index $\alpha = 1.46$ and normalization 
  $= 8.88 \times 10^{7}$ photons keV$^{-1}$ cm$^{-2}$ s$^{-1}$ at 1~keV.

  We simultaneously fit all of the EPIC-MOS spectra using
  the energy range $0.3-10.0$~keV.
  As with the {\it Chandra} spectral analysis, we use the 
  absorbed APEC model to 
  fit the cluster emission, employing the 
  extended C-statistic for our likelihood analysis
  within each sub-cluster region.

  From the {\it XMM} data, we 
  find $T_e = 10.8^{+0.5}_{-0.5}$~keV for sub-cluster B and
  $T_e = 18.6^{+1.6}_{-1.4}$~keV for sub-cluster C.
  We note that these values are $\simeq 25$\% lower than
  the electron temperatures we derive from the
  {\it Chandra} data.
  This systematic difference at high temperature
  is consistent with previous comparisons
  between {\it XMM} and {\it Chandra}
  (e.g., \citealt{nevalainen10, li12, mahdavi13}),
  although we note that, in our case, the statistical 
  significance of the difference is relatively
  small ($\lesssim 2\sigma$).
  We consequently choose to combine the temperature
  measurements from the two X-ray observatories,
  and we obtain maximum likelihood values of
  $T_e = 11.4^{+0.5}_{-0.5}$~keV for sub-cluster B
  and $T_e = 19.9^{+1.5}_{-1.4}$~keV for
  sub-cluster C.
  We explore the impact of using this joint temperature
  constraint, rather than the constraint from either
  {\it XMM} or {\it Chandra} individually, in
  Section~\ref{sec:limitations}.

\section{Model of the SZ Signal}
  \label{sec:model}

  In order to model the SZ signals from the 
  ICM of MACS J0717.5+3745, we employ
  the pseudo Compton-$y$ map 
  to describe the overall shape of the thermal SZ signal.
  The X-ray data we use to create the pseudo Compton-$y$
  map depend negligibly on the line-of-sight velocity
  of the ICM, and the map therefore provides a spatial template
  for the thermal SZ signal that is free from contamination
  from the kinetic SZ signal.
  We then convert this map to units of surface brightness
  in each Bolocam observing band according to the 
  thermal SZ equations \citep{sunyaev72},
  including relativistic corrections \citep{itoh98, nozawa98, nozawa98b, itoh04}.
  For this conversion we compute the responsivity-weighted
  average bandpass over all the Bolocam detectors, and
  from this spectrum we determine the effective band center for a thermal SZ spectrum.
  Due to relativistic corrections, this effective band center 
  depends on the ICM temperature.
  For the 140~GHz band, both the thermal and kinetic
  SZ band centers are $\simeq 140$~GHz,
  while the 268~GHz thermal SZ band center
  is $\simeq 275$~GHz and the kinetic SZ band center
  is $\simeq 268$~GHz.
  Then, for each Bolocam band,
  we convolve the pseudo Compton-$y$ map with both the Bolocam PSF and the
  transfer function of the data processing.

\begin{figure*}
    \centering \includegraphics[width=.76\textwidth]{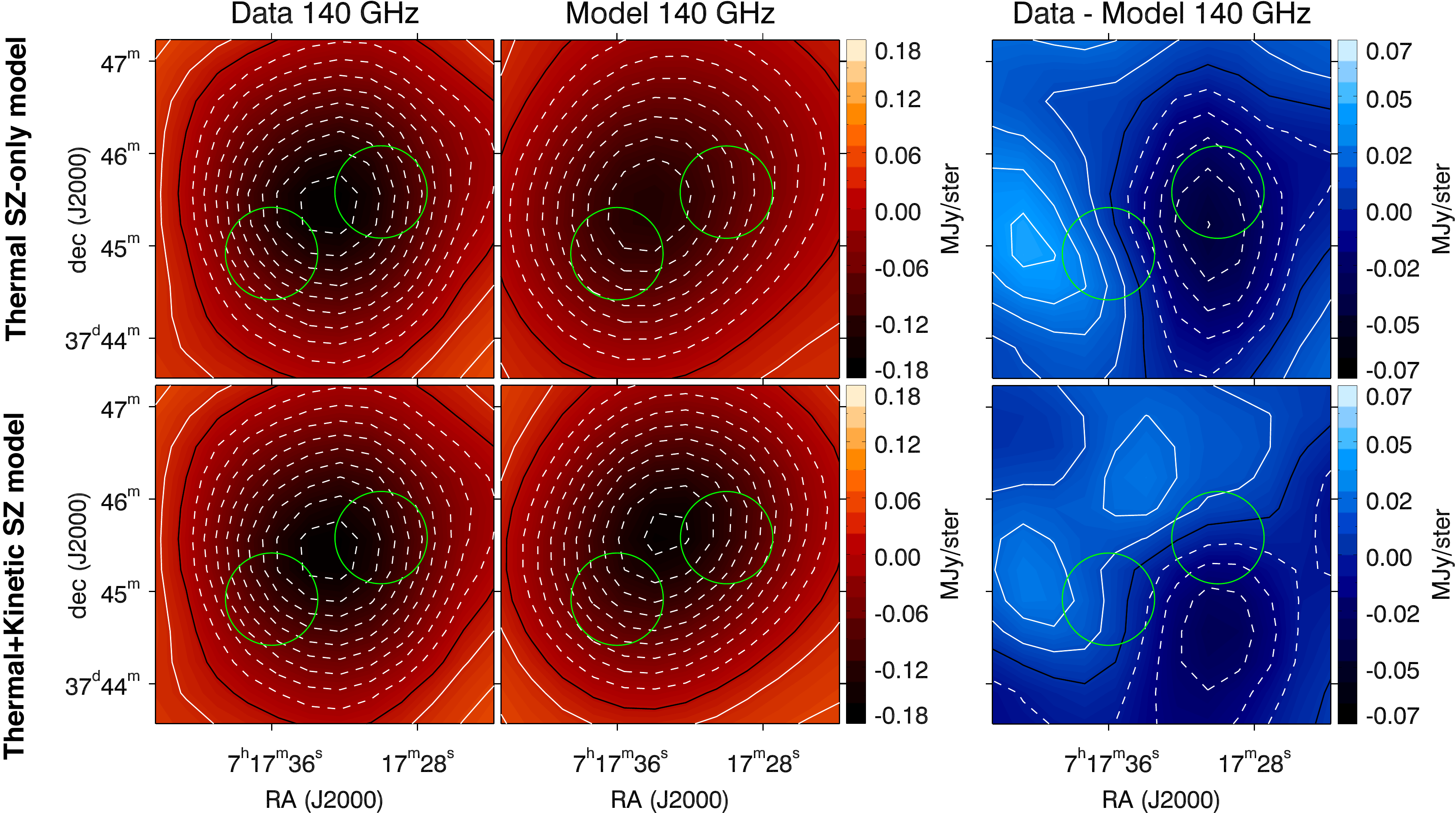}

    \vspace{15pt}

    \includegraphics[width=.76\textwidth]{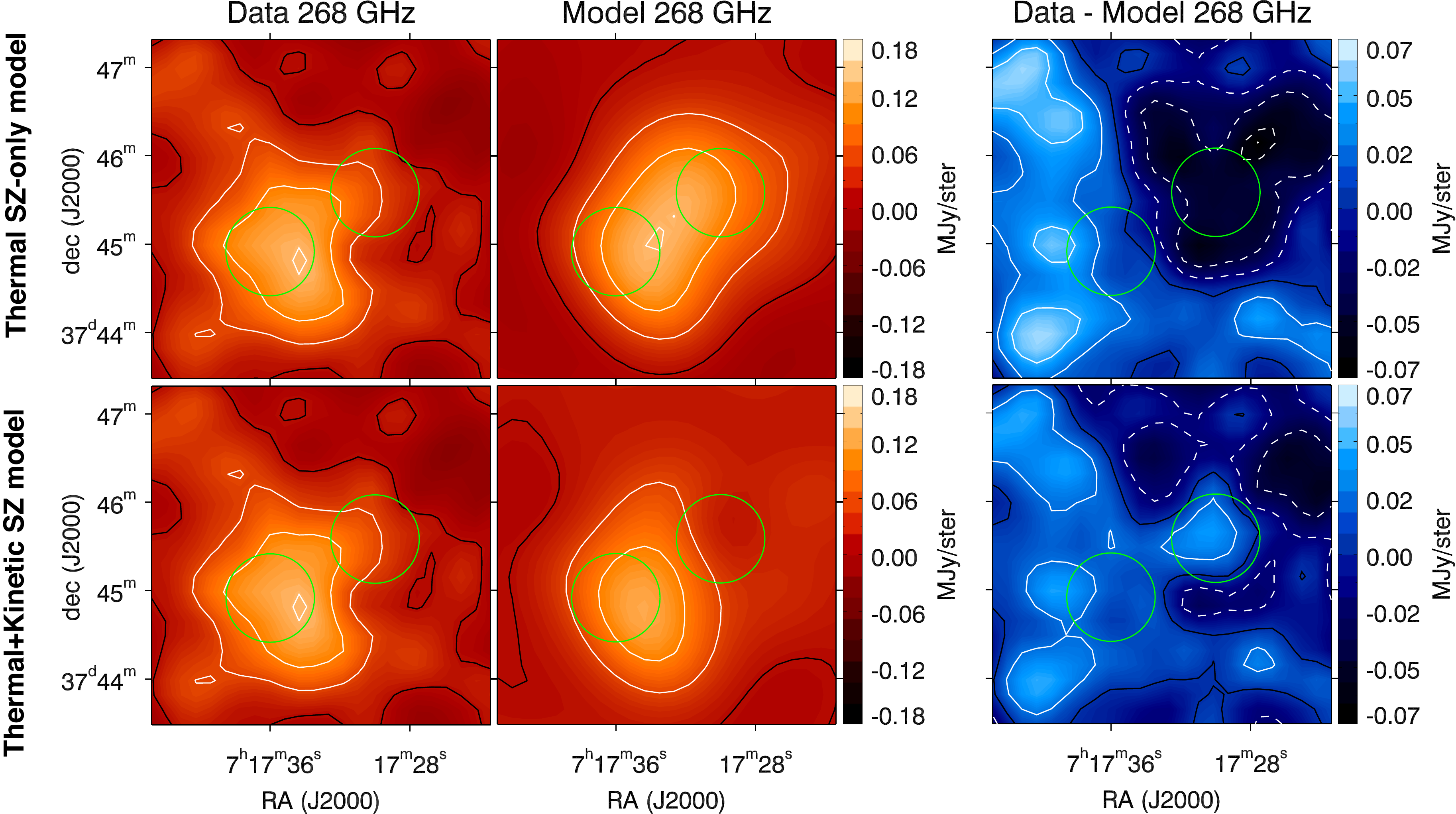}

    \vspace{15pt}

    \caption{Bolocam thumbnails showing the processed data
    within the $4\arcmin \times 4\arcmin$ region we use to constrain
    our model of the SZ signal. 
    From left to right the thumbnails show the Bolocam data, the
    best-fit model, and the difference 
    between the data and the best-fit model
    (i.e., the residual map). 
    The top block shows the 140~GHz data convolved with a $60\arcsec$ FWHM
    Gaussian, and the bottom block shows the 268~GHz
    data convolved with a $30\arcsec$ FWHM Gaussian.
    In the left plots, the
    contours are spaced by S/N~$=2$, with solid representing positive
    S/N, black representing 0, and dashed representing negative S/N.
    In the right plots the contours are spaced by S/N~$=1$, and
    the color stretch is reduced by a factor of 2.5 to better
    highlight the residuals between the data and the model.
    For each wavelength, the top row assumes a model composed of
    only the pseudo Compton-$y$ map with a single normalization
    (a purely thermal SZ signal).
    This model is not a good fit to the data,
    and there is a clear dipole residual from ESE to WNW,
    at a significance of $\simeq 5\sigma$ at 140~GHz
    and $\simeq 3\sigma$ at 268~GHz.
    The bottom row for each wavelength assumes our nominal model of
    the pseudo Compton-$y$ map with a single normalization
    plus an SZ template centered on sub-cluster B
    with different normalizations at 140 and 268~GHz 
    (a thermal plus kinetic SZ signal).
    This model does provide a good fit to the data,
    and the residual maps are consistent with noise.
    The green circles are centered on sub-cluster C (lower left)
    and sub-cluster B (upper right), with diameters of $60\arcsec$.}
    \label{fig:processed_thumbnails}
  \end{figure*}

  We first constrain the normalization of the pseudo Compton-$y$ map
  via a simultaneous fit to both the 140 and 268~GHz Bolocam data.
  Physically, this corresponds to a constraint on the effective
  line-of-sight extent of the ICM $l$, under the assumption of zero kinetic SZ signal.
  For this fit, we use only the data within a $4\arcmin \times 4\arcmin$
  square region approximately centered on the peak of the SZ
  signal at 140~GHz.
  We choose this region because it is large enough to contain the
  bulk of the SZ signal and it is small enough to mitigate the
  effects of the large-angular-scale atmospheric noise in the 268~GHz
  data.
  The quality of this fit is very poor, with a $\chi^2 = 853.4$ for
  717 degrees of freedom, indicating that the pseudo Compton-$y$ map
  alone is inadequate to describe our Bolocam SZ data (see 
  Figure~\ref{fig:processed_thumbnails}).

  Motivated by this poor fit, 
  along with the significant differences in the line-of-sight velocities
  measured by \citet{ma09} for the four identified sub-clusters in
  MACS J0717.5+3745, and the results from M12, we consider
  additional components to our model of the ICM.
  To determine which, if any, additional model components are required
  in order to describe the data,
  we perform a simulated F-test according to the procedure described in
  \citet{czakon13}.
  To perform this test we first insert the baseline model
  into each of our 1000 noise realizations (in this case
  the baseline model is the pseudo Compton-$y$ map with our best-fit
  single normalization).
  We then fit two models to each of these realizations, one consisting of
  only the baseline model, and one with an extension to the baseline model.
  We compute the value of $\Delta \chi^2$ from these two separate fits
  for each of the 1000 realizations,
  and the resulting values provide a measurement of
  the distribution of $\Delta \chi^2$
  for the null hypothesis that the 
  model extension is not required by the data.

  There are several possible model extensions to consider,
  and we therefore proceed according to the following decision tree:
  1) determine the value of $\Delta \chi^2$ separately for
  each possible model extension;
  2) perform the simulated F-test to determine which
  extension is most preferred by the data;
  3) if the most preferred model extension is preferred
  at a high enough significance, 
  which we quantify based on a probability to exceed (PTE)
  from the simulated F-test, 
  then the model extension is added to the baseline model.
  These steps are repeated until none of the possible model
  extensions have a simulated F-test PTE below our threshold,
  which we have chosen to be equal to 0.02.\footnote{
  As we describe below, we consider 5 independent potential model
  extensions, and consequently a possible total of 
  $2^5 = 32$ model permutations.
  Our PTE threshold, which is necessarily somewhat arbitrary,
  is therefore small enough to ensure that a random fluctuation
  among this set of 32 permutations is unlikely
  to produce a PTE small enough for us to 
  include an extension that is not justified by the data.
  We explore the sensitivity of our results to this PTE
  threshold in detail in Section~\ref{sec:systematics}.}

  As a first possible model extension,
  we consider a smooth template of the SZ signal
  centered on one of the four sub-clusters according to
  the positions given by \citet{ma09}, allowing for
  different normalizations of the template
  at 140 and 268~GHz.
  We construct the SZ template according to the
  average profile constrained by Bolocam for a sample of 45 clusters
  \citep{sayers13_pressure}, fixing the scale radius according
  to the estimated mass of each sub-cluster, 
  which we obtain by using
  the ratios of sub-cluster masses determined by \citet{limousin12},
  in combination with the whole-cluster value of $M_{500}$ determined by
  \citet{mantz10}.
  We note that this template represents a more physically motivated
  model than the Gaussian profile assumed by M12.
  We perform four separate fits, in each case fitting a single
  normalization to the pseudo Compton-$y$ map
  (i.e., assuming that the pseudo Compton-$y$ map contains
  only thermal SZ signal), along with separate
  normalizations at 140 and 268~GHz for an SZ template centered
  on one of the four sub-clusters
  (i.e., allowing the SZ template for that sub-cluster to be free to include
  any arbitrary mixture of thermal and kinetic SZ signal).
  We find values of $\Delta \chi^2$ equal to 48.3, 108.1, 9.5, and 26.3
  when the model contains an additional SZ template 
  coincident with sub-clusters A, B, C, and D, 
  respectively. These values of $\Delta \chi^2$ correspond to F-test PTEs of
  0.001, $4 \times 10^{-7}$, 0.213, and 0.026.
  We note that the second value is extrapolated, due to
  the fact that we only have 1000 realizations of $\Delta \chi^2$.

  As another possible extension to the baseline model of a pseudo Compton-$y$
  map with a single normalization in both Bolocam bands,
  we also explore the option of
  allowing for different normalizations of the 
  pseudo Compton-$y$ map at 140 and 268~GHz.
  Physically, this would represent a single bulk velocity
  for the entire cluster, with the cluster being isothermal so that
  the kinetic SZ signal
  has a spatial profile identical to the spatial profile of the thermal SZ signal.
  This fit results in a value of $\Delta \chi^2 = 52.1$, which
  corresponds to a simulated F-test PTE of $3 \times 10^{-4}$,
  where we have again used an extrapolation due to
  our finite number of realizations.
  Therefore, according to our F-test decision tree, we
  assume a new baseline model consisting of 
  the pseudo Compton-$y$ with a single normalization,
  along with a SZ template centered on sub-cluster B,
  as this is the model extension with the smallest
  simulated F-test PTE.

  To fully characterize the fit of this new baseline model, we insert the best-fit model into
  each of our 1000 noise realizations, and then fit the same model to 
  each of these realizations.
  The best-fit model has an overall $\chi^2 = 745.3$ for 715 degrees of freedom,
  which corresponds to a PTE of 0.29 based on the fits to the noise realizations,
  indicating that it provides an adequate description of our Bolocam SZ data.  

  Using this new baseline model,
  we again follow the F-test decision tree to
  determine whether the data still require an additional model extension.
  We perform three new fits that introduce an additional
  SZ component at A, C, and D to our baseline model.
  For these three fits, the value of $\Delta \chi^2$ is 23.9, 8.4, 
  and 21.1, which corresponds to a simulated F-test PTE of 0.033, 
  0.260, and 0.046.
  As before, we also perform a fit allowing the normalization of the
  pseudo Compton-$y$ map to be different at 140 and 268~GHz,
  and we find a $\Delta \chi^2 = 0.1$, with an associated 
  simulated F-test PTE of 0.963.
  All of these fits have PTEs larger than 0.02, and we therefore conclude
  that none of these additional degrees of freedom are required to describe
  our data.
  Consequently, our baseline model of the SZ signal includes a pseudo Compton-$y$
  map with a single normalization, along with an SZ template centered
  on sub-cluster B with separate normalizations at 140 and 268~GHz
  (see Figure~\ref{fig:processed_thumbnails}).

  \begin{figure*}
    \centering
    \includegraphics[width=.65\textwidth]{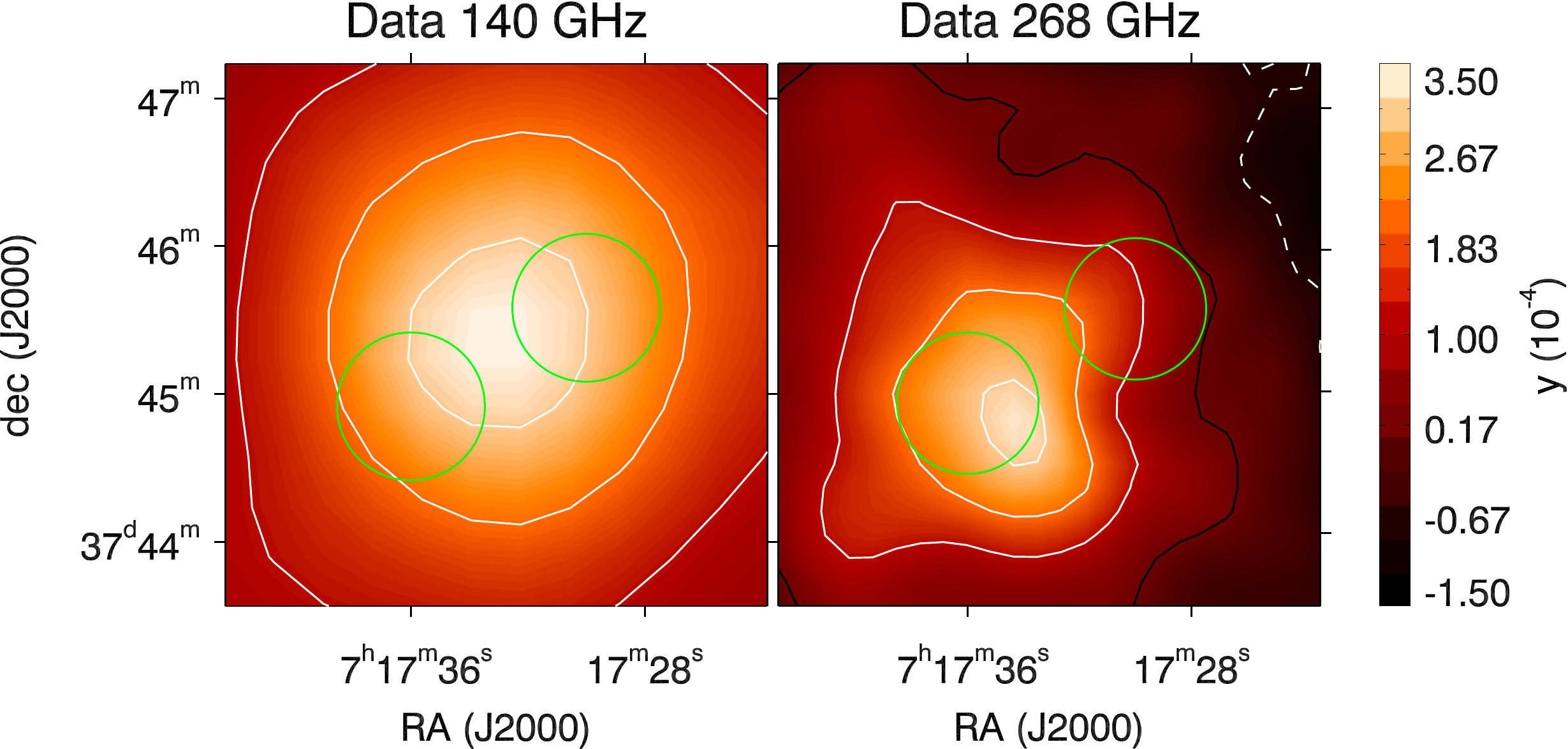}
    \caption{Thumbnails of the deconvolved Bolocam images at
    140 and 268~GHz. We have scaled both images to units of 
    Compton-$y$, including positionally dependent relativistic
    corrections based on the X-ray-determined temperature
    map. The relativistic corrections generally range from $8-15$\%
    at 140~GHz and from $20-40$\% at 268~GHz.
    The 140~GHz image is smoothed with a $60\arcsec$
    Gaussian, and the 268~GHz image is smoothed with a $30\arcsec$
    Gaussian. The contours are spaced by $1 \times 10^{-4}$, with
    solid showing positive $y$ and dashed showing negative $y$.
    The green circles show the $1\arcmin$ diameter apertures centered
    on sub-cluster C (lower left) and sub-cluster B (upper right).
    The total Compton-$y$ signal toward sub-cluster C is nearly identical
    at the two wavelengths, while there is a clear difference
    toward sub-cluster B.}
    \label{fig:deconv}
  \end{figure*}

  The data's strong lack of a preference for separate normalizations of the
  pseudo Compton-$y$ map at 140 and 268~GHz justifies our choice of 
  that model to describe the thermal component of the SZ signal.
  Furthermore, the best-fit normalization of the pseudo Compton-$y$ map
  is $1.08 \pm 0.11$.
  The pseudo Compton-$y$ map was normalized based on the integrated
  SZ signal measured at 31~GHz by the SZA as reported in M12.
  Compared to the Bolocam observing bands, the kinetic SZ signal is a 
  factor of $\simeq 2$ dimmer compared to the thermal SZ signal
  in the SZA observing band.
  Therefore, the consistent normalizations of the pseudo Compton-$y$
  map found by Bolocam and SZA further indicate that
  it provides a good description of the thermal SZ signal
  toward MACS J0717.5+3745.
  {As an additional cross-check, we also refit
  the normalization of the pseudo Compton-$y$ map using
  the Bolocam data, but excluding the data within a $1\arcmin$
  diameter aperture centered on sub-cluster B.
  This fit, which did not include any additional SZ components,
  results in a best-fit normalization of $1.13 \pm 0.08$ for 
  the pseudo Compton-$y$ map.
  The fit quality is good, with a PTE of 0.38,
  indicating that, outside sub-cluster B, the M12 pseudo
  Compton-$y$ map, which was normalized to SZA, 
  describes the Bolocam data well.}
  Although these results serve as additional evidence that our model choice
  is physically justified, we emphasize that our results described 
  below do not strictly depend on the pseudo Compton-$y$ map
  being a good template for the thermal SZ signal, only that
  our model is physically motivated and 
  provides an adequate description of the data.

\section{Measurement of the SZ Spectrum Toward Sub-Cluster B}  
  \label{sec:b}

  \subsection{Two-Band SZ Photometry}
  \label{sec:photometry}

  Based on the requirement of an additional model component
  centered on sub-cluster B to describe our data, we compute
  the SZ brightness at both 140 and 268~GHz toward that sub-cluster.
  In order to eliminate as much contamination from other regions
  of the cluster as possible, we use a circular
  aperture with a diameter of $1\arcmin$, which
  is slightly larger than the PSF FWHM at 140~GHz.
  We first compute the average surface brightnesses 
  within this aperture using
  the best-fit model from Section~\ref{sec:model},
  convolved with the Bolocam PSF to accurately represent
  the resolution of the measurement.
  To include all of the subtle effects of the noise,
  {such as the correlations between
  pixels due to residual atmospheric noise and primary CMB fluctuations,}
  we also compute the average surface brightness within the same aperture
  using the model fits to the 1000 noise realizations.
  Kolmogorov-Smirnov (KS) tests against Gaussians on the distributions
  of 1000 values at 140 and 268~GHz yield PTEs of 0.19 and 0.93,
  respectively, and therefore indicate that our noise is
  Gaussian within our ability to measure it.
  Using these model fits, we estimate the surface brightness
  of sub-cluster B to be $-0.344 \pm 0.028$~MJy sr$^{-1}$ at 140~GHz
  and $0.052 \pm 0.029$~MJy sr$^{-1}$ at 268~GHz, where the errors
  represent only measurement uncertainties.

  In addition to the best-fit model, we also compute the
  surface brightness toward sub-cluster B by directly integrating
  our deconvolved images, which are shown in Figure~\ref{fig:deconv}.
  As described in Section~\ref{sec:bolocam}, the deconvolved
  images have no sensitivity to the DC signal level.
  As a result, we determine the DC signal level of the deconvolved
  images using the best-fit model.
  Specifically, we add a signal offset to the
  deconvolved images so that the
  average signal level within the $4\arcmin \times 4\arcmin$ region
  we use to constrain the model is equal to the average signal level
  of the best-fit model within the same region.
  We exclude the $1\arcmin$ diameter aperture centered on 
  sub-cluster B in this calculation to avoid any potential bias
  in the surface brightness we derive within that aperture.
  This direct integration yields average surface brightnesses of
  $-0.341 \pm 0.027$~MJy sr$^{-1}$ and $0.095 \pm 0.049$~MJy sr$^{-1}$, respectively,
  where we have again estimated the uncertainties using the 1000
  noise realizations.
  As with the model derived results, we used a KS test to
  determine if the distribution of 1000 values is consistent
  with Gaussian, and we find PTEs of 0.75 and 0.57 at 140 and 268~GHz,
  respectively.
  We note that these surface brightness values are consistent 
  with those derived from the best-fit model,
  although there is significantly more measurement 
  uncertainty on the 268~GHz value.
  This additional uncertainty is a result of the significant large-angular-scale
  atmospheric noise in those data, which is amplified by 
  the deconvolution of the signal transfer function.

  \subsection{Systematic Uncertainties}
  \label{sec:systematics}

  First, we note that our flux calibration is accurate
  to 5\% at 140~GHz, and to 10\% at 268~GHz \citep{sayers12_planet}.
  We have included these uncertainties in our systematic
  error budget.

  To estimate the systematic errors due to the 
  model-dependence of our results, we repeat 
  our analysis of computing model-based and directly integrated
  surface brightnesses toward sub-cluster B at both 140 and 268~GHz 
  using a range of different models,
  with a summary of the results in Table~\ref{tab:systematics}.
  First, we replace the baseline pseudo Compton-$y$ map we use
  in our model with a set of 20 realizations of the pseudo
  Compton-$y$ map that we generate according to the X-ray measurement
  uncertainties on the mean $T_e$ for each {\it contbin} region 
  (see Section~\ref{sec:Chandra}).
  Next, we constrain our baseline model using 
  $3\arcmin \times 3\arcmin$ and $5\arcmin \times 5\arcmin$ regions of the 
  Bolocam images instead
  of the nominal $4\arcmin \times 4\arcmin$ region we use in Section~\ref{sec:model}.
  In addition, we consider an SZ model that does not include the
  pseudo Compton-$y$ map, and instead only includes
  SZ templates centered on the sub-clusters.
  We repeat the F-test decision tree described in 
  Section~\ref{sec:model} to determine which of the
  sub-clusters require an SZ template for this model.
  We find that, without
  the pseudo Compton-$y$ map, the data require SZ 
  templates centered on sub-clusters B, C, and D.
  This fit produces a PTE of 0.64,
  indicating that the data are adequately described
  by this model.\footnote{
    The adequacy of this somewhat simple and ad-hoc
    model in describing our data is likely due to Bolocam's
    coarse angular resolution, which largely blurs
    any sub-structures not well described by the smooth
    SZ templates. However, we note that this model
    requires twice as many free parameters as our
    baseline model in order to obtain an adequate fit
    according to our F-test decision tree.}
  Furthermore, we re-ran the F-test decision tree
  with the PTE threshold increased by a factor of 
  two to 0.04.
  With this new threshold, the model consists of the
  pseudo Compton-$y$ map with a single normalization,
  along with SZ templates centered on sub-clusters A and B
  (i.e., relative to the baseline model, an additional
  SZ template is required for sub-cluster A).
  Finally, we determine the effects of varying
  the scale radius of the profile used as a template
  of the SZ signal toward sub-cluster B.
  We vary the scale radius over a range of
  $0.67 - 1.5$ times its nominal value, which
  corresponds to a scaling of the assumed mass
  of sub-cluster B by a factor of $0.3 - 3.4$.

  \begin{deluxetable}{cccc}
    \tablecolumns{4}
    \tablewidth{\columnwidth}
    \tablecaption{Variations in Sub-Cluster B's Surface Brightness
      due to Possible Changes in Our Analysis Method}
    \tablehead{
      \multicolumn{2}{c}{Model-Derived}       & \multicolumn{2}{c}{Direct Integration} \\
      \colhead{140 GHz}  & \colhead{268 GHz}  & \colhead{140 GHz}  & \colhead{268 GHz} \\
      \colhead{MJy sr$^{-1}$} & \colhead{MJy sr$^{-1}$} & \colhead{MJy sr$^{-1}$} & \colhead{MJy sr$^{-1}$} }
    \startdata 
      \cutinhead{Nominal Values from Baseline Model}
      $-0.344 \pm 0.028$ & $0.052 \pm 0.029$ & $-0.341 \pm 0.027$ & $0.095 \pm 0.049$ \\
      \cutinhead{Variations due to Model Choice}
      \sidehead{vary pseudo Compton-$y$ within X-ray uncertainties}
      $\pm 0.012$ & $\pm 0.009$ & $\pm 0.006$ & $\pm 0.004$ \\
      ($\pm 0.4 \sigma$) & ($\pm 0.3 \sigma$) & ($\pm 0.2 \sigma$) & ($\pm 0.1 \sigma$) \\
      \sidehead{vary region used for fit from $3\arcmin$ to $5\arcmin$}
      $\le 0.003$ & $\le 0.019$ & $\le 0.003$ & $\le 0.024$ \\
      ($\le 0.1 \sigma$) & ($\le 0.7 \sigma$) & ($\le 0.1 \sigma$) & ($\le 0.5 \sigma$) \\
      \sidehead{model with no pseudo Compton-$y$; templates at B, C, and D}
      0.028 & 0.026 & 0.016 & 0.015 \\
      ($1.0 \sigma$) & ($0.9 \sigma$) & ($0.6 \sigma$) & ($0.3 \sigma$) \\
      \sidehead{F-test decision tree with PTE threshold equal to 0.04}
      0.007 & 0.017 & 0.001 & 0.026 \\
      ($0.3 \sigma$) & ($0.6 \sigma$) & ($0.0 \sigma$) & ($0.5 \sigma$) \\
      \sidehead{vary scale radius of B template by $0.67$ to $1.5$}
      $\le 0.017$ & $\le 0.026$ & $\le 0.017$ & $\le 0.022$ \\
      ($\le 0.6 \sigma$) & ($\le 0.9 \sigma$) & ($\le 0.6 \sigma$) & ($\le 0.4 \sigma$) \\
      \cutinhead{Variations due to Aperture Choice}
      \sidehead{aperture centered on \citet{limousin12} coords}
      0.018 & 0.019 & 0.009 & 0.003 \\
      ($0.6 \sigma$) & ($0.7 \sigma$) & ($0.3 \sigma$) & ($0.1 \sigma$) \\
      \sidehead{aperture centered on X-ray centroid}
      0.019 & 0.034 & 0.010 & 0.048 \\
      ($0.7 \sigma$) & ($1.2 \sigma$) & ($0.4 \sigma$) & ($1.0 \sigma$) \\
      \sidehead{vary aperture diameter from $0.67\arcmin$ to $1.5\arcmin$}
      $\le 0.029$ & $\le 0.013$ & $\le 0.024$ & $\le 0.011$ \\
      ($\le 1.0 \sigma$) & ($\le 0.4 \sigma$) & ($\le 0.9 \sigma$) & ($\le 0.2 \sigma$)
    \enddata
    \tablecomments{Top block: best-fit surface brightnesses from the
    baseline model described in Section~\ref{sec:model}, and associated
    $1\sigma$ uncertainties due to measurement noise only.
    Next block: variations in the surface brightness of sub-cluster B
    based on our choice of model. We consider five different
    model fits to describe the SZ data. These models are explained in detail
    in the text, and we refer the reader there for more details.
    From left to right, the columns give the change in surface brightness
    at 140 and 268~GHz for the model-derived and direct integration
    surface brightnesses.
    The top rows give these values in MJy sr$^{-1}$, and the bottom rows give
    these values relative to the measurement uncertainties in the top block.
    When noise variations to the models are considered, these values indicate the $1\sigma$
    range with a $\pm$ symbol, when a range of model inputs are considered, these
    values show the magnitude of the maximum change with a $\le$ symbol,
    and when a single alternative model is considered these values
    show the magnitude of the change with no symbol.
    Based on these results, we add a systematic uncertainty
    equal to 1.0 times the measurement uncertainty for the model-derived values
    and equal to 0.6 times the measurement uncertainty for the
    direct integration values.
    Bottom block: variations in the surface brightness of sub-cluster B for
    different choices of aperture.
    From top to bottom, the rows show the change relative to our
    nominal $1\arcmin$ diameter aperture centered on the \citet{ma09}
    coordinates for 1) an aperture centered on the \citet{limousin12}
    coordinates, 2) an aperture centered on the X-ray centroid,
    and 3) varying the aperture diameter between $0.67\arcmin$ and $1.5\arcmin$
    for the aperture centered on the \citet{ma09} coordinates.
    All of these differences are consistent with the expected
    measurement noise fluctuations for the different aperture choices.}
    \label{tab:systematics}
  \end{deluxetable}

  Considering this broad range of possible models that
  we could have chosen to describe our data, we find
  that the model-derived surface brightness of sub-cluster
  B never changes by more than 1.0 times the measurement
  uncertainties given in Section~\ref{sec:photometry}.
  For the surface brightness values obtained from
  direct integration of the deconvolved images, we
  find that the change is never larger than 0.6
  times the uncertainties given in Section~\ref{sec:photometry}.
  As expected, the model-derived surface brightnesses have a stronger
  model dependence compared to the directly integrated
  surface brightnesses, although the latter still have
  a noticeable model-dependence due to the method
  by which we constrain the DC signal level of the 
  deconvolved image.
  Based on these results, we conservatively include an additional
  systematic uncertainty of 1.0 times the measurement
  uncertainty for the model-derived surface brightnesses,
  and 0.6 times the measurement uncertainty for the
  directly integrated surface brightnesses.
  A full summary of our best-fit surface brightnesses, along
  with the full error budget on these values, 
  is given in Table~\ref{tab:sz_brightness}.

  \begin{deluxetable*}{cccccc}
    \tablecaption{SZ Surface Brightness}
    \tablehead{\colhead{Frequency} & \colhead{Best Fit} & \colhead{Measurement Err.} & 
      \colhead{Flux Err.} & \colhead{Modeling Err.} & \colhead{Total Err.} \\
      \colhead{GHz} & \colhead{MJy sr$^{-1}$} & \colhead{MJy sr$^{-1}$} & \colhead{MJy sr$^{-1}$} & 
      \colhead{MJy sr$^{-1}$} & \colhead{MJy sr$^{-1}$}}
    \startdata 
    \cutinhead{\bf Sub-Cluster B}
    \sidehead{Model Fits}
      140 & -0.344 & 0.028 & 0.017 & 0.028 & 0.043 \\
      268 & \phm{-}0.052 & 0.029 & 0.005 & 0.029 & 0.041 \\
    \sidehead{Direct Integration}
      140 & -0.341 & 0.027 & 0.017 & 0.016 & 0.036 \\
      268 & \phm{-}0.095 & 0.049 & 0.010 & 0.029 & 0.058 \\
    \cutinhead{\bf Sub-Cluster C}
    \sidehead{Model Fits}
      140 & -0.262 & 0.026 & 0.013 & 0.028 & 0.040 \\
      268 & \phm{-}0.217 & 0.039 & 0.022 & 0.029 & 0.053 \\
    \sidehead{Direct Integration}
      140 & -0.270 & 0.026 & 0.014 & 0.016 & 0.034 \\
      268 & \phm{-}0.220 & 0.059 & 0.022 & 0.029 & 0.069
    \enddata
    \tablecomments{The average surface brightness within a $1\arcmin$ diameter
      aperture centered on sub-clusters B and C. From left to right the columns give
      the observing frequency, the best-fit average surface brightness, 
      the measurement uncertainty on this value, the
      uncertainty on this value due to flux calibration,
      the uncertainty on this value due to the range of models
      we could have chosen to describe the data,
      and the total combined uncertainty which is the quadrature sum of the previous
      three columns.
      For each sub-cluster, 
      the top rows give the values we derive from the best-fit model of the
      SZ signal, and the bottom rows give values we derive from direct integration
      of the deconvolved images.}
    \label{tab:sz_brightness}
  \end{deluxetable*}

  In addition to exploring how our choice of model
  affects our results,
  we also examine the effects of varying the aperture
  we use to compute the average surface brightness toward
  sub-cluster B.
  First, we examine three possible choices for the
  location of sub-cluster B:
  1) the location given by \citet{ma09} based on the 
  distribution of galaxies (7:17:30.0, +37:45:35),
  2) the location given by \citet{limousin12} based on the
  matter distribution (7:17:30.2, +37:45:15),
  and 3) the location of the X-ray brightness centroid
  (7:17:31.4, +37:45:29).
  Our nominal analysis uses the \citet{ma09} coordinates,
  and we give the changes in surface brightness when we use
  the other two possible apertures in Table~\ref{tab:systematics}.
  Compared to the measurement uncertainties given in 
  Section~\ref{sec:photometry}, the surface brightnesses
  we measure in these new apertures differ by less
  than $\le 1.2\sigma$ with a median of $\simeq 0.7\sigma$.
  The three sets of 
  coordinates are separated from each other by $\simeq 20\arcsec$,
  which is a significant fraction of the aperture radius
  of $30\arcsec$, and means that less than 50\% of the area
  enclosed by one aperture is also enclosed by another aperture.
  Consequently, completely uncorrelated measurement noise
  between any given pair of apertures will produce 
  surface brightnesses that differ by $\simeq 1\sigma$.
  Therefore, the differences in surface brightness that we measure
  between these aperture locations are consistent with
  the expectation due to noise fluctuations.

  We also examine the effects of varying the diameter of the aperture
  from $0.67\arcmin$ to $1.5\arcmin$ (compared to the nominal diameter of $1\arcmin$),
  and again find results that are consistent within $1\sigma$.
  As with the different aperture locations, 
  this is consistent with the variations that we expect due
  to uncorrelated measurement noise between the aperture
  choices, and indicates that variations in the location, or
  diameter, of the aperture we use to measure the SZ surface brightness
  result in differences consistent with measurement noise.
  We therefore conclude that our results are not sensitive to the
  exact choice of aperture, and we do not include any additional systematic
  error in our overall noise budget.
  {We note that in all cases the apertures are comparable
  in size to the Bolocam PSF, and there is consequently some signal
  leakage from outside to inside the apertures and vice versa.
  Furthermore, the separation between the apertures centered on sub-clusters
  B and C is also comparable to the size of the Bolocam PSF, and
  so there is some signal leakage between apertures.
  Although we are not able to account for this signal leakage in our
  analysis, the consistency of our results using various aperture positions
  and diameters indicates that the leakage
  is below our measurement uncertainties.}

  \subsection{Comparison to Previous Results}  
  \label{sec:compare}

  We note that the 140~GHz surface brightnesses we find for
  sub-cluster B are 
  slightly different compared to the values reported in M12,
  although identical Bolocam data is used for both analyses.
  Our model-derived surface brightness of $-0.344 \pm 0.028$~MJy sr$^{-1}$
  is more than $1\sigma$ lower than the M12 value
  of $-0.293 \pm 0.030$~MJy sr$^{-1}$.\footnote{\label{foot:one}
    Table 3 of M12 lists a total flux density of $-19.5 \pm 2.0$~mJy, which corresponds
    to a surface brightness of $-0.293 \pm 0.030$~MJy sr$^{-1}$.}
  More than half of this difference is due to a minor error
  in the analysis presented in M12.
  The total flux densities given in Table 3 of M12 were mistakenly computed 
  from the average surface brightness within a $1.5\arcmin$ aperture,
  rather than the $1\arcmin$ aperture claimed in the text of M12
  and also used in our present analysis.
  The remaining difference between our current surface brightness
  values and the ones presented in M12
  is due to minor changes in our assumed model of the
  SZ signal.
  First, M12 constrained the normalization of the pseudo Compton-$y$
  map separately at 140 and 268~GHz, compared to the joint constraint
  we use in our present analysis.
  In addition, M12 assumed that the SZ template centered on sub-cluster B
  had a Gaussian profile, compared to the more physically motivated
  profile we use in this analysis, with a shape described by the
  best-fit profile to a sample of 45 clusters observed with Bolocam \citep{sayers13_pressure}.

  Furthermore, we note that, in our current analysis,
  the model-derived surface brightness
  agrees quite well with the surface brightness we obtain from
  a direct integration of the deconvolved image.
  This result is in contrast to the measurements presented in
  M12, where the two values differed by slightly more than
  $1\sigma$.
  This change is due to differences in how the DC signal
  offset of the deconvolved images is computed.
  M12 computed the DC signal offset based on a fit of
  the average profile determined by \citet{arnaud10} to
  the full 140~GHz Bolocam dataset.
  Although the fit quality of this single profile is not
  particularly poor, with a PTE of 0.07, the adequacy
  of using a single profile to describe a complex merging
  system like MACS J0717.5+3745 is questionable.
  Therefore, as described above,
  for this analysis we choose to constrain the DC signal
  offset of the 140~GHz deconvolved image using our
  nominal model of the SZ signal (a pseudo Compton-$y$ map with an additional
  SZ component centered on sub-cluster B).
  Not surprisingly, this change in our estimate of the
  DC signal level results in a better agreement between
  the model-derived and directly integrated
  surface brightnesses.

\section{Measurement of the SZ Spectrum Toward Sub-Cluster C}  
  \label{sec:c}

  M12 computed the SZ surface brightness toward 
  both sub-cluster B and sub-cluster C.
  The latter measurement was motivated primarily by
  the fact that \citet{ma09} identified sub-cluster C
  as the most massive component of MACS J0717.5+3745,
  along with the fact that sub-cluster C 
  is coincident with the highest surface brightness
  in the 268~GHz Bolocam image.
  Therefore, although our F-test decision tree indicates
  that our data do not require a component in addition
  to the thermal SZ template toward sub-cluster C, we
  again measure its SZ surface brightness.
  For these measurements we add an SZ template
  centered on sub-cluster C to our model, to ensure that
  the model has enough freedom to describe any
  possible deviations from a purely thermal SZ spectrum.
  We again estimate the SZ surface brightness
  using a $1\arcmin$ diameter aperture centered
  on the coordinates from \citet{ma09},
  with the results given in Table~\ref{tab:sz_brightness}.
  As with sub-cluster B, we estimate the uncertainties
  on these surface brightnesses using our 1000 noise
  realizations, and we again find that the distribution
  of values is consistent with Gaussian noise.
  In addition, we estimate the systematic
  uncertainty due to our choice of model using the
  same formalism described for sub-cluster B in
  Section~\ref{sec:systematics}.
  We find systematic errors consistent with those
  that we find for sub-cluster B, and we therefore
  adopt identical values for sub-cluster C.
  Finally, we note that the 140~GHz brightness values
  differ from those derived in M12 by roughly the
  same amounts as for sub-cluster B, with the differences
  due to the same reasons described in detail in 
  Section~\ref{sec:compare}.

  We do not attempt to constrain the SZ brightness
  toward either sub-cluster A or sub-cluster D.
  We do not consider sub-cluster A 
  due to the fact that it is not strongly
  detected in either Bolocam dataset.
  We do not consider sub-cluster D
  due to the fact that it is not separately resolved
  from sub-cluster C due to Bolocam's
  coarse angular resolution, and therefore
  any estimate of sub-cluster D's SZ brightness
  would be highly correlated with our estimate
  of sub-cluster C's SZ brightness.

\section{Peculiar Velocity Constraints}
  \label{sec:vpec}

  \begin{deluxetable*}{ccccc}
    \tablecolumns{5}
    \tablewidth{0pc}
    \tablecaption{Peculiar Velocity Constraints}
    \tablehead{
      \colhead{}  & \colhead{$T_e$}  & \colhead{optical} & \colhead{SZ model fit}  & 
      \colhead{SZ direct integration} \\
      \colhead{} & \colhead{keV} & \colhead{km s$^{-1}$} & \colhead{km s$^{-1}$} & \colhead{km s$^{-1}$} }
    \startdata 
      sub-cluster B & $11.4^{+0.5}_{-0.5}$ & $+3238^{+252}_{-242}$ & 
        $+3450^{+900\phn}_{-900\phn}$ ($1 - \textrm{Prob}[v_z \ge 0] = 1.3 \times 10^{-5}$) & 
        $+2550^{+1050}_{-1050}$ ($1 - \textrm{Prob}[v_z \ge 0] = 2.2 \times 10^{-3}$)  \\
      \\ \vspace{-11pt} \\
      sub-cluster C & $19.9^{+1.5}_{-1.4}$ & \phn$-733^{+486}_{-478}$ & 
        \phn$-550^{+1350}_{-1400}$ ($1 - \textrm{Prob}[v_z \le 0] = 3.7 \times 10^{-1}$) & 
        \phn$-500^{+1600}_{-1550}$ ($1 - \textrm{Prob}[v_z \le 0] = 4.3 \times 10^{-1}$) 
    \enddata
    \tablecomments{Line-of-sight velocity constraints from our
    analysis. From left to right the columns give the
    the X-ray-derived temperature from {\it Chandra} and {\it XMM}, the line-of-sight
    velocity derived by \citet{ma09} based on optical spectroscopy,
    and the line-of-sight velocity from our kinetic SZ
    constraints using the best-fit model and a
    direct integration of the deconvolved image.
    The top row shows the constraints for sub-cluster B,
    and the bottom row shows the constraints for sub-cluster C.
    For the fits we have used the best-fit SZ brightnesses
    given in Table~\ref{tab:sz_brightness}, with the total
    uncertainties listed in the far-right column of that
    table.
    Next to the kinetic SZ velocity constraints, we give the 
    probability that the line-of-sight velocity is greater
    than 0 for sub-cluster B, and the probability that
    the line-of-sight velocity is less than 0 for
    sub-cluster C.}
    \label{tab:vpec}
  \end{deluxetable*}

  Using our two-band measurements of the SZ surface brightness
  toward sub-clusters B and C, we are able to place constraints
  on the properties of the ICM of each sub-cluster.
  Based on the equations presented in Section~\ref{sec:sz},
  the total SZ brightness depends on four quantities
  related to the cluster ICM:
  $f(\nu,T_e)$, $y$, $\tau_e$, and $v_z$.
  Our two-band SZ surface brightness measurements 
  are insufficient to constrain all of these quantities, and so we therefore
  make the assumption that the ICM within each sub-cluster is
  isothermal and equal to the X-ray spectroscopic temperature
  determined within the same $1\arcmin$ diameter apertures
  that we use to measure the SZ surface brightness.
  As a result, $f(\nu,T_e)$ is fully constrained by
  the {\it Chandra}-and-{\it XMM}-measured $T_e$, and from
  Equations~\ref{eq:szy} and \ref{eq:tau} we have
  $y = \tau_e k_B T_e m_e^{-1} c^{-2}$, meaning that
  $y$ and $\tau_e$ are not independent.
  Therefore, we are left with two free parameters to constrain
  using the two-band Bolocam surface brightnesses,
  either $\tau_e$ and $v_z$ or $y$ and $v_z$
  (in practice we constrain $Y_{int}$ and $v_z$,
  where $Y_{int} = y \Delta \Omega$, and $\Delta \Omega$ is equal to
  the solid angle of our $1\arcmin$ aperture).
  In all of the fits, we compute the band-averaged values
  of $f(\nu,T_e)$ for a given $T_e$
  using the full Bolocam bandpasses rather than a single
  effective band center.

\begin{figure*}
    \centering
    \includegraphics[width=0.4\textwidth]{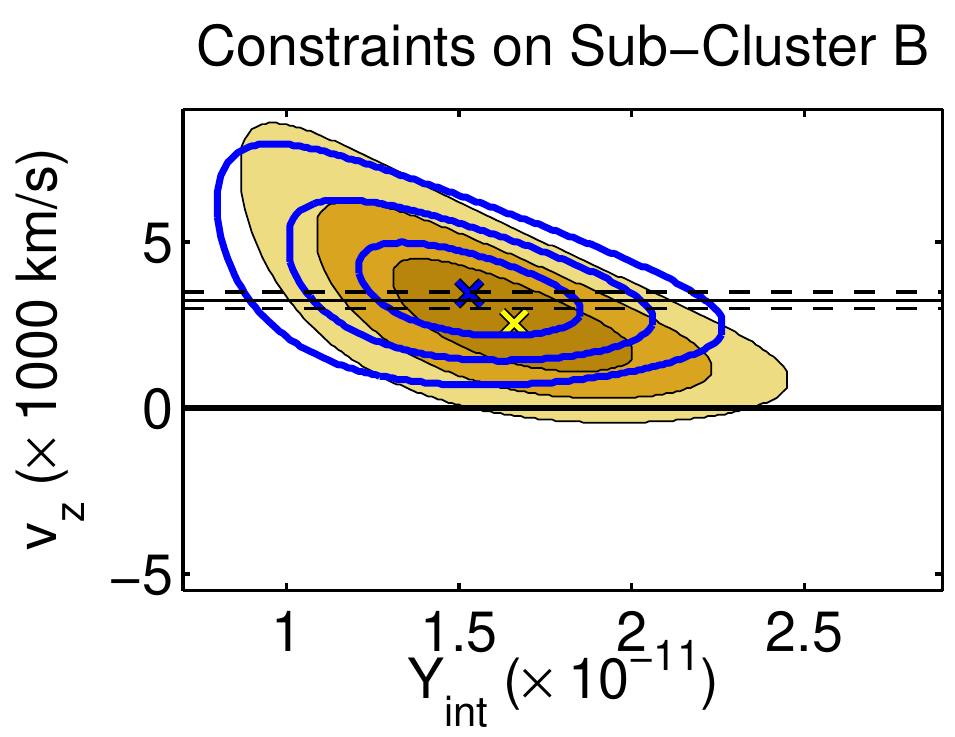}
    \hspace{.05\textwidth}
    \includegraphics[width=0.4\textwidth]{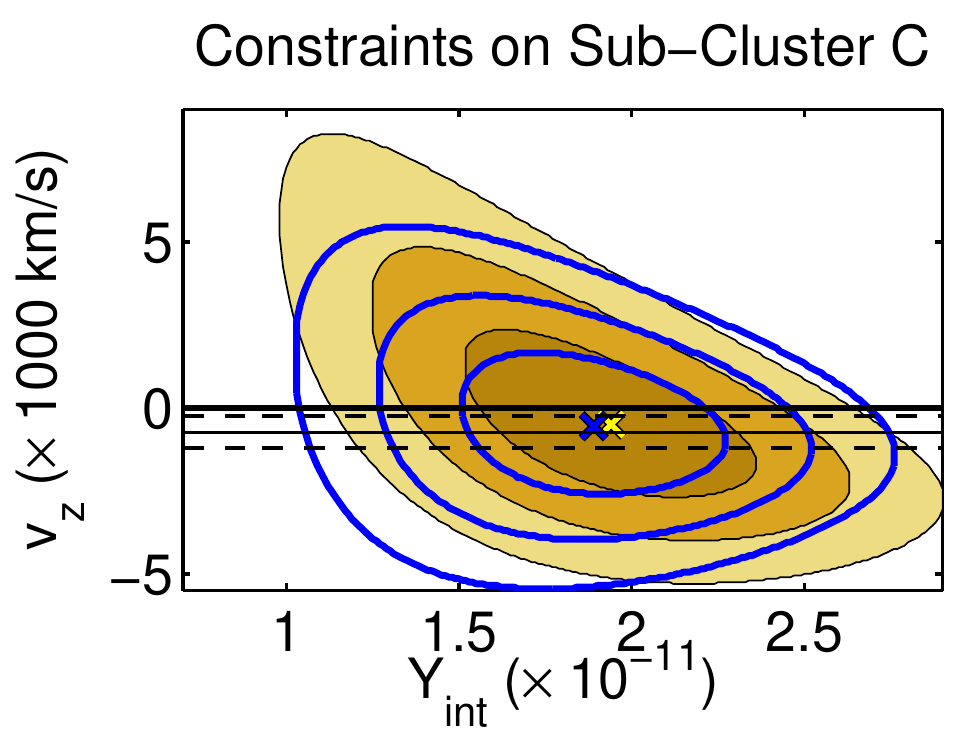}

    \centering
    \includegraphics[width=0.4\textwidth]{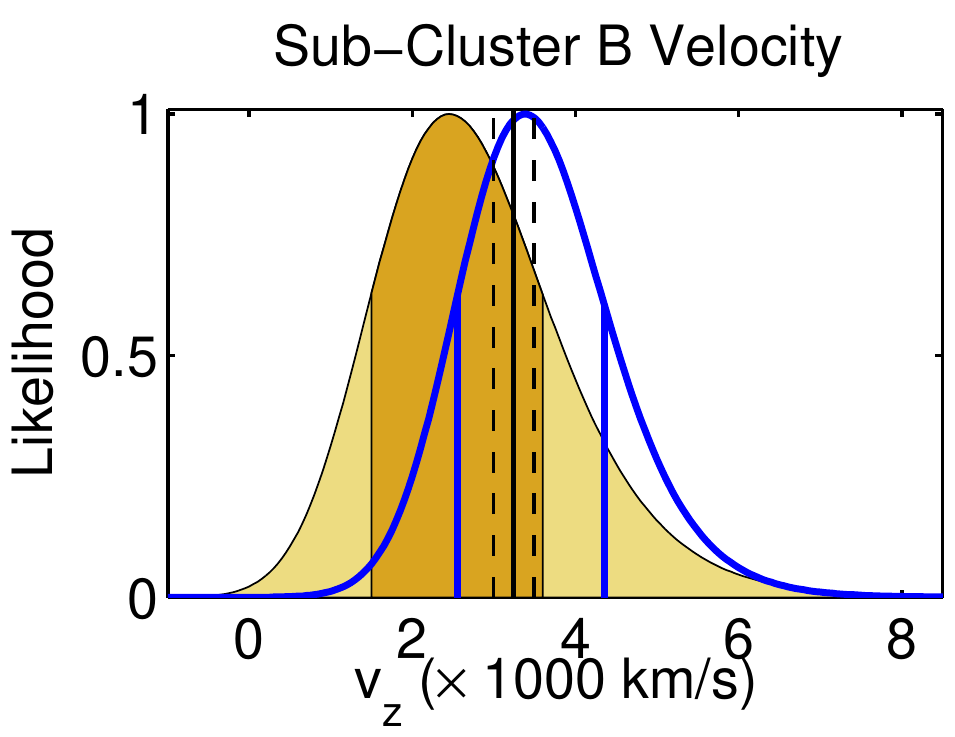}
    \hspace{.05\textwidth}
    \includegraphics[width=0.4\textwidth]{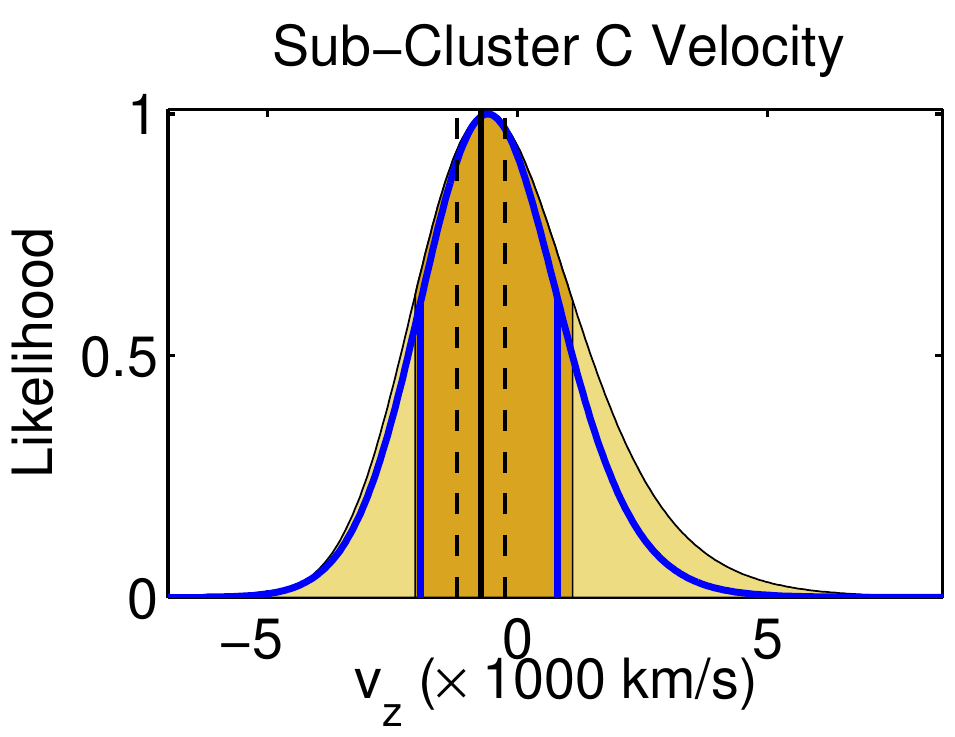}
    \caption{Our SZ-derived constraints on the ICM toward
    sub-cluster B (left) and sub-cluster C (right).
    The top row shows two-dimensional confidence regions
    for the values of $Y_{int}$ and $v_z$,
    with contours drawn at $1\sigma$, $2\sigma$, and $3\sigma$
    for a two-parameter likelihood (e.g., $1\sigma$
    corresponds to a $\Delta \chi^2 = 2.30$).
    The bottom row shows marginalized one-dimensional
    likelihoods for $v_z$, with vertical lines
    drawn at $\pm 1 \sigma$ (corresponding to $\Delta \chi^2 = 1$).
    In all cases blue corresponds to the constraints
    from the model-derived SZ surface brightnesses, and
    yellow corresponds to the constraints from the
    SZ surface brightnesses we derive from direct integration of
    the deconvolved images.
    The solid black line represents the best-fit velocity derived
    by \citet{ma09} based on optical spectroscopy,
    and the dashed lines show the corresponding
    $1\sigma$ confidence region around their best-fit.}
    \label{fig:vpec}
  \end{figure*}

\begin{figure*}
    \centering
    \includegraphics[width=.4\textwidth]{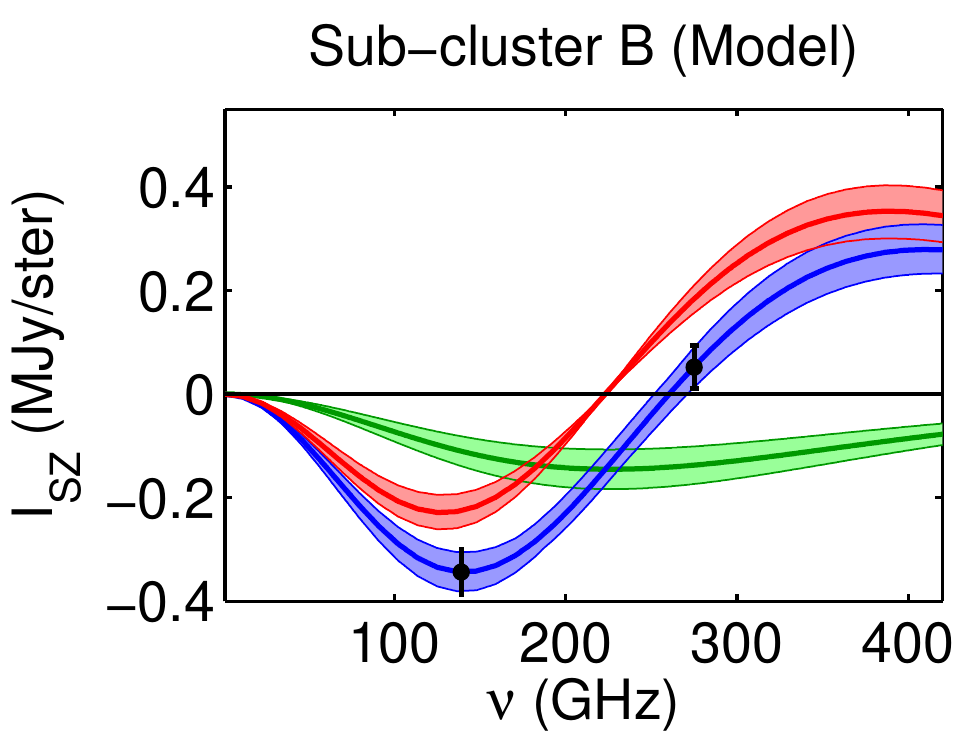}
    \hspace{.05\textwidth}
    \includegraphics[width=.4\textwidth]{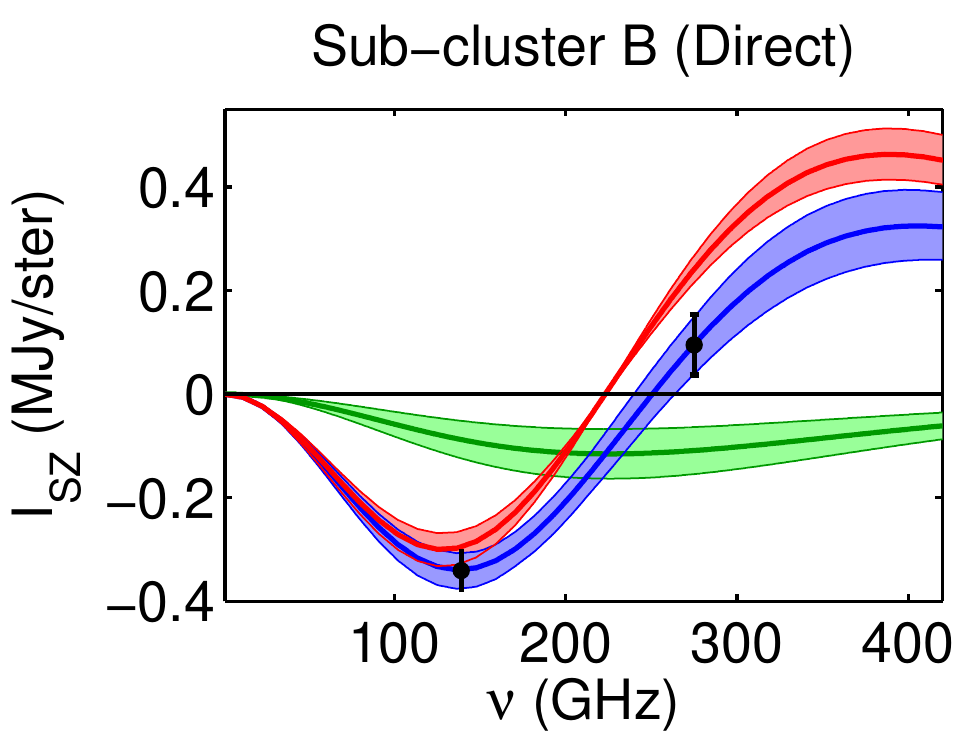}

    \centering
    \includegraphics[width=.4\textwidth]{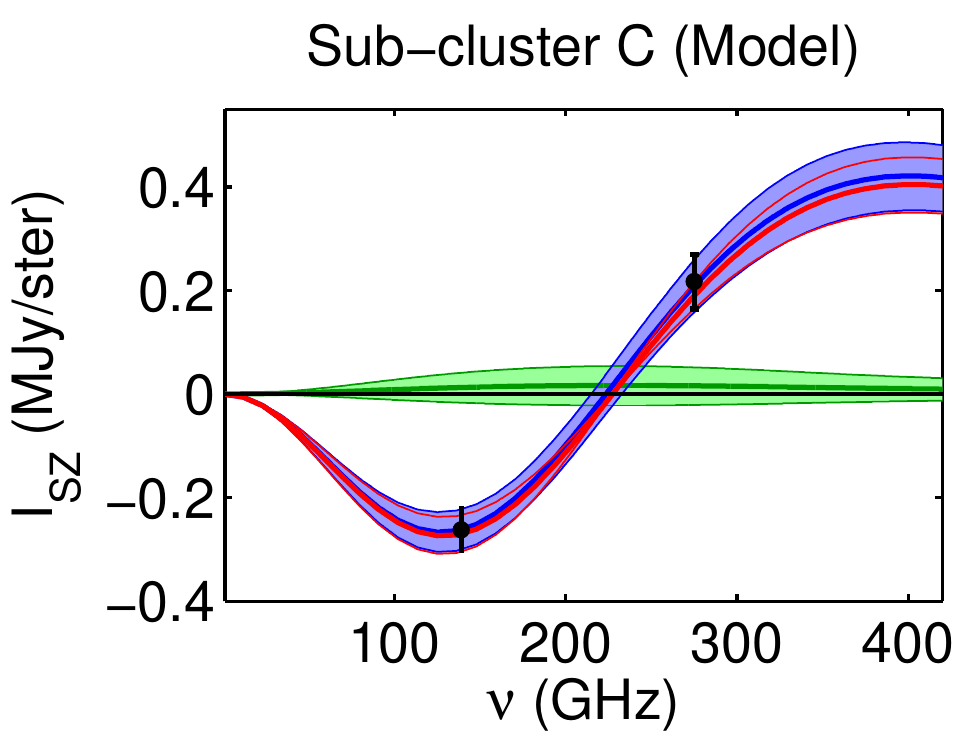}
    \hspace{.05\textwidth}
    \includegraphics[width=.4\textwidth]{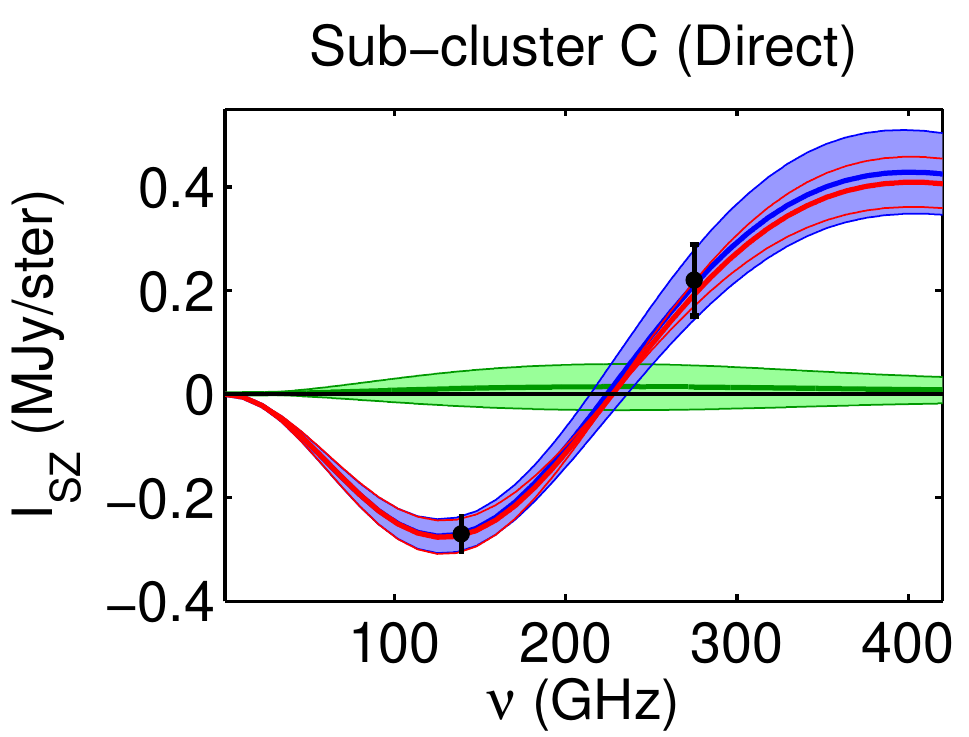}
    \caption{Our best-fit SZ spectra. The top row shows the fits to sub-cluster B,
    and the bottom row shows the fits to sub-cluster C.
    The left column shows the SZ surface brightnesses we determine from the 
    model fit, and the right column shows the SZ surface brightnesses
    we determine via direct integration of the deconvolved images.
    The best-fit thermal-SZ-only spectrum is shown in red,
    the best-fit kinetic SZ spectrum is shown in green, and
    the best-fit thermal plus kinetic SZ spectrum is shown in blue,
    with the widths showing the $1\sigma$ confidence region
    of the fits.
    We include relativistic corrections in all of the spectra.}
    \label{fig:sz_spectrum}
  \end{figure*}

  Using our X-ray measured $T_e$, along with our SZ
  surface brightnesses, we then
  perform a grid search
  to constrain the values of $v_z$ and $Y_{int}$
  for sub-clusters B and C.
  For these constraints, we use the best-fit SZ surface brightness
  values from Table~\ref{tab:sz_brightness}, along with
  the total uncertainties in the far-right column of 
  that table.
  Therefore, we fully include not only measurement
  uncertainties, but also flux calibration uncertainties,
  and possible systematic uncertainties due to our
  choice of model to describe the SZ signal.
  Because the noise in our SZ surface brightness measurements
  is indistinguishable from Gaussian,
  we compute likelihoods based on a Gaussian distribution.
  When fitting the SZ spectra, we marginalize over the range of $T_e$ 
  values allowed by the X-ray data,
  relying on the C statistic to give a likelihood 
  for each temperature in the range $2-40$~keV. 
  A summary of our results for both sub-clusters
  is given in Table~\ref{tab:vpec} and Figures~\ref{fig:vpec}
  and \ref{fig:sz_spectrum}, and we highlight
  some of these results below.

  For sub-cluster B we find a best-fit $v_z = +3450$~km s$^{-1}$
  using the SZ surface brightnesses we determine from the model
  fit to our data and a best-fit $v_z = +2550$~km s$^{-1}$
  using the SZ surface brightnesses we determine from direct
  integration of our deconvolved images.
  Both of these values are consistent with the value
  of $+3238$~km s$^{-1}$ determined by \citet{ma09} based
  on optical spectroscopy under the assumption 
  that the peculiar velocity of the entire cluster
  is 0 along the line-of-sight (see Figure~\ref{fig:vpec}).
  The $1\sigma$ uncertainties about these best-fit velocities
  are similar for both the model-derived
  and direct integration results, and
  are $\lesssim 1000$~km s$^{-1}$.
  We also compute the probability of $v_z \ge 0$,
  and we obtain values of $(1 - \textrm{Prob}[v_z \ge 0]) = 1.3 \times 10^{-5}$ and $2.2 \times 10^{-3}$
  for the model-derived and direct integration SZ surface brightnesses, respectively
  (see the bottom panels of Figure~\ref{fig:vpec}).
  For a Gaussian distribution, these one-sided probabilities 
  correspond to a difference from $v_z = 0$
  of 4.2$\sigma$ and 2.9$\sigma$, respectively.
  For sub-cluster C we find a best-fit $v_z$ of
  $\simeq -500$~km s$^{-1}$ from both the model fit and
  direct integration of the deconvolved image, which
  is fully consistent with both the value of $-733$~km s$^{-1}$
  determined by \citet{ma09} and with zero velocity.
  We note that the uncertainties on the value of $v_z$ for sub-cluster
  C are $\simeq 50$\% larger compared to sub-cluster B.
  This increase is due entirely to the higher temperature
  of sub-cluster C.
  This higher temperature produces a smaller value of $\tau_e$ for a fixed value
  of $Y_{int}$ and therefore a correspondingly lower
  kinetic SZ signal for a fixed value of $v_z$.

  {We note that the difference between our best-fit $v_z$
  and the best fit $v_z$ from \citet{ma09} is quite small
  for both sub-clusters ($0.23\sigma$ and $0.13\sigma$
  for the model-derived results for sub-clusters B and C,
  and $0.64\sigma$ and $0.14\sigma$ for the direct-integration
  results for sub-clusters B and C).
  The random probability of obtaining such results
  from two independent measurements
  of two independent parameters is 2\% for our model-derived
  results and 6\% for our direct-integration results.
  These probabilities are small, but they are not small
  enough to cause significant concern.
  In addition, our intentionally conservative estimates
  of the uncertainties due to our choice of SZ model
  have likely resulted in over-estimated errors on the 
  SZ brightness, thus rendering the good agreement 
  between our results and those of \citet{ma09} more likely.}

\section{Discussion}
  \label{sec:discussion}

 \subsection{Differences Compared to the Results in M12}

  Compared to the results presented in M12, our best-fit
  values of $v_z$ for sub-cluster B are somewhat lower
  ($+3450$~km s$^{-1}$ and $+2550$~km s$^{-1}$ compared to $+4640$~km s$^{-1}$
  and $+3600$~km s$^{-1}$ for the model-derived
  and direct integration surface brightnesses, respectively).
  This is mainly due to an increase in the best-fit
  surface brightness at 268~GHz as a result of the additional
  data we use in our present analysis.
  In contrast, our best-fit values of $v_z$ for sub-cluster
  C are smaller in magnitude compared to M12
  ($-550$~km s$^{-1}$ and $-500$~km s$^{-1}$ compared to $-3720$~km s$^{-1}$
  and $-4120$~km s$^{-1}$ for the model-derived
  and direct integration surface brightnesses, respectively).
  These differences are again driven by the
  additional 268~GHz data we use in our current analysis,
  which indicates that sub-cluster C is dimmer
  compared to the analysis of M12.
  However, we emphasize that all of our measured values 
  of $v_z$ are consistent to within $1\sigma$
  of the values presented in M12,
  and there is no tension between the two results.
 
  Our uncertainties on the value of $v_z$ for
  sub-cluster B are a factor of $\simeq 3$ smaller than
  the uncertainties reported in M12.
  This improvement is almost entirely due to the significant amount of 
  additional 268~GHz data we use in this analysis
  and the corresponding factor of 
  $\simeq 2.5$ decrease in the uncertainties on
  the 268~GHz surface brightness.
  A small additional improvement is driven by
  the lower value of $T_e$ found in our present
  analysis, which for a fixed $Y_{int}$ 
  corresponds to a larger $\tau_e$
  and therefore a larger kinetic SZ signal
  for a fixed $v_z$.
  In contrast, our uncertainties on $v_z$ 
  for sub-cluster C have only
  decreased by a factor of $\simeq 2$ compared to M12
  even though there is a similar reduction in the
  268~GHz measurement uncertainties.
  This difference relative to sub-cluster B is driven by our 
  best-fit value of $v_z$, which is significantly
  larger (less negative) than the results
  in M12.
  As a result, the best-fit value of $Y_{int}$ is
  smaller, and therefore the best-fit value of $\tau_e$
  is smaller.
  Consequently, for a given change in $v_z$, the 
  corresponding change in the kSZ surface
  brightness is also smaller, resulting in
  less constraining power on the value of $v_z$.

  In contrast to the analysis presented in M12, note that we
  include additional systematic uncertainties in our
  derived SZ surface brightnesses due to differences based
  on the range of models we could have chosen to
  describe the SZ signal.
  These systematic uncertainties increase the
  total error estimate on the model-derived and 
  directly integrated SZ surface brightnesses
  by $\simeq 40$\% and $\simeq 20$\%, respectively.
  We perform fits of $v_z$ without
  including this additional systematic error,
  and verify that the fractional improvement
  in our constraints matches these values.
  Therefore, the model-dependence of our SZ data
  results in a non-negligible degradation of our
  constraining power on $v_z$.

  \subsection{Limitations to Our Kinetic SZ Constraints}
  \label{sec:limitations}

  Given the range of multi-wavelength data that we use to
  place constraints on $v_z$, we also estimate how each
  of these datasets contribute to our overall uncertainties.
  First, as noted in Section~\ref{sec:xmm}, previous results
  have indicated that there is
  a systematic difference in temperatures derived 
  from {\it Chandra} and {\it XMM}, and the temperatures we measure
  are in general agreement with this systematic difference.
  At this point, the cause of this difference has not been 
  conclusively demonstrated.
  Due to this lack of a conclusive understanding of the difference, 
  combined with the fact that 
  the difference between the X-ray temperatures we derive
  from the two observatories
  is of modest statistical significance, we choose
  to constrain the electron temperatures via the 
  joint likelihood from the {\it Chandra} and {\it XMM} data.
  If we instead adopt the {\it XMM}-only values of $T_e$, then
  we find best-fit values of $v_z$ equal to
  $+3300$~km s$^{-1}$ and $+2450$~km s$^{-1}$ for the model fit
  and direct integration of sub-cluster B
  and $-450$~km s$^{-1}$ and $-400$~km s$^{-1}$ for the model fit
  and direct integration of sub-cluster C.
  If we instead adopt the {\it Chandra}-only values of $T_e$, then
  we find best-fit values $v_z$ equal to
  $+4000$~km s$^{-1}$ and $+2900$~km s$^{-1}$ for the model fit
  and direct integration of sub-cluster B
  and $-550$~km s$^{-1}$ and $-450$~km s$^{-1}$ for the model fit
  and direct integration of sub-cluster C.
  For sub-cluster B, the {\it Chandra}-only temperatures
  yield line-of-sight velocities that differ
  by $\simeq 0.5\sigma$,
  but all of the other values are statistically 
  indistinguishable from our results in Table~\ref{tab:vpec}.
  Therefore, we conclude that X-ray calibration uncertainties
  do not strongly affect our constraints on $v_z$.
  We further note that the significance of our kinetic 
  SZ measurement from $v_z = 0$ is nearly independent of the exact value
  of $T_e$ and the slight differences in 
  $v_z$ for the different temperatures are due
  to the inverse relationship between $T_e$ and $\tau_e$
  for a fixed $y$, coupled with the inverse relationship
  between $v_z$ and $\tau_e$ for a fixed kinetic SZ 
  surface brightness.

  To assess the impact of the X-ray uncertainties on $T_e$, we also rerun all of 
  our fits with vanishing uncertainties on the
  X-ray derived $T_e$.
  Even in the case of sub-cluster C, when using the {\it Chandra}-only
  measurement with uncertainties of $T_e$ $^{+7.8}_{-3.8}$~keV, 
  the derived uncertainties on 
  $v_z$ increase by only $\simeq 10$\% when using the measured uncertainties
  instead of assuming that the uncertainty on $T_e$ is equal to 0.
  Therefore, the X-ray uncertainties are not significant
  in our overall error budget on $v_z$.

  To determine the effect of the CIB on our measurement of $v_z$,
  we also compute the SZ brightness under the assumption
  that the CIB is completely and noiselessly subtracted from
  the data.
  Specifically, compared to our default noise realizations,
  we remove the noise from the undetected CIB, along with our
  uncertainties on the subtracted CIB (see the Appendix).
  This results in a negligible change in the 140~GHz surface
  brightness uncertainties, and a $\simeq 10-20$\% reduction
  in the 268~GHz surface brightness uncertainties.
  There is a corresponding $\simeq 10-20$\% reduction in
  our derived uncertainties on $v_z$.
  In addition, we estimate the potential bias that would
  result from not subtracting any of the Bolocam or 
  SPIRE-detected galaxies from our 268~GHz data.
  We find that our best-fit 268~GHz surface brightness
  values change
  by $\simeq 10$\%, indicating that the
  bright sources in the CIB produce a non-negligible
  bias in the SZ surface brightnesses we measure.
  Therefore, at our sensitivities,
  the CIB has a noticeable effect on our kinetic SZ
  measurement, and is more significant than
  uncertainties on the X-ray-derived electron temperature.

  We also perform fits under the assumption that
  primary CMB fluctuations are perfectly subtracted
  from our data by removing them from our noise realizations.
  This does not produce a noticeable change in the 
  268~GHz surface brightness constraints, but does
  improve the 140~GHz surface brightness constraints
  by $\simeq 5$\%, with a corresponding
  improvement in our derived constraints on $v_z$.
  Therefore, noise from primary CMB fluctuations
  has an effect on our kinetic SZ measurements that
  is smaller than, but comparable to, noise from CIB fluctuations.
  This mild sensitivity to primary CMB fluctuations
  is due to the relative shallowness of our 140~GHz
  data, which have an rms of $\simeq 30$~$\mu$K$_{CMB}$-arcmin
  (see Table 1 of \citealt{sayers13_pressure}).

  Examining the error budget in Table~\ref{tab:sz_brightness},
  the dominant uncertainties are associated with
  SZ measurement noise and the 
  exact choice of model used to describe the SZ data,
  although we note that uncertainties due to absolute flux calibration
  are only a factor of $\simeq 2$ smaller.
  As detailed in Section~\ref{sec:bolocam}, a model is required 
  to interpret our SZ data because
  the large-angular scale atmospheric noise necessitates
  high-pass filtering of the data, which removes
  signal on large angular scales.
  As a result, a spatial model of the SZ is the only
  way to recover this large-scale signal in order to obtain
  an absolute surface brightness.
  Therefore, this modeling uncertainty is a direct
  result of the measurement noise in the SZ data,
  and is not a fundamental limitation.
  Consequently, deeper SZ data would provide
  a significant improvement to our kinetic SZ 
  measurement, although these deeper SZ data
  will require better absolute flux calibration,
  are likely to require an improved subtraction
  of the CIB (and possibly the primary CMB fluctuations), 
  and may require an improved
  understanding the the X-ray temperature calibration
  or the line-of-sight temperature structure
  (e.g., \citealt{chluba12, prokhorov12}).

  \subsection{Additional Potential Sources of Bias}
  \label{sec:add_bias}

  Our analysis constrains the line-of-sight peculiar velocities of 
  two of the sub-clusters of MACS J0717.5+3745 via a measurement of the
  SZ surface brightnesses within small apertures centered
  on these sub-clusters.
  Due to the complex dynamics in MACS J0717.5+3745, the SZ signal
  within these apertures may not be sourced by gas bound to
  a single sub-cluster with a single coherent bulk velocity.
  However, as described in Section~\ref{sec:m0717}, 
  the X-ray data show that sub-cluster
  B does appear to have a relatively intact core region.
  Therefore, at least for sub-cluster B, the assumption
  of a single bound ICM appears to be justified.
  Sub-cluster C seems to be more disturbed, and this
  assumption may not be valid for that region.

  In addition to possible merger-induced gas inhomogeneities,
  there are also
  likely to be line-of-sight projection effects that
  cause the SZ signal within a single aperture to be 
  sourced by the ICMs of multiple sub-clusters.
  In part to answer this question, \citet{ruan13} studied
  the SZ signal from a simulated triple-merger system in detail.
  Their simulated cluster is similar to MACS J0717.5+3745,
  and contains one sub-cluster with a velocity of 2500~km s$^{-1}$.
  They used kinetic SZ measurements at 90 and 268~GHz to 
  constrain the line-of-sight velocities of sub-clusters
  within the merger and found best-fit velocities that
  are consistent with the true velocities of the sub-clusters
  to within $\simeq 10$\%.
  This is partly due to the fact that the SZ signal from the
  core of the sub-cluster of interest is significantly brighter
  than the SZ signal away from the core of other sub-clusters
  in projection.
  However, the merging sub-cluster also induces a small
  kinetic SZ signal in the ICMs of the other sub-clusters
  in projection with it, and this induced signal 
  serves to bring the SZ-measured velocity into better
  agreement with the true velocity.
  These results indicate that at our current measurement
  precision, the bias in our measured velocities due
  to interactions and projection effects from other
  sub-clusters is likely to be statistically insignificant.

  \subsection{Cosmological Implications}

  Our kinetic SZ measurements are in good agreement
  with the spectroscopic measurements of \citet{ma09}
  and indicate that sub-cluster B is
  moving with a line-of-sight velocity of $\simeq 3000$~km s$^{-1}$
  compared to the center of mass of the system.
  \cite{ma09} note that this value is close to the 
  maximum expected velocity due to infall from infinity.
  For example, if sub-cluster B starts from rest at infinity,
  and if the main cluster has a mass of 
  $1.5 \times 10^{15}$~M$_{\odot}$, then sub-cluster B would need
  to be within $\simeq 1.5$~Mpc of the main cluster to reach
  an infall velocity of 3000~km s$^{-1}$.
  This is in fairly good agreement with
  N-body simulations, which indicate that a relative velocity
  of $\simeq 3000$~km s$^{-1}$ is possible for a MACS J0717.5+3745-like
  cluster within the framework of the standard
  cosmological model.
  For example, \citet{lee10} showed that mergers with main
  cluster masses above $1 \times 10^{15}$~M$_{\odot}$
  at $z = 0.5$ have a non-negligible probability
  of producing velocities larger than 3000~km s$^{-1}$
  when the sub-cluster is within the virial radius
  of the main cluster.
  In addition, the cluster studied by \citet[][see Section~\ref{sec:add_bias}]{ruan13}
  was
  selected from a cosmological simulation of a 400~$h^{-1}$Mpc cube, and one of its
  sub-clusters has a line-of-sight velocity of 2500~km s$^{-1}$.
  Therefore, we conclude that while the velocity of sub-cluster
  B is large, it is not in any tension with the standard
  cosmological models.

\section{Summary}
  \label{sec:summary}

  We detect an extended SZ signal toward MACS J0717.5+3745
  at high significance in two observing bands
  with Bolocam (140 and 268~GHz).
  The 268~GHz data also contain significant emission from
  dusty star forming galaxies.
  We subtract all of the galaxies brighter than
  $\simeq 1$~mJy using a combination of {\it Herschel}-SPIRE
  and Bolocam data, although both this subtraction,
  and the un-subtracted population of dimmer galaxies, 
  produce a non-negligible amount of noise in our
  measurement of the SZ signal (see Section~\ref{sec:limitations}).
  Using a rigorous decision tree based on application of the 
  F-test, we find that a physically-motivated model composed of a
  {\it Chandra}-derived pseudo Compton-$y$ map to describe
  the thermal SZ signal, plus an additional template centered
  on sub-cluster B with different normalizations at 140 and
  268~GHz, is the minimum model that is adequate to describe our data.
  We note that sub-cluster B has a measured spectroscopic line-of-sight
  velocity of $+3200$~km s$^{-1}$ \citep{ma09}.

  From this best-fit model, we compute the two-band 
  SZ surface brightness toward sub-cluster B, along with the
  most massive sub-cluster, C.
  We also compute the SZ surface brightness by directly
  integrating the Bolocam images, although the best-fit
  model is required to constrain the DC signal level
  of these images, which is filtered away by our data
  processing.
  For both the model-derived and directly integrated SZ
  surface brightnesses, we include uncertainties due
  to measurement noise and absolute flux calibration.
  In addition, we include an uncertainty due to the 
  variations in derived surface brightnesses
  for a range of physically motivated
  models that we could have chosen to describe our data,
  and we find that this uncertainty is similar to 
  our measurement uncertainty.

  Using our measured SZ surface brightnesses toward sub-clusters
  B and C, along with our X-ray-derived electron
  temperatures for each sub-cluster, we constrain
  a spectral model consisting of thermal and kinetic 
  SZ components (see Figure~\ref{fig:sz_spectrum}).
  For these fits, we assume that the ICM is isothermal
  within small apertures centered on each sub-cluster,
  and we include corrections for relativistic effects.
  We find that a thermal SZ signal is adequate to
  describe the SZ surface brightnesses of sub-cluster C,
  but that an additional kinetic SZ signal is 
  required for sub-cluster B.
  From our model-derived SZ surface brightnesses,
  this kinetic SZ signal implies a line-of-sight
  velocity of $v_z = +3450$~km s$^{-1}$, while the directly
  integrated SZ surface brightnesses imply
  a line-of-sight velocity of $v_z = +2550$~km s$^{-1}$,
  both of which are in good agreement with the
  spectroscopic measurement of \citet[][See Figure~\ref{fig:vpec}]{ma09}.
  From the model fit we find that 
  $(1 - \textrm{Prob}[v_z \ge 0])$ is $1.3 \times 10^{-5}$, which
  corresponds to being $4.2\sigma$ from 0
  for a Gaussian distribution.
  Similarly, from the direct integration of the SZ surface brightness, 
  we find that 
  $(1 - \textrm{Prob}[v_z \ge 0])$ is $2.2 \times 10^{-3}$, which
  corresponds to being $2.9\sigma$ from 0
  for a Gaussian distribution.

  We consider potential biases in our derived 
  values of $v_z$ due to possible systematics
  in the X-ray derived $T_e$, and due to merger and projection
  effects as a result of the complex dynamics of this cluster,
  and we find that neither bias is likely to be
  significant compared to our measurement uncertainties.
  We find that raw SZ measurement sensitivity
  limits our constraints on $v_z$, and uncertainties
  from the X-ray data, the CIB, the CMB, and flux calibration are sub-dominant,
  although deeper SZ measurements will likely be limited
  by some combination of these factors.
  Our data, combined with the results from \citet{ma09},
  indicate that sub-cluster B is moving with a line-of-sight 
  velocity of $\simeq +3000$~km s$^{-1}$,
  a value that is high, but not in tension with standard
  cosmological models.

\section{Acknowledgments}

We acknowledge the assistance of: 
the day crew and Hilo
staff of the Caltech Submillimeter Observatory, who provided
invaluable assistance during data-taking for this
data set; 
Kathy Deniston, Barbara Wertz, and Diana Bisel, who provided effective
administrative support at Caltech and in Hilo;
the Bolocam observations were partially supported by the Gordon and Betty
Moore Foundation.
JS was supported by 
NSF/AST-0838261, NASA/NNX11AB07G, and the Norris Foundation
CCAT Postdoctoral Fellowship;
support for TM was provided by NASA through Einstein Fellowship Program grant
number PF0-110077 awarded by the {\it Chandra} X-ray Center,
which is operated by the Smithsonian Astrophysical Observatory for NASA
under contract NAS8-03060;
PMK was supported by a NASA Postdoctoral Program Fellowship;
NC was partially supported by a NASA Graduate Student
Research Fellowship;
AM was partially supported by NSF/AST-0838187 and NSF/AST-1140019;
EP and JAS were partially supported by NASA/NNX07AH59G;
SS was supported by NASA Earth and Space Science Fellowship
NASA/NNX12AL62H;
KU acknowledges partial support from the National Science Council
of Taiwan grant NSC100-2112-M-001-008-MY3 and from the Academia Sinica Career
Development Award.
A portion of this research was carried out at the Jet Propulsion
Laboratory, California Institute of Technology, under a contract
with the National Aeronautics and Space Administration.
This research made use of the Caltech Submillimeter Observatory,
which was operated at the time by the California Institute of Technology
under cooperative agreement with the National Science Foundation 
(NSF/AST-0838261).
This work is based in part on observations made with {\it Herschel},
a European Space Agency Cornerstone Mission with a significant
participation by NASA. Partial support for this work was provided
by NASA through an award issued by JPL/Caltech.

{\it Facilities:} \facility{CSO}, \facility{{\it Chandra}}, \facility{{\it Herschel}-SPIRE}, \facility{{\it XMM-Newton}}.  


\begin{thebibliography}{102}

\bibitem[{{Arnaud} {et~al.}(2010){Arnaud}, {Pratt}, {Piffaretti},
  {B{\"o}hringer}, {Croston}, \& {Pointecouteau}}]{arnaud10}
{Arnaud}, M., {Pratt}, G.~W., {Piffaretti}, R., {B{\"o}hringer}, H., {Croston},
  J.~H., \& {Pointecouteau}, E. 2010, \aap, 517, A92

\bibitem[{{Austermann} {et~al.}(2009){Austermann}, {Aretxaga}, {Hughes},
  {Kang}, {Kim}, {Lowenthal}, {Perera}, {Sanders}, {Scott}, {Scoville},
  {Wilson}, \& {Yun}}]{austermann09}
{Austermann}, J.~E., {et~al.} 2009, \mnras, 393, 1573

\bibitem[{{Bai} {et~al.}(2007){Bai}, {Marcillac}, {Rieke}, {Rieke}, {Tran},
  {Hinz}, {Rudnick}, {Kelly}, \& {Blaylock}}]{bai07}
{Bai}, L., {et~al.} 2007, \apj, 664, 181

\bibitem[{{Bennett} {et~al.}(2012){Bennett}, {Larson}, {Weiland}, {Jarosik},
  {Hinshaw}, {Odegard}, {Smith}, {Hill}, {Gold}, {Halpern}, {Komatsu}, {Nolta},
  {Page}, {Spergel}, {Wollack}, {Dunkley}, {Kogut}, {Limon}, {Meyer}, {Tucker},
  \& {Wright}}]{bennet12}
{Bennett}, C.~L., {et~al.} 2012, ArXiv:1212.5225

\bibitem[{{Benson} {et~al.}(2003){Benson}, {Church}, {Ade}, {Bock}, {Ganga},
  {Hinderks}, {Mauskopf}, {Philhour}, {Runyan}, \& {Thompson}}]{benson03}
{Benson}, B.~A., {et~al.} 2003, \apj, 592, 674

\bibitem[{{B{\'e}thermin} {et~al.}(2011){B{\'e}thermin}, {Dole}, {Lagache}, {Le
  Borgne}, \& {Penin}}]{bethermin11}
{B{\'e}thermin}, M., {Dole}, H., {Lagache}, G., {Le Borgne}, D., \& {Penin}, A.
  2011, \aap, 529, A4

\bibitem[{{Bhattacharya} \& {Kosowsky}(2008)}]{bhattacharya08}
{Bhattacharya}, S., \& {Kosowsky}, A. 2008, \prd, 77, 083004

\bibitem[{{Birkinshaw}(1999)}]{birkinshaw99}
{Birkinshaw}, M. 1999, \physrep, 310, 97

\bibitem[{{Blain} {et~al.}(2002){Blain}, {Smail}, {Ivison}, {Kneib}, \&
  {Frayer}}]{blain02}
{Blain}, A.~W., {Smail}, I., {Ivison}, R.~J., {Kneib}, J.-P., \& {Frayer},
  D.~T. 2002, \physrep, 369, 111

\bibitem[{{Bonafede} {et~al.}(2009){Bonafede}, {Feretti}, {Giovannini},
  {Govoni}, {Murgia}, {Taylor}, {Ebeling}, {Allen}, {Gentile}, \&
  {Pihlstr{\"o}m}}]{bonafede09}
{Bonafede}, A., {et~al.} 2009, \aap, 503, 707

\bibitem[{{Bulbul} {et~al.}(2012){Bulbul}, {Smith}, {Foster}, {Cottam},
  {Loewenstein}, {Mushotzky}, \& {Shafer}}]{bulbul2012}
{Bulbul}, G.~E., {Smith}, R.~K., {Foster}, A., {Cottam}, J., {Loewenstein}, M.,
  {Mushotzky}, R., \& {Shafer}, R. 2012, \apj, 747, 32

\bibitem[{{Carlstrom} {et~al.}(2002){Carlstrom}, {Holder}, \&
  {Reese}}]{carlstrom02}
{Carlstrom}, J.~E., {Holder}, G.~P., \& {Reese}, E.~D. 2002, \araa, 40, 643

\bibitem[{{Chluba} {et~al.}(2012){Chluba}, {Nagai}, {Sazonov}, \&
  {Nelson}}]{chluba12}
{Chluba}, J., {Nagai}, D., {Sazonov}, S., \& {Nelson}, K. 2012, \mnras, 426,
  510

\bibitem[{{Conley} {et~al.}(2011){Conley}, {Guy}, {Sullivan}, {Regnault},
  {Astier}, {Balland}, {Basa}, {Carlberg}, {Fouchez}, {Hardin}, {Hook},
  {Howell}, {Pain}, {Palanque-Delabrouille}, {Perrett}, {Pritchet}, {Rich},
  {Ruhlmann-Kleider}, {Balam}, {Baumont}, {Ellis}, {Fabbro}, {Fakhouri},
  {Fourmanoit}, {Gonz{\'a}lez-Gait{\'a}n}, {Graham}, {Hudson}, {Hsiao},
  {Kronborg}, {Lidman}, {Mourao}, {Neill}, {Perlmutter}, {Ripoche}, {Suzuki},
  \& {Walker}}]{conley11}
{Conley}, A., {et~al.} 2011, \apjs, 192, 1

\bibitem[{{Czakon}(2013)}]{czakon13}
{Czakon}, N.~G. 2013, {\it in prep}

\bibitem[{{Dorman} \& {Arnaud}(2001)}]{Dorman2001}
{Dorman}, B., \& {Arnaud}, K.~A. 2001, in Astronomical Society of the Pacific
  Conference Series, Vol. 238, Astronomical Data Analysis Software and Systems
  X, ed. F.~R. {Harnden}, Jr., F.~A. {Primini}, \& H.~E. {Payne}, 415

\bibitem[{{Downes}(2009)}]{downes_thesis}
{Downes}, T.~P. 2009, PhD thesis, University of Chicago

\bibitem[{{Downes} {et~al.}(2012){Downes}, {Welch}, {Scott}, {Austermann},
  {Wilson}, \& {Yun}}]{downes12}
{Downes}, T.~P., {Welch}, D., {Scott}, K.~S., {Austermann}, J., {Wilson},
  G.~W., \& {Yun}, M.~S. 2012, \mnras, 423, 529

\bibitem[{{Ebeling} {et~al.}(2004){Ebeling}, {Barrett}, \&
  {Donovan}}]{ebeling04}
{Ebeling}, H., {Barrett}, E., \& {Donovan}, D. 2004, \apjl, 609, L49

\bibitem[{{Ebeling} {et~al.}(2007){Ebeling}, {Barrett}, {Donovan}, {Ma},
  {Edge}, \& {van Speybroeck}}]{ebeling07}
{Ebeling}, H., {Barrett}, E., {Donovan}, D., {Ma}, C.-J., {Edge}, A.~C., \&
  {van Speybroeck}, L. 2007, \apjl, 661, L33

\bibitem[{{Ebeling} {et~al.}(2001){Ebeling}, {Edge}, \& {Henry}}]{ebeling01}
{Ebeling}, H., {Edge}, A.~C., \& {Henry}, J.~P. 2001, \apj, 553, 668

\bibitem[{{Edge} {et~al.}(2003){Edge}, {Ebeling}, {Bremer}, {R{\"o}ttgering},
  {van Haarlem}, {Rengelink}, \& {Courtney}}]{edge03}
{Edge}, A.~C., {Ebeling}, H., {Bremer}, M., {R{\"o}ttgering}, H., {van
  Haarlem}, M.~P., {Rengelink}, R., \& {Courtney}, N.~J.~D. 2003, \mnras, 339,
  913

\bibitem[{{Feldman} {et~al.}(2010){Feldman}, {Watkins}, \&
  {Hudson}}]{feldman10}
{Feldman}, H.~A., {Watkins}, R., \& {Hudson}, M.~J. 2010, \mnras, 407, 2328

\bibitem[{{Finn} {et~al.}(2010){Finn}, {Desai}, {Rudnick}, {Poggianti}, {Bell},
  {Hinz}, {Jablonka}, {Milvang-Jensen}, {Moustakas}, {Rines}, \&
  {Zaritsky}}]{finn10}
{Finn}, R.~A., {et~al.} 2010, \apj, 720, 87

\bibitem[{{Fruscione} {et~al.}(2006){Fruscione}, {McDowell}, {Allen},
  {Brickhouse}, {Burke}, {Davis}, {Durham}, {Elvis}, {Galle}, {Harris},
  {Huenemoerder}, {Houck}, {Ishibashi}, {Karovska}, {Nicastro}, {Noble},
  {Nowak}, {Primini}, {Siemiginowska}, {Smith}, \& {Wise}}]{Fruscione2006}
{Fruscione}, A., {et~al.} 2006, in Society of Photo-Optical Instrumentation
  Engineers (SPIE) Conference Series, Vol. 6270, Society of Photo-Optical
  Instrumentation Engineers (SPIE) Conference Series

\bibitem[{{Geach} {et~al.}(2006){Geach}, {Smail}, {Ellis}, {Moran}, {Smith},
  {Treu}, {Kneib}, {Edge}, \& {Kodama}}]{geach06}
{Geach}, J.~E., {et~al.} 2006, \apj, 649, 661

\bibitem[{{Glenn} {et~al.}(2002){Glenn}, {Knowles}, {Rownd}, {Edgington},
  {Golwala}, {Lange}, {Bock}, {Goldin}, {Nguyen}, {Ade}, {Haig}, \&
  {Mauskopf}}]{glenn02}
{Glenn}, J., {et~al.} 2002, in Astronomical Society of the Pacific Conference
  Series, Vol. 283, A New Era in Cosmology, ed. N.~{Metcalfe} \& T.~{Shanks},
  398

\bibitem[{{Griffin} \& {Orton}(1993)}]{griffin93}
{Griffin}, M.~J., \& {Orton}, G.~S. 1993, Icarus, 105, 537

\bibitem[{{Haig} {et~al.}(2004){Haig}, {Ade}, {Aguirre}, {Bock}, {Edgington},
  {Enoch}, {Glenn}, {Goldin}, {Golwala}, {Heng}, {Laurent}, {Maloney},
  {Mauskopf}, {Rossinot}, {Sayers}, {Stover}, \& {Tucker}}]{haig04}
{Haig}, D.~J., {et~al.} 2004, in Society of Photo-Optical Instrumentation
  Engineers (SPIE) Conference Series, Vol. 5498, Society of Photo-Optical
  Instrumentation Engineers (SPIE) Conference Series, ed. C.~M. {Bradford},
  P.~A.~R. {Ade}, J.~E. {Aguirre}, J.~J. {Bock}, M.~{Dragovan}, L.~{Duband},
  L.~{Earle}, J.~{Glenn}, H.~{Matsuhara}, B.~J. {Naylor}, H.~T. {Nguyen},
  M.~{Yun}, \& J.~{Zmuidzinas}, 78--94

\bibitem[{{Hall} {et~al.}(2010){Hall}, {Keisler}, {Knox}, {Reichardt}, {Ade},
  {Aird}, {Benson}, {Bleem}, {Carlstrom}, {Chang}, {Cho}, {Crawford}, {Crites},
  {de Haan}, {Dobbs}, {George}, {Halverson}, {Holder}, {Holzapfel}, {Hrubes},
  {Joy}, {Lee}, {Leitch}, {Lueker}, {McMahon}, {Mehl}, {Meyer}, {Mohr},
  {Montroy}, {Padin}, {Plagge}, {Pryke}, {Ruhl}, {Schaffer}, {Shaw},
  {Shirokoff}, {Spieler}, {Stalder}, {Staniszewski}, {Stark}, {Switzer},
  {Vanderlinde}, {Vieira}, {Williamson}, \& {Zahn}}]{hall10}
{Hall}, N.~R., {et~al.} 2010, \apj, 718, 632

\bibitem[{{Hand} {et~al.}(2012){Hand}, {Addison}, {Aubourg}, {Battaglia},
  {Battistelli}, {Bizyaev}, {Bond}, {Brewington}, {Brinkmann}, {Brown}, {Das},
  {Dawson}, {Devlin}, {Dunkley}, {Dunner}, {Eisenstein}, {Fowler}, {Gralla},
  {Hajian}, {Halpern}, {Hilton}, {Hincks}, {Hlozek}, {Hughes}, {Infante},
  {Irwin}, {Kosowsky}, {Lin}, {Malanushenko}, {Malanushenko}, {Marriage},
  {Marsden}, {Menanteau}, {Moodley}, {Niemack}, {Nolta}, {Oravetz}, {Page},
  {Palanque-Delabrouille}, {Pan}, {Reese}, {Schlegel}, {Schneider}, {Sehgal},
  {Shelden}, {Sievers}, {Sif{\'o}n}, {Simmons}, {Snedden}, {Spergel}, {Staggs},
  {Swetz}, {Switzer}, {Trac}, {Weaver}, {Wollack}, {Yeche}, \&
  {Zunckel}}]{hand12}
{Hand}, N., {et~al.} 2012, Physical Review Letters, 109, 041101

\bibitem[{{Hasselfield} {et~al.}(2013){Hasselfield}, {Moodley}, {Bond}, {Das},
  {Devlin}, {Dunkley}, {Dunner}, {Fowler}, {Gallardo}, {Gralla}, {Hajian},
  {Halpern}, {Hincks}, {Marriage}, {Marsden}, {Niemack}, {Nolta}, {Page},
  {Partridge}, {Schmitt}, {Sehgal}, {Sievers}, {Staggs}, {Swetz}, {Switzer}, \&
  {Wollack}}]{hasselfield13_planet}
{Hasselfield}, M., {et~al.} 2013, ArXiv:1303.4714

\bibitem[{{Hayashi} \& {White}(2006)}]{hayashi06}
{Hayashi}, E., \& {White}, S.~D.~M. 2006, \mnras, 370, L38

\bibitem[{{Hinshaw} {et~al.}(2012){Hinshaw}, {Larson}, {Komatsu}, {Spergel},
  {Bennett}, {Dunkley}, {Nolta}, {Halpern}, {Hill}, {Odegard}, {Page}, {Smith},
  {Weiland}, {Gold}, {Jarosik}, {Kogut}, {Limon}, {Meyer}, {Tucker}, {Wollack},
  \& {Wright}}]{hinshaw12}
{Hinshaw}, G., {et~al.} 2012, ArXiv:1212.5226

\bibitem[{{Holzapfel} {et~al.}(1997){Holzapfel}, {Ade}, {Church}, {Mauskopf},
  {Rephaeli}, {Wilbanks}, \& {Lange}}]{holzapfel97}
{Holzapfel}, W.~L., {Ade}, P.~A.~R., {Church}, S.~E., {Mauskopf}, P.~D.,
  {Rephaeli}, Y., {Wilbanks}, T.~M., \& {Lange}, A.~E. 1997, \apj, 481, 35

\bibitem[{{Itoh} {et~al.}(1998){Itoh}, {Kohyama}, \& {Nozawa}}]{itoh98}
{Itoh}, N., {Kohyama}, Y., \& {Nozawa}, S. 1998, \apj, 502, 7

\bibitem[{{Itoh} \& {Nozawa}(2004)}]{itoh04}
{Itoh}, N., \& {Nozawa}, S. 2004, \aap, 417, 827

\bibitem[{{Jauzac} {et~al.}(2012){Jauzac}, {Jullo}, {Kneib}, {Ebeling},
  {Leauthaud}, {Ma}, {Limousin}, {Massey}, \& {Richard}}]{jauzac12}
{Jauzac}, M., {et~al.} 2012, \mnras, 426, 3369

\bibitem[{{Kitayama} {et~al.}(2004){Kitayama}, {Komatsu}, {Ota}, {Kuwabara},
  {Suto}, {Yoshikawa}, {Hattori}, \& {Matsuo}}]{kitayama04}
{Kitayama}, T., {Komatsu}, E., {Ota}, N., {Kuwabara}, T., {Suto}, Y.,
  {Yoshikawa}, K., {Hattori}, M., \& {Matsuo}, H. 2004, \pasj, 56, 17

\bibitem[{{Kosowsky} \& {Bhattacharya}(2009)}]{kosowsky09}
{Kosowsky}, A., \& {Bhattacharya}, S. 2009, \prd, 80, 062003

\bibitem[{{Kuntz} \& {Snowden}(2008)}]{kuntz2008}
{Kuntz}, K.~D., \& {Snowden}, S.~L. 2008, \apj, 674, 209

\bibitem[{{Lee} \& {Komatsu}(2010)}]{lee10}
{Lee}, J., \& {Komatsu}, E. 2010, \apj, 718, 60

\bibitem[{{Levenson} {et~al.}(2010){Levenson}, {Marsden}, {Zemcov}, {Amblard},
  {Blain}, {Bock}, {Chapin}, {Conley}, {Cooray}, {Dowell}, {Ellsworth-Bowers},
  {Franceschini}, {Glenn}, {Griffin}, {Halpern}, {Nguyen}, {Oliver}, {Page},
  {Papageorgiou}, {P{\'e}rez-Fournon}, {Pohlen}, {Rangwala}, {Rowan-Robinson},
  {Schulz}, {Scott}, {Serra}, {Shupe}, {Valiante}, {Vieira}, {Vigroux},
  {Wiebe}, {Wright}, \& {Xu}}]{levenson10}
{Levenson}, L., {et~al.} 2010, \mnras, 409, 83

\bibitem[{{Li} {et~al.}(2012){Li}, {Jia}, {Chen}, {Xiang}, {Wang}, \&
  {Zhao}}]{li12}
{Li}, C.~K., {Jia}, S.~M., {Chen}, Y., {Xiang}, F., {Wang}, Y.~S., \& {Zhao},
  H.~H. 2012, \aap, 545, A100

\bibitem[{{Limousin} {et~al.}(2012){Limousin}, {Ebeling}, {Richard},
  {Swinbank}, {Smith}, {Jauzac}, {Rodionov}, {Ma}, {Smail}, {Edge}, {Jullo}, \&
  {Kneib}}]{limousin12}
{Limousin}, M., {et~al.} 2012, \aap, 544, A71

\bibitem[{{Ma} {et~al.}(2009){Ma}, {Ebeling}, \& {Barrett}}]{ma09}
{Ma}, C.-J., {Ebeling}, H., \& {Barrett}, E. 2009, \apjl, 693, L56

\bibitem[{{Ma} {et~al.}(2012){Ma}, {Branchini}, \& {Scott}}]{ma12}
{Ma}, Y.-Z., {Branchini}, E., \& {Scott}, D. 2012, \mnras, 425, 2880

\bibitem[{{Mahdavi} {et~al.}(2013){Mahdavi}, {Hoekstra}, {Babul}, {Bildfell},
  {Jeltema}, \& {Henry}}]{mahdavi13}
{Mahdavi}, A., {Hoekstra}, H., {Babul}, A., {Bildfell}, C., {Jeltema}, T., \&
  {Henry}, J.~P. 2013, \apj, 767, 116

\bibitem[{{Mak} {et~al.}(2011){Mak}, {Pierpaoli}, \& {Osborne}}]{mak11}
{Mak}, D.~S.~Y., {Pierpaoli}, E., \& {Osborne}, S.~J. 2011, \apj, 736, 116

\bibitem[{{Mantz} {et~al.}(2010){Mantz}, {Allen}, {Ebeling}, {Rapetti}, \&
  {Drlica-Wagner}}]{mantz10}
{Mantz}, A., {Allen}, S.~W., {Ebeling}, H., {Rapetti}, D., \& {Drlica-Wagner},
  A. 2010, \mnras, 406, 1773

\bibitem[{{Marcillac} {et~al.}(2007){Marcillac}, {Rigby}, {Rieke}, \&
  {Kelly}}]{marcillac07}
{Marcillac}, D., {Rigby}, J.~R., {Rieke}, G.~H., \& {Kelly}, D.~M. 2007, \apj,
  654, 825

\bibitem[{{Markwardt}(2009)}]{markwardt09}
{Markwardt}, C.~B. 2009, in Astronomical Society of the Pacific Conference
  Series, Vol. 411, Astronomical Data Analysis Software and Systems XVIII, ed.
  D.~A. {Bohlender}, D.~{Durand}, \& P.~{Dowler}, 251

\bibitem[{{Mauskopf} {et~al.}(2012){Mauskopf}, {Horner}, {Aguirre}, {Bock},
  {Egami}, {Glenn}, {Golwala}, {Laurent}, {Nguyen}, \& {Sayers}}]{mauskopf12}
{Mauskopf}, P.~D., {et~al.} 2012, \mnras, 421, 224

\bibitem[{{Medezinski} {et~al.}(2013){Medezinski}, {Umetsu}, {Nonino},
  {Merten}, {Zitrin}, {Broadhurst}, {Donahue}, {Sayers}, {Waizmann},
  {Koekemoer}, {Coe}, {Molino}, {Melchior}, {Mroczkowski}, {Czakon}, {Postman},
  {Meneghetti}, {Lemze}, {Ford}, {Grillo}, {Kelson}, {Bradley}, {Moustakas},
  {Bartelmann}, {Ben{\'{\i}}tez}, {Biviano}, {Bouwens}, {Golwala}, {Graves},
  {Infante}, {Jim{\'e}nez-Teja}, {Jouvel}, {Lahav}, {Moustakas}, {Ogaz},
  {Rosati}, {Seitz}, \& {Zheng}}]{medezinski13}
{Medezinski}, E., {et~al.} 2013, ArXiv:1304.1223

\bibitem[{{Meneghetti} {et~al.}(2011){Meneghetti}, {Fedeli}, {Zitrin},
  {Bartelmann}, {Broadhurst}, {Gottl{\"o}ber}, {Moscardini}, \&
  {Yepes}}]{meneghetti11}
{Meneghetti}, M., {Fedeli}, C., {Zitrin}, A., {Bartelmann}, M., {Broadhurst},
  T., {Gottl{\"o}ber}, S., {Moscardini}, L., \& {Yepes}, G. 2011, \aap, 530,
  A17

\bibitem[{{Mroczkowski} {et~al.}(2012){Mroczkowski}, {Dicker}, {Sayers},
  {Reese}, {Mason}, {Czakon}, {Romero}, {Young}, {Devlin}, {Golwala},
  {Korngut}, {Sarazin}, {Bock}, {Koch}, {Lin}, {Molnar}, {Pierpaoli}, {Umetsu},
  \& {Zemcov}}]{mroczkowski12}
{Mroczkowski}, T., {et~al.} 2012, \apj, 761, 47

\bibitem[{{Nevalainen} {et~al.}(2010){Nevalainen}, {David}, \&
  {Guainazzi}}]{nevalainen10}
{Nevalainen}, J., {David}, L., \& {Guainazzi}, M. 2010, \aap, 523, A22

\bibitem[{{Nguyen} {et~al.}(2010){Nguyen}, {Schulz}, {Levenson}, {Amblard},
  {Arumugam}, {Aussel}, {Babbedge}, {Blain}, {Bock}, {Boselli}, {Buat},
  {Castro-Rodriguez}, {Cava}, {Chanial}, {Chapin}, {Clements}, {Conley},
  {Conversi}, {Cooray}, {Dowell}, {Dwek}, {Eales}, {Elbaz}, {Fox},
  {Franceschini}, {Gear}, {Glenn}, {Griffin}, {Halpern}, {Hatziminaoglou},
  {Ibar}, {Isaak}, {Ivison}, {Lagache}, {Lu}, {Madden}, {Maffei}, {Mainetti},
  {Marchetti}, {Marsden}, {Marshall}, {O'Halloran}, {Oliver}, {Omont}, {Page},
  {Panuzzo}, {Papageorgiou}, {Pearson}, {Perez Fournon}, {Pohlen}, {Rangwala},
  {Rigopoulou}, {Rizzo}, {Roseboom}, {Rowan-Robinson}, {Scott}, {Seymour},
  {Shupe}, {Smith}, {Stevens}, {Symeonidis}, {Trichas}, {Tugwell}, {Vaccari},
  {Valtchanov}, {Vigroux}, {Wang}, {Ward}, {Wiebe}, {Wright}, {Xu}, \&
  {Zemcov}}]{nguyen10}
{Nguyen}, H.~T., {et~al.} 2010, \aap, 518, L5

\bibitem[{{Nozawa} {et~al.}(1998{\natexlab{a}}){Nozawa}, {Itoh}, \&
  {Kohyama}}]{nozawa98b}
{Nozawa}, S., {Itoh}, N., \& {Kohyama}, Y. 1998{\natexlab{a}}, \apj, 508, 17

\bibitem[{{Nozawa} {et~al.}(1998{\natexlab{b}}){Nozawa}, {Itoh}, \&
  {Kohyama}}]{nozawa98}
---. 1998{\natexlab{b}}, \apj, 507, 530

\bibitem[{{Nozawa} {et~al.}(2006){Nozawa}, {Itoh}, {Suda}, \&
  {Ohhata}}]{nozawa06}
{Nozawa}, S., {Itoh}, N., {Suda}, Y., \& {Ohhata}, Y. 2006, Nuovo Cimento B
  Serie, 121, 487

\bibitem[{{Nusser} \& {Davis}(2011)}]{nusser11}
{Nusser}, A., \& {Davis}, M. 2011, \apj, 736, 93

\bibitem[{{Osborne} {et~al.}(2011){Osborne}, {Mak}, {Church}, \&
  {Pierpaoli}}]{osborne11}
{Osborne}, S.~J., {Mak}, D.~S.~Y., {Church}, S.~E., \& {Pierpaoli}, E. 2011,
  \apj, 737, 98

\bibitem[{{Ott}(2010)}]{ott10}
{Ott}, S. 2010, in Astronomical Society of the Pacific Conference Series, Vol.
  434, Astronomical Data Analysis Software and Systems XIX, ed. Y.~{Mizumoto},
  K.-I. {Morita}, \& M.~{Ohishi}, 139

\bibitem[{{Ott} {et~al.}(2006){Ott}, {Bakker}, {Brumfitt}, {de Candussio},
  {Dwedari}, {Heras}, {Leeks}, {Marston}, {Mathieu}, {Pizarro}, {Siddiqui},
  {Ali}, {Latter}, {Morris}, {Rector}, {Schulz}, {Corrales Garcia}, {De
  Meester}, {Huygen}, {Vandenbussche}, {Guest}, {Kemp}, {Kester}, {Shipman},
  {Zaal}, {Lorenzani}, {Sturm}, {Wetzstein}, \& {Wieprecht}}]{ott06}
{Ott}, S., {et~al.} 2006, in Astronomical Society of the Pacific Conference
  Series, Vol. 351, Astronomical Data Analysis Software and Systems XV, ed.
  C.~{Gabriel}, C.~{Arviset}, D.~{Ponz}, \& S.~{Enrique}, 516

\bibitem[{{Planck Collaboration} {et~al.}(2013{\natexlab{a}}){Planck
  Collaboration}, {Ade}, {Aghanim}, {Armitage-Caplan}, {Arnaud}, {Ashdown},
  {Atrio-Barandela}, {Aumont}, {Baccigalupi}, {Banday}, \&
  et~al.}]{planck13_calibration}
{Planck Collaboration} {et~al.} 2013{\natexlab{a}}, ArXiv:1303.5069

\bibitem[{{Planck Collaboration} {et~al.}(2013{\natexlab{b}}){Planck
  Collaboration}, {Ade}, {Aghanim}, {Armitage-Caplan}, {Arnaud}, {Ashdown},
  {Atrio-Barandela}, {Aumont}, {Baccigalupi}, {Banday}, \&
  et~al.}]{planck13_cmb1}
---. 2013{\natexlab{b}}, ArXiv:1303.5075

\bibitem[{{Planck Collaboration} {et~al.}(2013{\natexlab{c}}){Planck
  Collaboration}, {Ade}, {Aghanim}, {Armitage-Caplan}, {Arnaud}, {Ashdown},
  {Atrio-Barandela}, {Aumont}, {Baccigalupi}, {Banday}, \&
  et~al.}]{planck13_cmb2}
---. 2013{\natexlab{c}}, ArXiv:1303.5076

\bibitem[{{Planck Collaboration} {et~al.}(2013{\natexlab{d}}){Planck
  Collaboration}, {Ade}, {Aghanim}, {Arnaud}, {Ashdown}, {Aumont},
  {Baccigalupi}, {Balbi}, {Banday}, {Barreiro}, {Battaner}, {Benabed},
  {Benoit-Levy}, {Bernard}, {Bersanelli}, {Bielewicz}, {Bikmaev}, {Bobin},
  {Bock}, {Bonaldi}, {Bond}, {Borrill}, {Bouchet}, {Burigana}, {Butler},
  {Cabella}, {Cardoso}, {Catalano}, {Chamballu}, {Chiang}, {Chon},
  {Christensen}, {Clements}, {Colombi}, {Colombo}, {Crill}, {Cuttaia}, {Da
  Silva}, {Dahle}, {Davies}, {Davis}, {de Bernardis}, {de Gasperis}, {de
  Zotti}, {Delabrouille}, {Democles}, {Diego}, {Dolag}, {Dole}, {Donzelli},
  {Dore}, {Doerl}, {Douspis}, {Dupac}, {Ensslin}, {Finelli}, {Flores-Cacho},
  {Forni}, {Frailis}, {Frommert}, {Galeotta}, {Ganga}, {Genova-Santos},
  {Giard}, {Giardino}, {Gonzalez-Nuevo}, {Gregorio}, {Gruppuso}, {Hansen},
  {Harrison}, {Hernandez-Monteagudo}, {Herranz}, {Hildebrandt}, {Hivon},
  {Holmes}, {Hovest}, {Huffenberger}, {Hurier}, {Jaffe}, {Jaffe}, {Jasche},
  {Jones}, {Juvela}, {Keihanen}, {Keskitalo}, {Khamitov}, {Kisner}, {Knoche},
  {Kunz}, {Kurki-Suonio}, {Lagache}, {Lahteenmaki}, {Lamarre}, {Lasenby},
  {Lawrence}, {Le Jeune}, {Leonardi}, {Lilje}, {Linden-Vornle},
  {Lopez-Caniego}, {Macias-Perez}, {Maino}, {Mak}, {Mandolesi}, {Maris},
  {Marleau}, {Martinez-Gonzalez}, {Masi}, {Matarrese}, {Mazzotta},
  {Melchiorri}, {Melin}, {Mendes}, {Mennella}, {Migliaccio}, {Mitra},
  {Miville-Deschenes}, {Moneti}, {Montier}, {Morgante}, {Mortlock}, {Moss},
  {Munshi}, {Murphy}, {Naselsky}, {Nati}, {Natoli}, {Netterfield},
  {Norgaard-Nielsen}, {Noviello}, {Novikov}, {Novikov}, {Osborne}, {Pagano},
  {Paoletti}, {Perdereau}, {Perrotta}, {Piacentini}, {Piat}, {Pierpaoli},
  {Pietrobon}, {Plaszczynski}, {Pointecouteau}, {Polenta}, {Popa}, {Poutanen},
  {Pratt}, {Prunet}, {Puget}, {Puisieux}, {Rachen}, {Rebolo}, {Reinecke},
  {Remazeilles}, {Renault}, {Ricciardi}, {Roman}, {Rubino-Martin}, {Rusholme},
  {Sandri}, {Savini}, {Scott}, {Spencer}, {Sunyaev}, {Sutton}, {Suur-Uski},
  {Sygnet}, {Tauber}, {Terenzi}, {Toffolatti}, {Tomasi}, {Tristram}, {Tucci},
  {Valenziano}, {Valiviita}, {Van Tent}, {Vielva}, {Villa}, {Vittorio}, {Wade},
  {Welikala}, {Yvon}, {Zacchei}, {Zibin}, \& {Zonca}}]{planck13_kSZ}
---. 2013{\natexlab{d}}, ArXiv:1303.5090

\bibitem[{{Prokhorov} \& {Colafrancesco}(2012)}]{prokhorov12}
{Prokhorov}, D.~A., \& {Colafrancesco}, S. 2012, \mnras, 424, L49

\bibitem[{{Rawle} {et~al.}(2012){Rawle}, {Edge}, {Egami}, {Rex}, {Smith},
  {Altieri}, {Fiedler}, {Haines}, {Pereira}, {P{\'e}rez-Gonz{\'a}lez},
  {Portouw}, {Valtchanov}, {Walth}, {van der Werf}, \& {Zemcov}}]{rawle12}
{Rawle}, T.~D., {et~al.} 2012, \apj, 747, 29

\bibitem[{{Reese} {et~al.}(2010){Reese}, {Kawahara}, {Kitayama}, {Ota},
  {Sasaki}, \& {Suto}}]{reese10}
{Reese}, E.~D., {Kawahara}, H., {Kitayama}, T., {Ota}, N., {Sasaki}, S., \&
  {Suto}, Y. 2010, \apj, 721, 653

\bibitem[{{Reichardt} {et~al.}(2012){Reichardt}, {Shaw}, {Zahn}, {Aird},
  {Benson}, {Bleem}, {Carlstrom}, {Chang}, {Cho}, {Crawford}, {Crites}, {de
  Haan}, {Dobbs}, {Dudley}, {George}, {Halverson}, {Holder}, {Holzapfel},
  {Hoover}, {Hou}, {Hrubes}, {Joy}, {Keisler}, {Knox}, {Lee}, {Leitch},
  {Lueker}, {Luong-Van}, {McMahon}, {Mehl}, {Meyer}, {Millea}, {Mohr},
  {Montroy}, {Natoli}, {Padin}, {Plagge}, {Pryke}, {Ruhl}, {Schaffer},
  {Shirokoff}, {Spieler}, {Staniszewski}, {Stark}, {Story}, {van Engelen},
  {Vanderlinde}, {Vieira}, \& {Williamson}}]{reichardt12}
{Reichardt}, C.~L., {et~al.} 2012, \apj, 755, 70

\bibitem[{{Rephaeli}(1995{\natexlab{a}})}]{rephaeli95}
{Rephaeli}, Y. 1995{\natexlab{a}}, \araa, 33, 541

\bibitem[{{Rephaeli}(1995{\natexlab{b}})}]{rephaeli95b}
---. 1995{\natexlab{b}}, \apj, 445, 33

\bibitem[{{Roseboom} {et~al.}(2013){Roseboom}, {Dunlop}, {Cirasuolo}, {Geach},
  {Smail}, {Halpern}, {van der Werf}, {Almaini}, {Arumugam}, {Asboth}, {Auld},
  {Blain}, {Bremer}, {Bock}, {Bowler}, {Buitrago}, {Chapin}, {Chapman},
  {Chrysostomou}, {Clarke}, {Conley}, {Coppin}, {Danielson}, {Farrah}, {Glenn},
  {Hatziminaoglou}, {Ibar}, {Ivison}, {Jenness}, {van Kampen}, {Karim},
  {Mackenzie}, {Marsden}, {Meijerink}, {Micha{\l}owski}, {Oliver}, {Page},
  {Pearson}, {Scott}, {Simpson}, {Smith}, {Spaans}, {Swinbank}, {Symeonidis},
  {Targett}, {Valiante}, {Viero}, {Wang}, {Willott}, \& {Zemcov}}]{roseboom13}
{Roseboom}, I.~G., {et~al.} 2013, ArXiv:1308.4443

\bibitem[{{Ruan} {et~al.}(2013){Ruan}, {Quinn}, \& {Babul}}]{ruan13}
{Ruan}, J.~J., {Quinn}, T.~R., \& {Babul}, A. 2013, \mnras

\bibitem[{{Sanders}(2006)}]{Sanders2006}
{Sanders}, J.~S. 2006, \mnras, 371, 829

\bibitem[{{Sayers} {et~al.}(2012){Sayers}, {Czakon}, \&
  {Golwala}}]{sayers12_planet}
{Sayers}, J., {Czakon}, N.~G., \& {Golwala}, S.~R. 2012, \apj, 744, 169

\bibitem[{{Sayers} {et~al.}(2013{\natexlab{a}}){Sayers}, {Czakon}, {Mantz},
  {Golwala}, {Ameglio}, {Downes}, {Koch}, {Lin}, {Maughan}, {Molnar},
  {Moustakas}, {Mroczkowski}, {Pierpaoli}, {Shitanishi}, {Siegel}, {Umetsu}, \&
  {Van der Pyl}}]{sayers13_pressure}
{Sayers}, J., {et~al.} 2013{\natexlab{a}}, \apj, 768, 177

\bibitem[{{Sayers} {et~al.}(2011){Sayers}, {Golwala}, {Ameglio}, \&
  {Pierpaoli}}]{sayers11}
{Sayers}, J., {Golwala}, S.~R., {Ameglio}, S., \& {Pierpaoli}, E. 2011, \apj,
  728, 39

\bibitem[{{Sayers} {et~al.}(2013{\natexlab{b}}){Sayers}, {Mroczkowski},
  {Czakon}, {Golwala}, {Mantz}, {Ameglio}, {Downes}, {Koch}, {Lin}, {Molnar},
  {Moustakas}, {Muchovej}, {Pierpaoli}, {Shitanishi}, {Siegel}, \&
  {Umetsu}}]{sayers13_radio}
{Sayers}, J., {et~al.} 2013{\natexlab{b}}, \apj, 764, 152

\bibitem[{{Sazonov} \& {Sunyaev}(1998)}]{sazonov98}
{Sazonov}, S.~Y., \& {Sunyaev}, R.~A. 1998, \apj, 508, 1

\bibitem[{{Smith} {et~al.}(2012){Smith}, {Wang}, {Oliver}, {Auld}, {Bock},
  {Brisbin}, {Burgarella}, {Chanial}, {Chapin}, {Clements}, {Conversi},
  {Cooray}, {Dowell}, {Eales}, {Farrah}, {Franceschini}, {Glenn}, {Griffin},
  {Ivison}, {Mortier}, {Page}, {Papageorgiou}, {Pearson}, {P{\'e}rez-Fournon},
  {Pohlen}, {Rawlings}, {Raymond}, {Rodighiero}, {Roseboom}, {Rowan-Robinson},
  {Savage}, {Scott}, {Seymour}, {Symeonidis}, {Tugwell}, {Vaccari},
  {Valtchanov}, {Vigroux}, {Ward}, {Wright}, \& {Zemcov}}]{smith12}
{Smith}, A.~J., {et~al.} 2012, \mnras, 419, 377

\bibitem[{{Smith} {et~al.}(2001){Smith}, {Brickhouse}, {Liedahl}, \&
  {Raymond}}]{Smith2001}
{Smith}, R.~K., {Brickhouse}, N.~S., {Liedahl}, D.~A., \& {Raymond}, J.~C.
  2001, \apjl, 556, L91

\bibitem[{{Snowden} {et~al.}(2008){Snowden}, {Mushotzky}, {Kuntz}, \&
  {Davis}}]{snowden2008}
{Snowden}, S.~L., {Mushotzky}, R.~F., {Kuntz}, K.~D., \& {Davis}, D.~S. 2008,
  \aap, 478, 615

\bibitem[{{Story} {et~al.}(2012){Story}, {Reichardt}, {Hou}, {Keisler}, {Aird},
  {Benson}, {Bleem}, {Carlstrom}, {Chang}, {Cho}, {Crawford}, {Crites}, {de
  Haan}, {Dobbs}, {Dudley}, {Follin}, {George}, {Halverson}, {Holder},
  {Holzapfel}, {Hoover}, {Hrubes}, {Joy}, {Knox}, {Lee}, {Leitch}, {Lueker},
  {Luong-Van}, {McMahon}, {Mehl}, {Meyer}, {Millea}, {Mohr}, {Montroy},
  {Padin}, {Plagge}, {Pryke}, {Ruhl}, {Sayre}, {Schaffer}, {Shaw}, {Shirokoff},
  {Spieler}, {Staniszewski}, {Stark}, {van Engelen}, {Vanderlinde}, {Vieira},
  {Williamson}, \& {Zahn}}]{story12}
{Story}, K.~T., {et~al.} 2012, ArXiv:1210.7231

\bibitem[{{Sunyaev} \& {Zel'dovich}(1972)}]{sunyaev72}
{Sunyaev}, R.~A., \& {Zel'dovich}, Y.~B. 1972, Comments on Astrophysics and
  Space Physics, 4, 173

\bibitem[{{Suzuki} {et~al.}(2012){Suzuki}, {Rubin}, {Lidman}, {Aldering},
  {Amanullah}, {Barbary}, {Barrientos}, {Botyanszki}, {Brodwin}, {Connolly},
  {Dawson}, {Dey}, {Doi}, {Donahue}, {Deustua}, {Eisenhardt}, {Ellingson},
  {Faccioli}, {Fadeyev}, {Fakhouri}, {Fruchter}, {Gilbank}, {Gladders},
  {Goldhaber}, {Gonzalez}, {Goobar}, {Gude}, {Hattori}, {Hoekstra}, {Hsiao},
  {Huang}, {Ihara}, {Jee}, {Johnston}, {Kashikawa}, {Koester}, {Konishi},
  {Kowalski}, {Linder}, {Lubin}, {Melbourne}, {Meyers}, {Morokuma}, {Munshi},
  {Mullis}, {Oda}, {Panagia}, {Perlmutter}, {Postman}, {Pritchard}, {Rhodes},
  {Ripoche}, {Rosati}, {Schlegel}, {Spadafora}, {Stanford}, {Stanishev},
  {Stern}, {Strovink}, {Takanashi}, {Tokita}, {Wagner}, {Wang}, {Yasuda},
  {Yee}, \& {Supernova Cosmology Project}}]{suzuki12}
{Suzuki}, N., {et~al.} 2012, \apj, 746, 85

\bibitem[{{Thompson} \& {Nagamine}(2012)}]{thompson12}
{Thompson}, R., \& {Nagamine}, K. 2012, \mnras, 419, 3560

\bibitem[{{Tully} \& {Fisher}(1977)}]{tully77}
{Tully}, R.~B., \& {Fisher}, J.~R. 1977, \aap, 54, 661

\bibitem[{{van Weeren} {et~al.}(2009){van Weeren}, {R{\"o}ttgering},
  {Br{\"u}ggen}, \& {Cohen}}]{vanweeren09}
{van Weeren}, R.~J., {R{\"o}ttgering}, H.~J.~A., {Br{\"u}ggen}, M., \& {Cohen},
  A. 2009, \aap, 505, 991

\bibitem[{{Viero} {et~al.}(2013){Viero}, {Wang}, {Zemcov}, {Addison},
  {Amblard}, {Arumugam}, {Aussel}, {B{\'e}thermin}, {Bock}, {Boselli}, {Buat},
  {Burgarella}, {Casey}, {Clements}, {Conley}, {Conversi}, {Cooray}, {De
  Zotti}, {Dowell}, {Farrah}, {Franceschini}, {Glenn}, {Griffin},
  {Hatziminaoglou}, {Heinis}, {Ibar}, {Ivison}, {Lagache}, {Levenson},
  {Marchetti}, {Marsden}, {Nguyen}, {O'Halloran}, {Oliver}, {Omont}, {Page},
  {Papageorgiou}, {Pearson}, {P{\'e}rez-Fournon}, {Pohlen}, {Rigopoulou},
  {Roseboom}, {Rowan-Robinson}, {Schulz}, {Scott}, {Seymour}, {Shupe}, {Smith},
  {Symeonidis}, {Vaccari}, {Valtchanov}, {Vieira}, {Wardlow}, \&
  {Xu}}]{viero12}
{Viero}, M.~P., {et~al.} 2013, \apj, 772, 77

\bibitem[{{Waizmann} {et~al.}(2012){Waizmann}, {Redlich}, \&
  {Bartelmann}}]{waizmann12}
{Waizmann}, J.-C., {Redlich}, M., \& {Bartelmann}, M. 2012, \aap, 547, A67

\bibitem[{{Wardlow} {et~al.}(2010){Wardlow}, {Smail}, {Wilson}, {Yun},
  {Coppin}, {Cybulski}, {Geach}, {Ivison}, {Aretxaga}, {Austermann}, {Edge},
  {Fazio}, {Huang}, {Hughes}, {Kodama}, {Kang}, {Kim}, {Mauskopf}, {Perera}, \&
  {Scott}}]{wardlow10}
{Wardlow}, J.~L., {et~al.} 2010, \mnras, 401, 2299

\bibitem[{{Weiland} {et~al.}(2011){Weiland}, {Odegard}, {Hill}, {Wollack},
  {Hinshaw}, {Greason}, {Jarosik}, {Page}, {Bennett}, {Dunkley}, {Gold},
  {Halpern}, {Kogut}, {Komatsu}, {Larson}, {Limon}, {Meyer}, {Nolta}, {Smith},
  {Spergel}, {Tucker}, \& {Wright}}]{weiland11}
{Weiland}, J.~L., {et~al.} 2011, \apjs, 192, 19

\bibitem[{{Wu} {et~al.}(2012){Wu}, {Tsai}, {Sayers}, {Benford}, {Bridge},
  {Blain}, {Eisenhardt}, {Stern}, {Petty}, {Assef}, {Bussmann}, {Comerford},
  {Cutri}, {Evans}, {Griffith}, {Jarrett}, {Lake}, {Lonsdale}, {Rho},
  {Stanford}, {Weiner}, {Wright}, \& {Yan}}]{wu12}
{Wu}, J., {et~al.} 2012, \apj, 756, 96

\bibitem[{{Zahn} {et~al.}(2012){Zahn}, {Reichardt}, {Shaw}, {Lidz}, {Aird},
  {Benson}, {Bleem}, {Carlstrom}, {Chang}, {Cho}, {Crawford}, {Crites}, {de
  Haan}, {Dobbs}, {Dor{\'e}}, {Dudley}, {George}, {Halverson}, {Holder},
  {Holzapfel}, {Hoover}, {Hou}, {Hrubes}, {Joy}, {Keisler}, {Knox}, {Lee},
  {Leitch}, {Lueker}, {Luong-Van}, {McMahon}, {Mehl}, {Meyer}, {Millea},
  {Mohr}, {Montroy}, {Natoli}, {Padin}, {Plagge}, {Pryke}, {Ruhl}, {Schaffer},
  {Shirokoff}, {Spieler}, {Staniszewski}, {Stark}, {Story}, {van Engelen},
  {Vanderlinde}, {Vieira}, \& {Williamson}}]{zahn12}
{Zahn}, O., {et~al.} 2012, \apj, 756, 65

\bibitem[{{Zemcov} {et~al.}(2012){Zemcov}, {Aguirre}, {Bock}, {Bradford},
  {Czakon}, {Glenn}, {Golwala}, {Lupu}, {Maloney}, {Mauskopf}, {Million},
  {Murphy}, {Naylor}, {Nguyen}, {Rosenman}, {Sayers}, {Scott}, \&
  {Zmuidzinas}}]{zemcov12}
{Zemcov}, M., {et~al.} 2012, \apj, 749, 114

\bibitem[{{Zemcov} {et~al.}(2013){Zemcov}, {Blain}, {Cooray}, {B{\'e}thermin},
  {Bock}, {Clements}, {Conley}, {Conversi}, {Dowell}, {Farrah}, {Glenn},
  {Griffin}, {Halpern}, {Jullo}, {Kneib}, {Marsden}, {Nguyen}, {Oliver},
  {Richard}, {Roseboom}, {Schulz}, {Scott}, {Shupe}, {Smith}, {Valtchanov},
  {Viero}, {Wang}, \& {Wardlow}}]{zemcov13}
---. 2013, \apjl, 769, L31

\bibitem[{{Zhuravleva} {et~al.}(2013){Zhuravleva}, {Churazov}, {Kravtsov},
  {Lau}, {Nagai}, \& {Sunyaev}}]{Zhuravleva2013}
{Zhuravleva}, I., {Churazov}, E., {Kravtsov}, A., {Lau}, E.~T., {Nagai}, D., \&
  {Sunyaev}, R. 2013, \mnras, 428, 3274

\bibitem[{{Zitrin} {et~al.}(2009){Zitrin}, {Broadhurst}, {Rephaeli}, \&
  {Sadeh}}]{zitrin09}
{Zitrin}, A., {Broadhurst}, T., {Rephaeli}, Y., \& {Sadeh}, S. 2009, \apjl,
  707, L102

\end{thebibliography}

\appendix
\label{sec:appen}

\begin{figure}
  \centering
  \includegraphics[width=.4\textwidth]{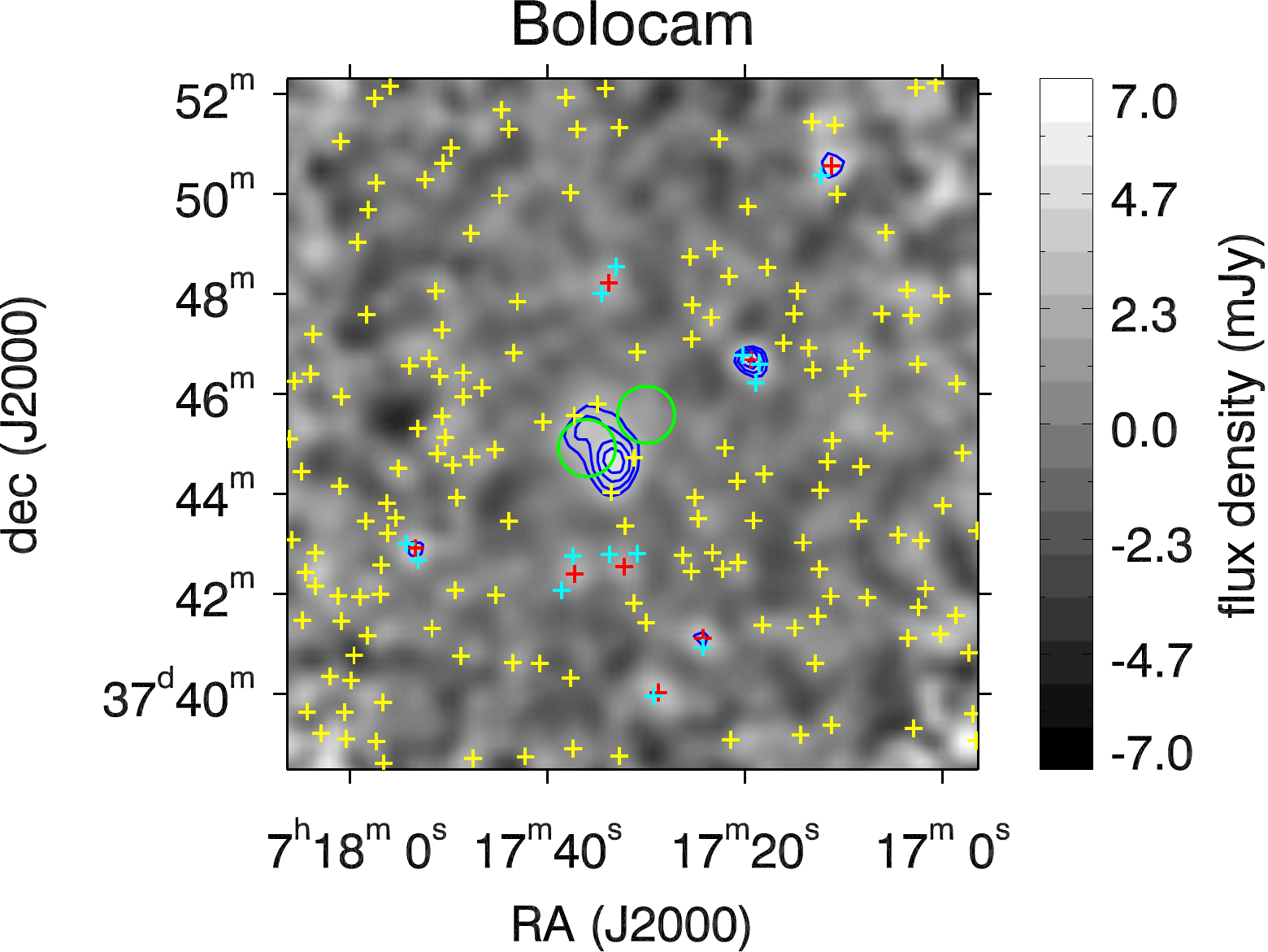}
  \hspace{.05\textwidth}
  \includegraphics[width=.4\textwidth]{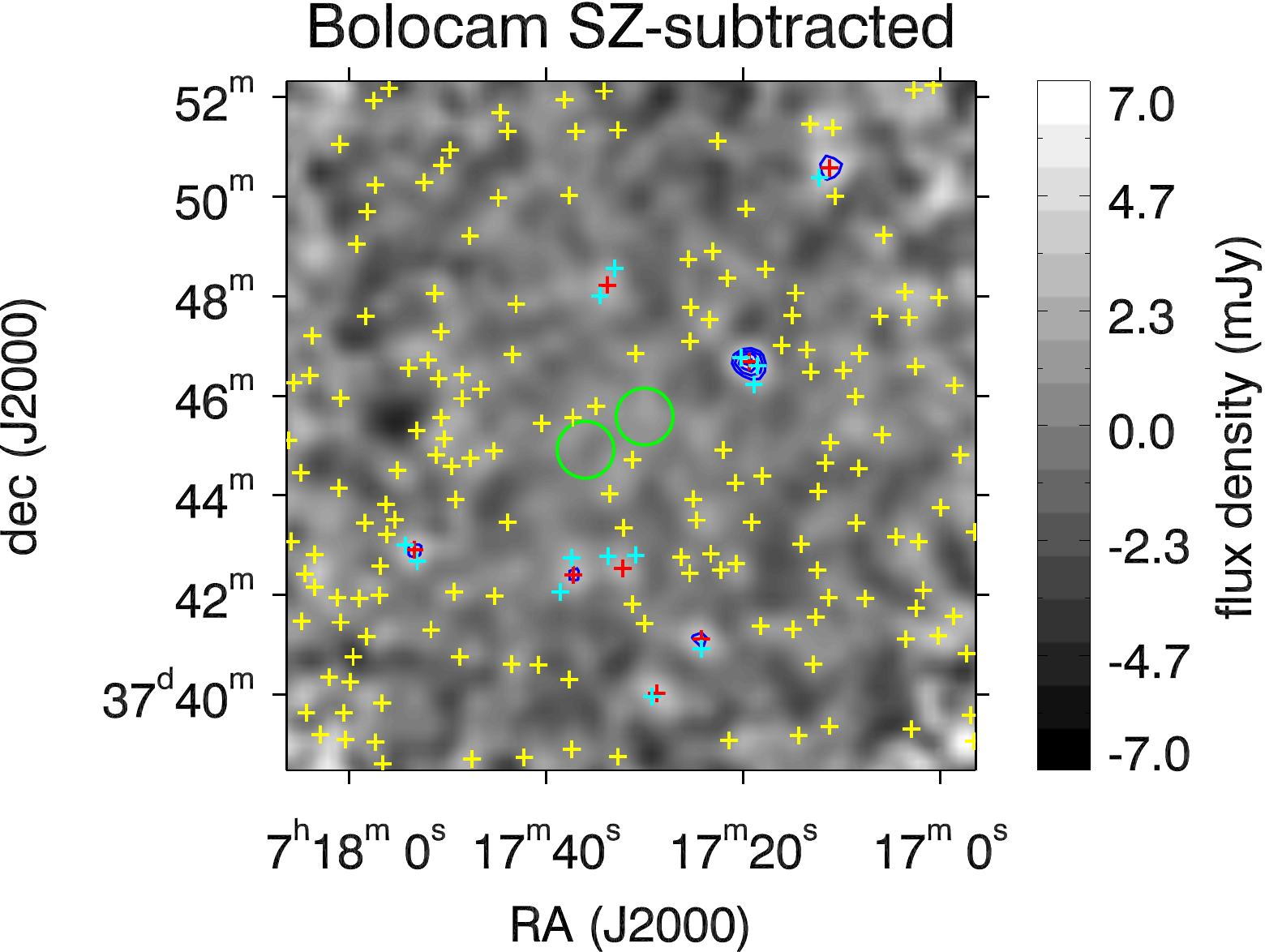}

  \vspace{15pt}
  
  \centering
  \includegraphics[width=.4\textwidth]{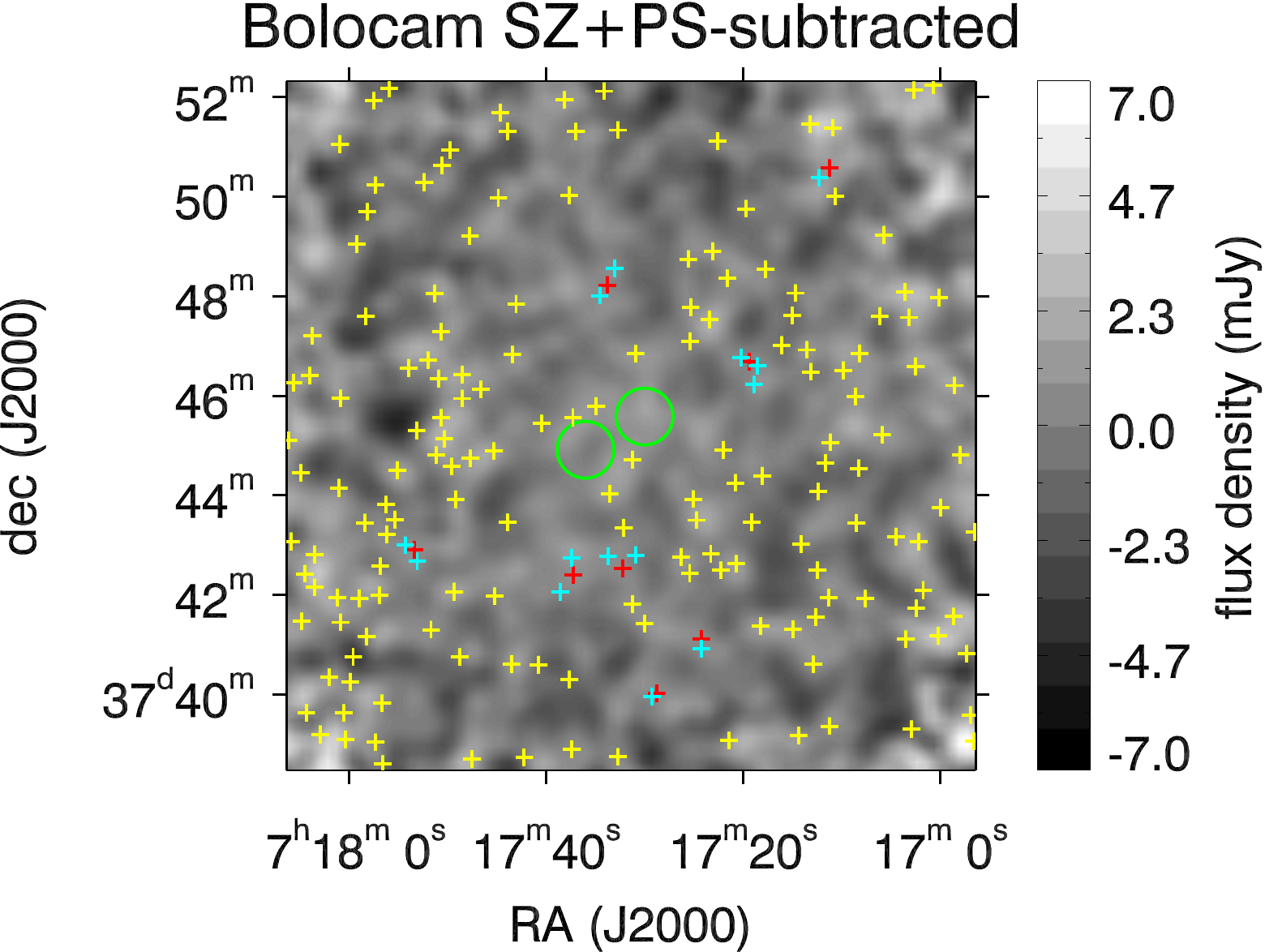}
  \hspace{.05\textwidth}
  \includegraphics[width=.4\textwidth]{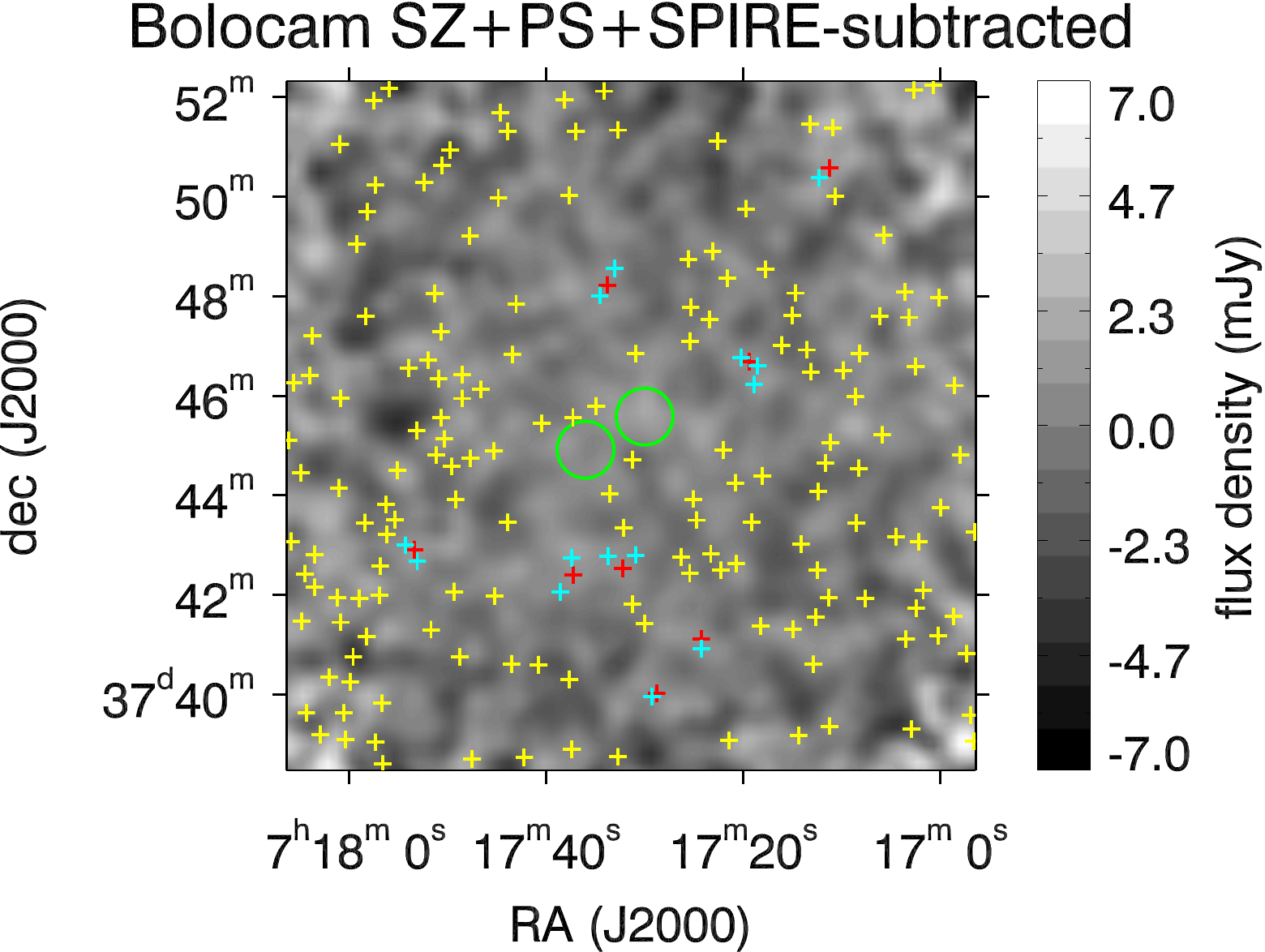}
  \caption{Bolocam 268~GHz images we obtain from the adaptive-PCA reduction. 
    Clockwise from upper left:
    nominal image, image after subtracting the best-fit extended-SZ template,
    image after also subtracting the 8 unresolved sources detected by Bolocam,
    and image after also subtracting the 162 sources detected by SPIRE
    and extrapolated to Bolocam's band using a greybody fit.
    The blue contours show S/N, starting at 4 and separated by 1.
    There are no contours around the bright regions in the corners 
    of the images, due to the higher noise in these regions as a result
    of reduced integration time relative to the central region.
    Note that even in this heavily filtered image, an extended SZ
    signal is detected at high significance.
    The crosses show:
    Bolocam detections (red), likely SPIRE counterparts to those detections (cyan),
    and all other SPIRE detections with an extrapolated S/N $>2$ (yellow).
    The green circles show the $1\arcmin$ diameter apertures centered 
    on sub-cluster C (lower left) and sub-cluster B (upper right).}
  \label{fig:ps_image}
\end{figure}

There is a significant amount of signal from unresolved dusty
star-forming galaxies (e.g., \citealt{blain02}) in our 268~GHz Bolocam map,
and we describe our treatment of these galaxies in this Appendix.
Unlike the resolved SZ signal we seek to measure, all of the
dusty star-forming galaxies in our Bolocam image are unresolved.
Therefore, to maximize our sensitivity to these unresolved sources,
we process the 268~GHz data using an adaptive principal component analysis (PCA)
algorithm in place of the common-mode subtraction we use for our SZ analysis.
For brevity, we refer to the maps generated by these reductions as the
adaptive-PCA map and the common-mode-subtraction map.
The details of the adaptive-PCA algorithm we use for this analysis are given in \citet{wu12}.
This processing results in an adaptive-PCA map with a noise rms of 0.7 mJy/beam, which
is approximately equal to the confusion noise from unresolved
star-forming galaxies.
We then subtract a template of the extended SZ signal from the
adaptive-PCA map by fitting a gNFW profile to the common-mode-subtraction
map, processing the gNFW model through the adaptive-PCA reduction,
and subtracting the processed model from the adaptive-PCA map (see Figure~\ref{fig:ps_image}).
The resulting extended-SZ-subtracted adaptive-PCA
map contains a total of 8 unresolved galaxy candidates with a 
S/N $>4$, and we measure the best-fit flux density for
each of these candidates after accounting for the filtering effects
of the adaptive-PCA reduction (see Table~\ref{tab:point_sources}).
Using these best-fit flux densities and positions, we then
process these 8 candidates through the common-mode reduction and
subtract them from the common-mode-subtraction map used for
our SZ analysis.
In addition, we generate 1000 random realizations of each of the 8 candidates
based on the measurement uncertainties on the best-fit flux densities,
and we add one realization of the uncertainty for
each candidate to each of the 1000 noise realizations
described in Section~\ref{sec:bolocam}.

In addition to our Bolocam data, we also search for
dusty star-forming galaxies in
three-band (250, 350, and 500~$\mu$m) observations obtained
with the SPIRE photometer.  
All of the SPIRE images are dominated by
confusion noise from unresolved galaxies \citep{nguyen10}, and the effective rms is
7.2, 5.3, and 5.8~mJy/beam at 250, 350, and 500~$\mu$m.
The SPIRE data are reduced using the {\it Herschel}
Interactive Processing Environment HIPE \citep{ott06, ott10},
along with the HerMES SMAP package \citep{levenson10, viero12}.
Source catalogs are then compiled using the SCAT procedure
\citep{smith12}, and we identify a total of 200 source candidates
within the $14\arcmin \times 14\arcmin$ Bolocam coverage with
S/N $>3$ in any of the SPIRE bands.
We note that very few of the SPIRE
detections are located within the extended SZ signal
detected with Bolocam, and none of the sources located within
the extended SZ signal are particularly bright,
indicating that there is little
contamination of the extended SZ signal due to bright dusty star-forming galaxies.

For each candidate SPIRE galaxy we fit 
{the three-band SPIRE data to} a greybody spectral
energy distribution of the form
\begin{equation}
  S(\nu) = A \times \nu^{1.7} \times B(\nu, T),
\end{equation}
where $B(\nu,T)$ is the Planck blackbody equation and
the normalization $A$ and temperature $T$ are free parameters.
In performing these fits, we make the assumption that this 
greybody parameterization describes the emission within the 
SPIRE PSF regardless of whether the emission is sourced
by a single galaxy or many galaxies.
Consequently, we do not include the effects of 
confusion noise in these fits.
Using the greybody fits to the SPIRE data, we estimate
the 268~GHz flux density centered on each of the 200
SPIRE candidates.
{We note that, probably due to the fact that multiple
sources above the SPIRE measurement noise RMS
are likely to be present within the extent of the SPIRE PSF,
this simple greybody model does not provide an adequate
fit to all of the SPIRE source candidates.
Specifically, 1/3 of candidates produce a fit PTE~$<0.05$,
indicating that a greybody fit does not describe the 
emission detected within the SPIRE PSF for those 
candidates.
In addition, even if we discard these 1/3 of the candidates, 
the distribution of PTE values for the remaining 2/3 of 
the candidates is still marginally inconsistent with a uniform distribution,
quantified by a KS test PTE of 0.03.
This implies that a greybody fit is inadequate to describe a significant
fraction of the SPIRE candidates.
Unfortunately, it is not practical to fit a more complicated
model to the 3-band SPIRE data alone due to the lack
of spectral information, and so there is no clear model
extension to obtain a better fit quality for the 
SPIRE candidates that are not adequately described
using a greybody model.
We discuss the implications of this modeling limitation
in more detail below.}

\begin{deluxetable}{cccccc}
  \tablecaption{Unresolved Sources Detected by Bolocam at 268 GHz}
  \tablehead{\colhead{RA (J2000)} & \colhead{dec (J2000)} & 
             \colhead{flux density (mJy)} & \colhead{de-boosted (mJy)} & 
             \colhead{SPIRE extrapolated (mJy)} & \colhead{SPIRE distance ($\arcsec$)}} \\
  \startdata
    7:17:19.4 & 37:46:41 & $7.6 \pm 0.9$ & $6.9 \pm 0.8$ & $6.2 \pm 0.5$ & 11 \\
              &          &                 &               & $1.6 \pm 0.2$ & 11 \\
              &          &                 &               & $0.4 \pm 0.1$ & 28 \\ 
    7:17:53.4 & 37:42:55 & $5.2 \pm 1.2$ & $3.7 \pm 1.4$ & $2.2 \pm 0.1$ & 12 \\ 
              &          &                 &               & $1.4 \pm 0.1$ & 15 \\ 
    7:17:24.3 & 37:41:07 & $5.3 \pm 1.1$ & $4.0 \pm 1.2$ & $3.7 \pm 0.6$ & 11 \\ 
    7:17:37.3 & 37:42:24 & $3.9 \pm 0.9$ & $3.2 \pm 1.0$ & $1.2 \pm 0.1$ & 21 \\
              &          &                 &               & $2.5 \pm 0.5$ & 25 \\ 
    7:17:32.2 & 37:42:32 & $3.6 \pm 0.9$ & $3.0 \pm 1.0$ & $0.5 \pm 0.1$ & 22 \\ 
              &          &                 &               & $0.8 \pm 0.1$ & 23 \\ 
    7:17:11.2 & 37:50:34 & $6.3 \pm 1.4$ & $4.0 \pm 1.7$ & $3.0 \pm 0.1$ & 17 \\ 
    7:17:28.8 & 37:40:02 & $5.3 \pm 1.3$ & $3.5 \pm 1.6$ & $5.3 \pm 0.3$ & 7 \\ 
    7:17:33.8 & 37:48:13 & $3.5 \pm 0.9$ & $2.9 \pm 1.0$ & $2.9 \pm 0.5$ & 16 \\
              &          &                 &               & $0.6 \pm 0.1$ & 22
  \enddata
  \tablecomments{Unresolved sources that we detect with a S/N~$\ge 4$ in the 
    Bolocam 268~GHz data. 
    From left
    to right, the columns give the Bolocam RA, Bolocam dec, Bolocam flux density,
    noise-de-boosted Bolocam flux density, extrapolated flux density at 268~GHz
    based on our greybody fit to measurements from the 3 SPIRE bands,
    and distance from the Bolocam detection to the SPIRE detection.
    Note that the SPIRE measurements are dominated by confusion noise,
    which is highly correlated between the bands and is not included
    in the greybody fit uncertainties.
    Consequently, the uncertainties on the greybody fits assume that
    there is either a single source contributing all of the signal
    in the three SPIRE bands, or that the superposition of 
    sources can be adequately described by a single greybody.
    For each of these eight candidate sources, the sum of the
    SPIRE greybody fits is consistent with the de-boosted
    Bolocam flux density, indicating that the Bolocam candidates
    are likely robust detections.}
  \label{tab:point_sources}
\end{deluxetable}

\begin{figure}
  \centering
  \includegraphics[width=0.4\textwidth]{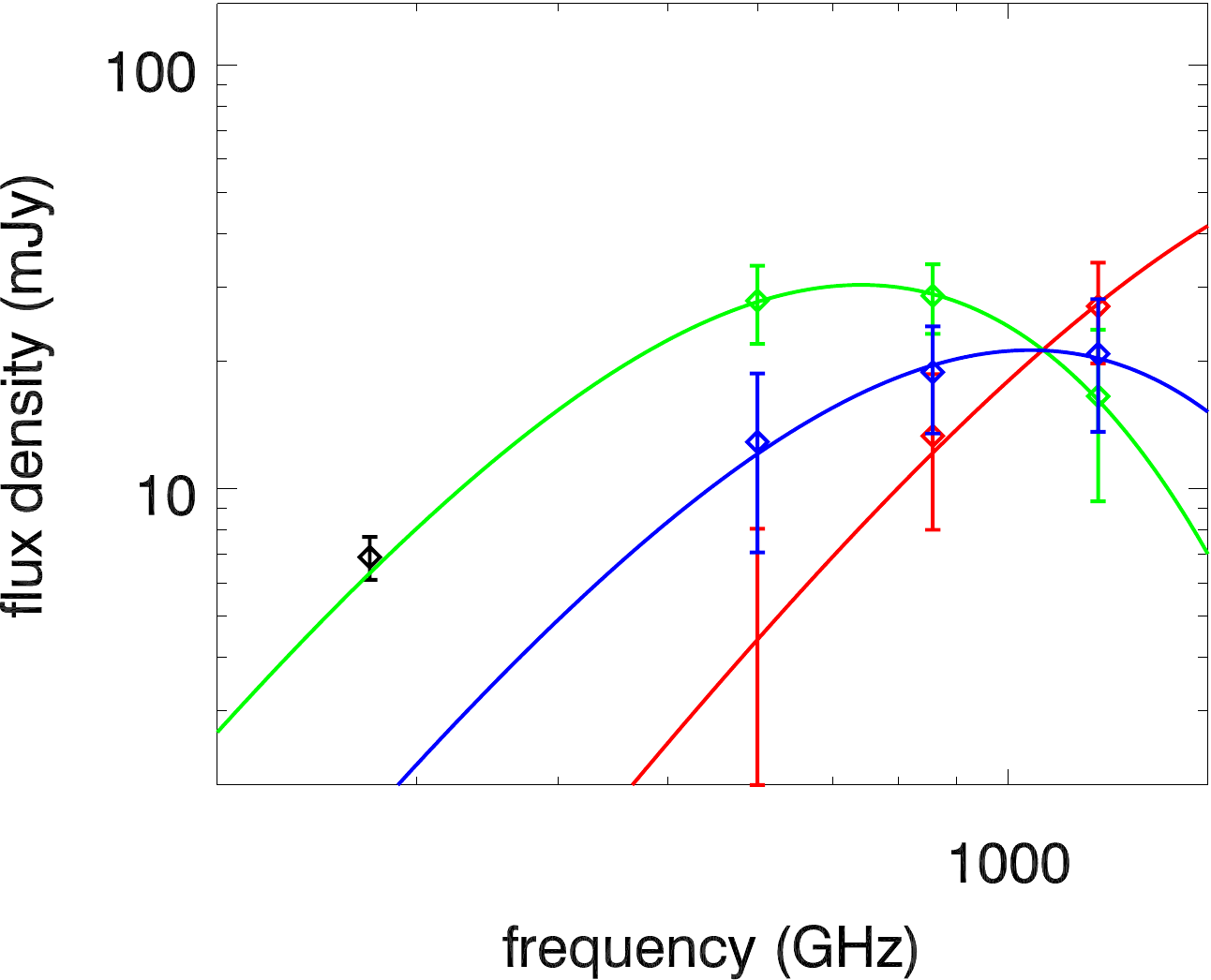}
  \hspace{.05\textwidth}
  \includegraphics[width=0.4\textwidth]{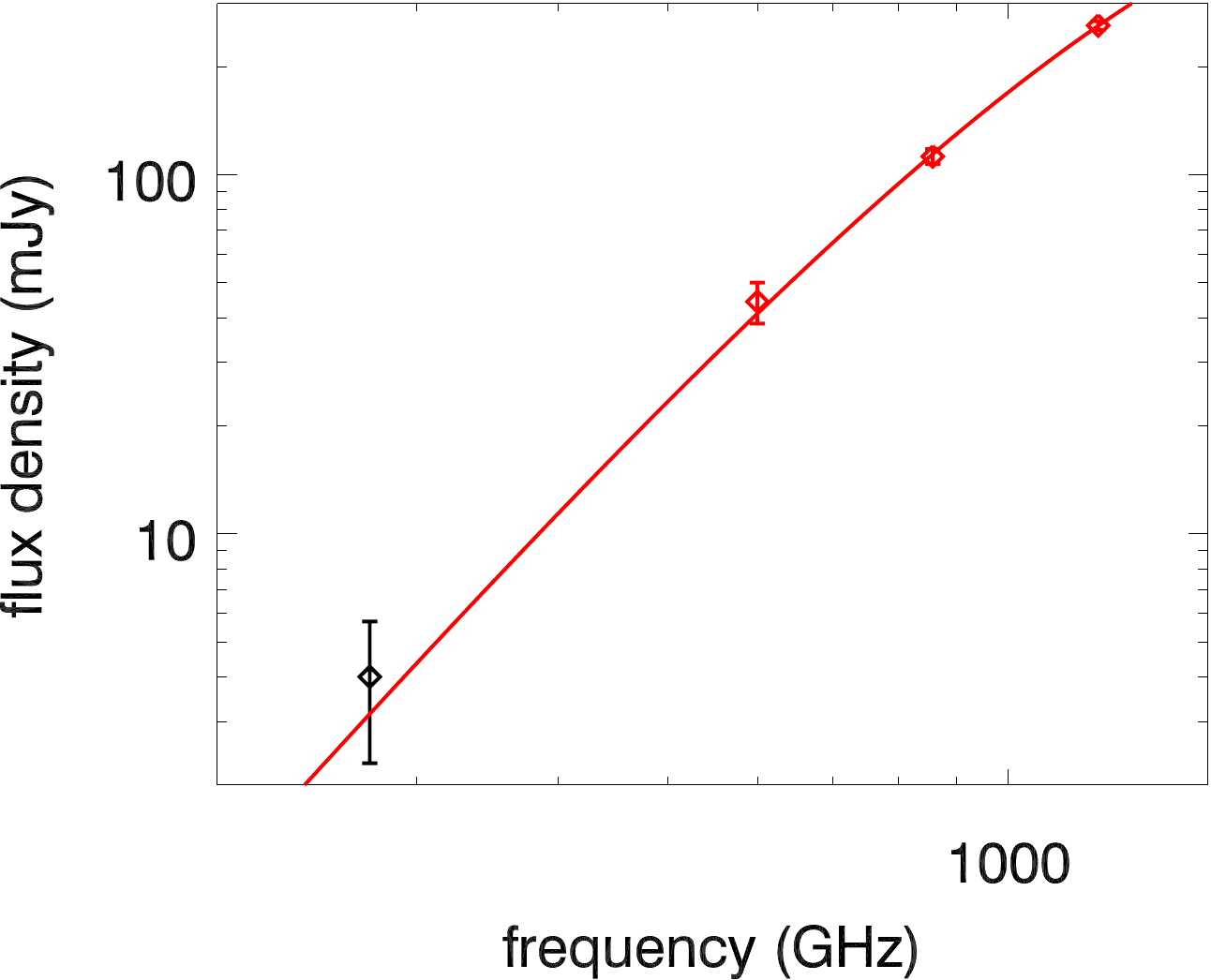}
  \caption{Two of the eight candidate galaxies detected by Bolocam.
  In each plot the de-boosted Bolocam flux density is shown as a 
  black diamond, and the SPIRE flux densities are shown
  as colored diamonds with a different color for each
  possible counterpart.
  The lines show greybody SED fits to the SPIRE data
  only, with a separate fit for each possible counterpart.}
  \label{fig:SED}
\end{figure}

We find a total of 14 of these SPIRE candidates are located 
within $30\arcsec$ of the 8 Bolocam candidates, indicating
that they are possible counterparts.
To compare the Bolocam and SPIRE measurements, 
we first estimate the best-fit noise-de-boosted flux density 
for the 8 candidates detected by Bolocam 
according to the AzTEC de-boosting algorithm presented in
\citet{austermann09} by interpolating the tabulated values
presented in \citet{downes12}.
As shown in Figure~\ref{fig:SED} and Table~\ref{tab:point_sources},
we in general find good agreement between the de-boosted Bolocam
flux densities and the sum of the extrapolated SPIRE flux
densities for all of the likely counterparts.
Therefore, the Bolocam candidates are
robust detections.

Discarding the 14 SPIRE candidates that are likely
counterparts of the Bolocam candidates because they have
already been subtracted from the common-mode-subtraction map,
along with 24 SPIRE
candidates that have an extrapolated 268~GHz S/N $<2$,
we then generate
an image from the remaining 162 SPIRE candidates 
using the extrapolated 268~GHz SPIRE flux densities
convolved with the Bolocam PSF.
This image is then processed through the adaptive-PCA reduction,
and we subtract it from the corresponding 
extended-SZ-and-Bolocam-candidate-subtracted adaptive-PCA map
(see Figure~\ref{fig:ps_image}).
Removal of this SPIRE template, which we generate independently from 
our Bolocam data, results in a significant
reduction in the rms of the Bolocam image, with a $\Delta \chi^2 = 166$.
To put this value in context, the total $\chi^2$ of the SPIRE
template is 240, and 
therefore a perfect correlation between
the SPIRE extrapolations and the Bolocam data would have resulted
in a $\Delta \chi^2 = 240$.
As a further quality check, we compute the normalization of the SPIRE
template that best fits our 268~GHz Bolocam 
extended-SZ-and-Bolocam-candidate-subtracted adaptive-PCA map
and find a value of $0.84 \pm 0.09$
including flux calibration uncertainties.  
The consistency of this value with unity indicates that the SPIRE template
is a good description of the CIB in the 268~GHz Bolocam data.

\begin{figure}
  \centering
  \includegraphics[width=.45\textwidth]{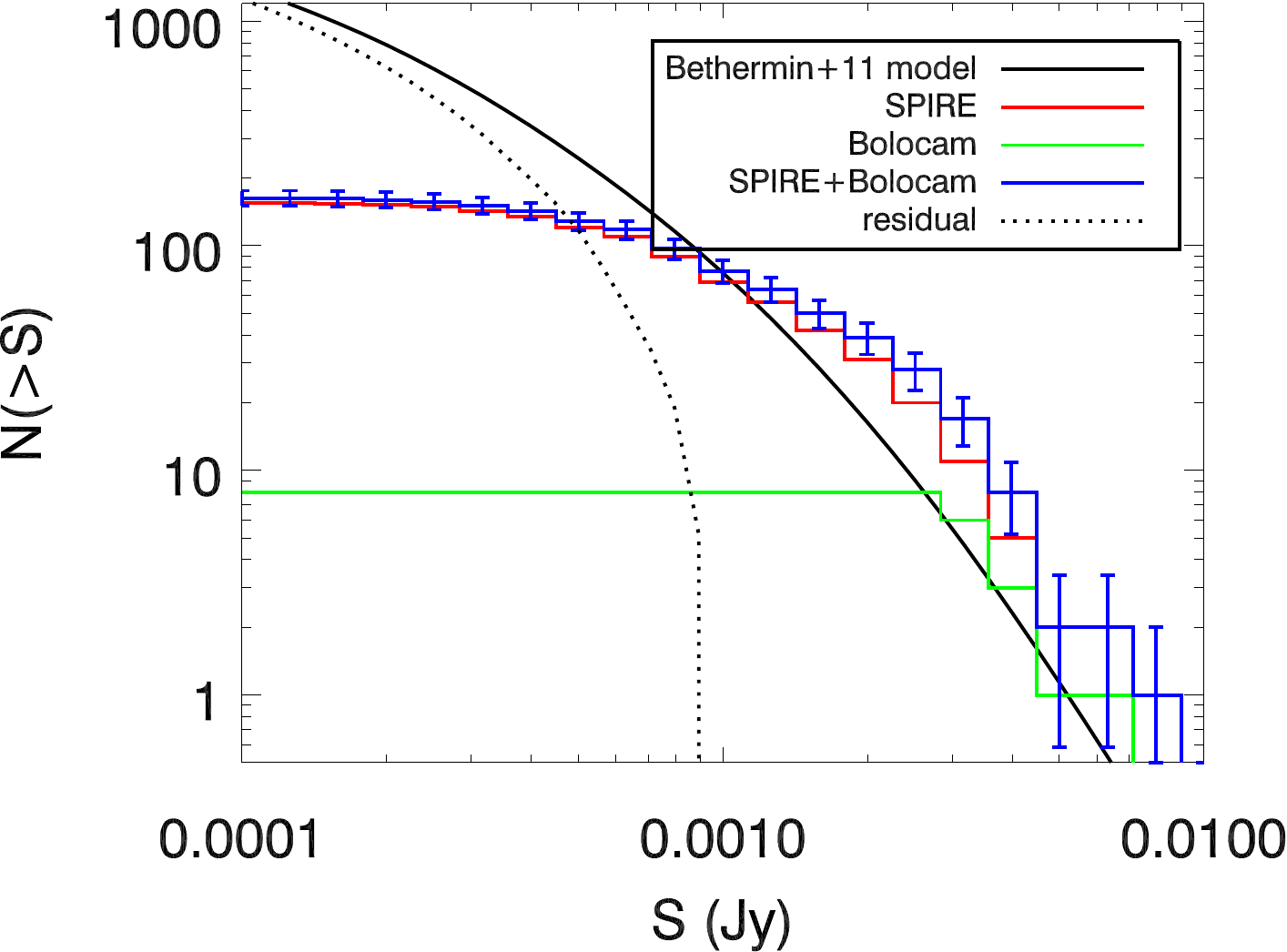}
  \hspace{.05\textwidth}
  \includegraphics[width=.45\textwidth]{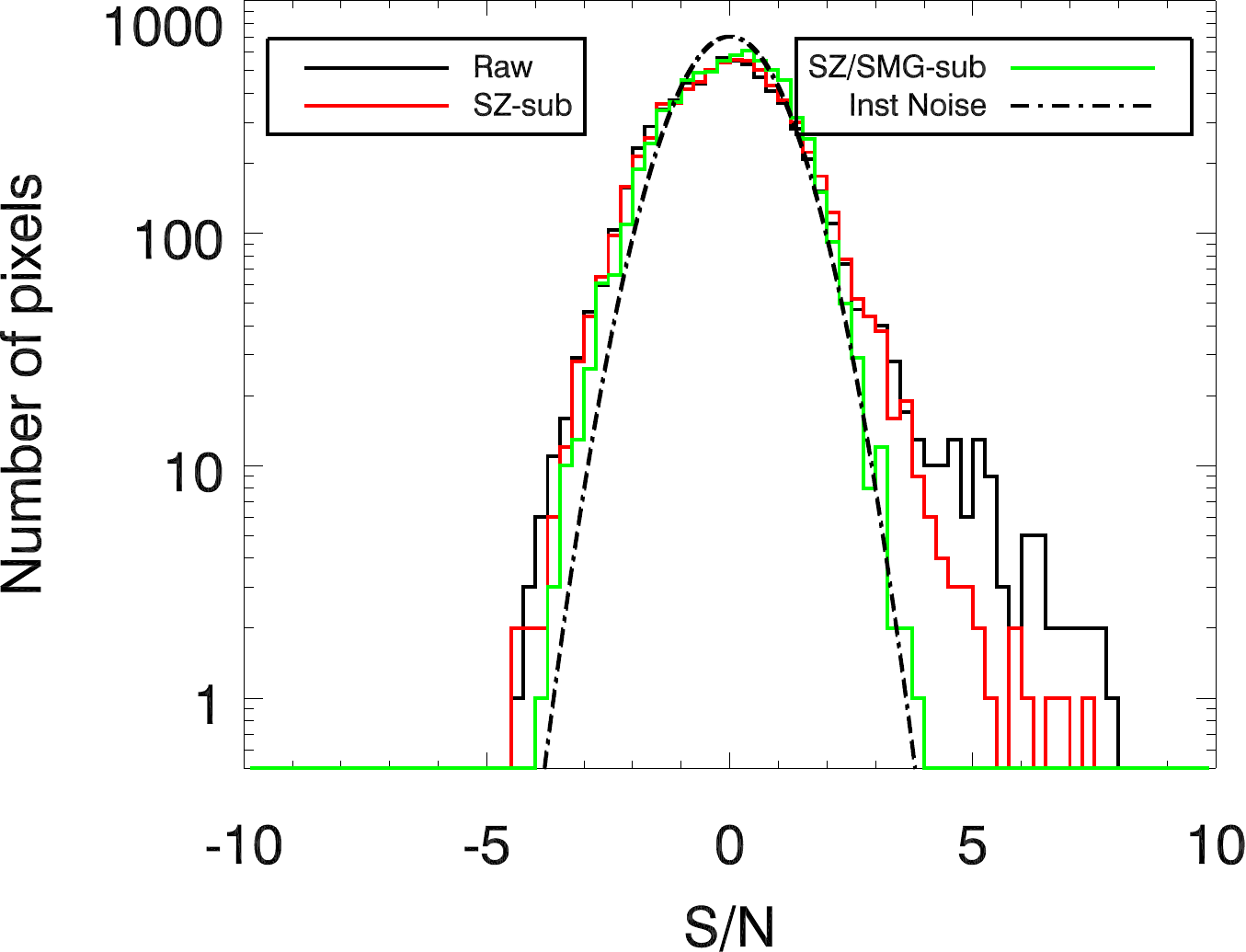}
  \caption{Left: number of galaxies above a given 268~GHz flux density
    within the $14\arcmin \times 14\arcmin$ Bolocam image.
    The solid black line denotes the \cite{bethermin11}
    model prediction, green denotes the Bolocam detections, 
    red denotes the extrapolated SPIRE detections
    after removing possible counterparts to the Bolocam detections
    along with S/N~$<2$ extrapolations,
    blue denotes all Bolocam and SPIRE detections, and 
    the dashed black line is the difference between the
    model and our total detections.
    Right: S/N histogram for the 268~GHz Bolocam 
    adaptive-PCA map.
    Black shows the histogram prior to any signal subtraction,
    red shows the histogram after subtraction of the best-fit
    extended-SZ model, green shows the histogram after further subtraction
    of the eight sources detected by Bolocam and the 
    162 sources detected by SPIRE and extrapolated to
    268~GHz, and the dot-dashed black line shows the
    histogram of our jackknife realizations,
    which contain all of the non-astronomical noise
    present in our data.
    The difference in width between the green histogram
    and the dot-dashed black histogram matches the
    prediction of the \citet{bethermin11} model,
    indicating that it provides an adequate description of our data.}
  \label{fig:n_s}
\end{figure}

To subtract these 162 SPIRE-detected sources from our 
common-mode-subtraction map, 
we process the SPIRE template through the common-mode reduction
and subtract the resulting map from the corresponding common-mode-subtraction
map we use for SZ analysis.
We also generate 1000 map realizations with flux densities for each source
randomly distributed according to the uncertainty on the extrapolation
and add one such realization to each of the 1000 noise realizations.
As described above, 
the likely Bolocam counterparts are not included in this 
SPIRE-source template
to avoid potential double subtractions, and the S/N~$<2$
extrapolations are removed to prevent the subtraction
of possible false SPIRE detections.
We estimate that, between the Bolocam and SPIRE detections, we 
remove all of the dusty galaxies with a 268~GHz flux
density $\gtrsim 1$~mJy along with $\simeq 100$ galaxies
with lower flux densities, significantly reducing the contamination
from these sources on our measurement of the SZ signal
(see Figure~\ref{fig:n_s}, left).

{Due to the fact that the SPIRE candidates are 
in general not well described by the greybody model, the derived 
uncertainty on the extrapolation is likely to be underestimated.
Therefore, to determine the effect that this underestimate might
have on our SZ results, we 
artificially increase the uncertainties on the SPIRE
photometry for each candidate until the reduced $\chi^2$ of the
greybody fit is equal to 1.
For candidates with a reduced $\chi^2 < 1$, the uncertainties are
left unchanged.
Although this procedure is unphysical, it does provide a reasonable
basis for estimating the uncertainty on the extrapolation for
these sources.
We find that the difference between the estimated uncertainties on the 
268~GHz SZ brightnesses
for sub-clusters B and C with and without including this artificial increase
on the extrapolation uncertainty is $< 4 \times 10^{-4}$~MJy sr$^{-1}$,
or $\lesssim 1$\% of the nominal uncertainties.
Therefore, we have not accounted for the potentially underestimated
uncertainties on the SPIRE candidates that are poorly fit
by the greybody model.}

{In addition, we also
estimate the uncertainty on the extrapolation due to our particular
choice of greybody model. Specifically, we fixed the spectral
index of the greybody to 1.7, while the measured 
spectral indices from large source catalogs vary between $\simeq 1 - 2$
(e.g., \citealt{roseboom13}).
To bound the maximum possible change in the extrapolated CIB template due
to this source variation, we reran our analysis using spectral
indices of 1.1 and 2.1, which correspond to the $1\sigma$ bounds
determined by \citet{roseboom13}.
Using the templates determined from these spectral indices, the 268~GHz
SZ brightnesses toward sub-clusters B and C change by  
$0.001 - 0.005$~MJy sr$^{-1}$, or by as much as 10\% of their
nominal uncertainties.
When added in quadrature with our overall uncertainties this amount
is negligible, and we therefore have not included it in our
uncertainty estimates.}

{Furthermore, we note that the best-fit amplitude of our nominal CIB
template using Bolocam data is $0.84 \pm 0.09$, hinting that
the SPIRE data may be producing a template that is slightly
over-predicting the CIB at 268~GHz.
This could be caused by, for example, noise-boosted flux densities
in the SPIRE data or an incorrect choice of SED template.
To estimate the effects of such a potential bias, we re-estimated the
268~GHz SZ brightnesses of sub-clusters B and C after subtracting
an extrapolated SPIRE template with the best-fit Bolocam
normalization of 0.84 (rather than 1).
Using this renormalized template results in 
SZ brightnesses that differ by $\lesssim 0.001$~MJy sr$^{-1}$
compared to those found with the default normalization of the template,
or by approximately 2\% of our nominal uncertainties.
Therefore, the bias associated with a potential over-estimate
of the CIB template is negligible, and we have not attempted
to account for such a bias in our analysis.}

Bolocam and SPIRE do not individually detect the faint galaxies that comprise most of the
CIB. Consequently, we use the 
CIB model determined by \cite{bethermin11}
to account for the noise fluctuations from these
undetected galaxies, as it provides a reasonable estimate for the behavior of the
component of the CIB below the SPIRE detection limit.
We model the undetected CIB using a number counts distribution obtained
by subtracting the number counts already detected by SPIRE and
Bolocam from the \citet{bethermin11} number counts model.
We then generate 1000
random sky realizations of the sources in this population,
process each such realization through the common-mode reduction,
and add one realization of this faint CIB model to each of our 1000 noise realizations.

We note that the \cite{bethermin11} model was calibrated using observations
of blank sky and therefore is not necessarily an accurate
description of the CIB toward a massive cluster like 
MACS J0717.5+3745.
Although the emission from cluster-member galaxies at 268~GHz is likely
to be negligible compared to the background CIB,\footnote{
  For example, see the arguments presented in 
  Section 4 of \citet{sayers13_radio} based on the results presented in
  \citet{geach06, bai07, marcillac07, finn10, rawle12}.}
the significant magnification of the background due to gravitational
lensing can distort the number counts.
In particular, lensing preserves the total surface brightness of the CIB,
but causes a significant and spatially dependent change in the number counts \citep{zemcov13}.
Our data show hints of this change as an excess of sources at bright flux densities,
which is consistent with measurements toward massive clusters using AzTEC
at the same wavelength \citep{wardlow10, downes_thesis}. It is
therefore not clear how well the unlensed \citet{bethermin11}
model describes the faint population of dusty star-forming galaxies
toward MACS J0717.5+3745.

To test the validity of the \cite{bethermin11} model in describing our
MACS J0717.5+3745 data, we add a random sky
realization to each jackknife realization of the adaptive-PCA map, 
where the sky realizations are based on the aforementioned difference
between the \citet{bethermin11} model and our detected number counts.
We find that adding these random sky realizations increases the 
noise rms by 12.3\%.
We then fit a Gaussian to the distributions of pixel S/N values
for the adaptive-PCA map jackknife realizations, and to the actual
data after subtraction of the extended-SZ template, the Bolocam detections,
and the extrapolated SPIRE detections
(see Figure~\ref{fig:n_s}, right).
We find that the Gaussian standard deviations returned
by the fits differ by $11.9 \pm 0.8$\%, in excellent agreement
with the prediction based on the \citet{bethermin11} model. 
Therefore, the \citet{bethermin11} model provides a good
description of the global noise fluctuations due to dim unresolved
galaxies at the level we are able to ascertain with our Bolocam data.

{In principal, we could subtract some of the signal from these
dim unresolved galaxies via e.g., a cross correlation analysis between the
SPIRE and Bolocam maps. Although these dim galaxies are not detected in
any single SPIRE band, their signal will be correlated across the
ensemble of SPIRE and Bolocam bands.
However, the SPIRE data, particularly at 500~$\mu$m, contains
a non-negligible amount of diffuse SZ signal. This signal would also be
correlated across the multiple bands, and therefore such an analysis
could subtract SZ signal from the Bolocam data in addition to CIB signal.
Furthermore, the effective reddening of the CIB due to the SZ signal in the
SPIRE data could potentially cause a significant over-estimate of the
signal when extrapolated to the Bolocam bands.
Consequently, to mitigate these effects, 
we have chosen to subtract only the unresolved bright galaxies
individually detected by SPIRE, which should not be significantly
contaminated by the diffuse SZ signal.
Although beyond the scope of this work, an optimal analysis would jointly
constrain a model of the SZ and CIB signals via a simultaneous fit
to both the SPIRE and Bolocam data. However, given the relative
dimness of the SZ signal in the SPIRE bands, along with the
relative dimness of the CIB signal in the Bolocam bands, the 
improvement from such a joint fit is likely to be minimal compared
to our analysis procedure.}

We note that the fluctuations from the CIB are accounted for
in our 140~GHz data by adding noise realizations generated
according to the power spectrum measurements from SPT \citep{hall10}.
We do not attempt to subtract any individual galaxies due to the
large, and consequently potentially untrustworthy, spectral extrapolation
that would be required from the SPIRE measurements.
In addition, as noted above, there are very
few galaxies detected within the extended region containing bright SZ signal, and none of those
galaxies are particularly bright.
We note that this is in general true for massive clusters, as 
shown in \citet{zemcov13}.
{However, to verify that the CIB is sufficiently
dim that we can neglect subtracting it at 140~GHz, we 
extrapolated the SPIRE detections to 140~GHz and 
removed them from the Bolocam data.
Even if this large extrapolation is potentially untrustworthy,
it should provide a reasonable estimate of the potential 
brightness of the CIB at 140~GHz.
When we subtract this extrapolated CIB, we find that
the best-fit 140~GHz SZ brightnesses toward both sub-cluster B and 
sub-cluster C changes by $\simeq 5 \times 10^{-4}$~MJy sr$^{-1}$, or by approximately
1\% of the total uncertainty on each sub-cluster's SZ brightness.
Therefore, neglecting to subtract the CIB at 140~GHz has
a negligible effect on our overall results.}

\end{document}